\documentclass[
    prd, twocolumn, superscriptaddress, nofootinbib, amsfont, amsmath, amssymb, preprintnumbers,
]{revtex4-2}
\usepackage{booktabs}
\usepackage{threeparttable}
\usepackage{graphicx}
\usepackage{dcolumn}
\usepackage{bm}
\usepackage{mathtools}
\usepackage{ulem}
\usepackage{color,txfonts}
\usepackage{xcolor}
\definecolor{blue0}{rgb}{0,0,0.6}
\usepackage[colorlinks,linkcolor=blue0,anchorcolor=blue0,citecolor=blue0,urlcolor=blue0]{hyperref}

\begin{document}

\title{Constraints on axionlike particles from 16.5 years of Fermi-LAT data and prospects for VLAST}

\author{Zhi-Qi Guo}
\email{zhiqiguo@njnu.edu.cn}
\affiliation{Department of Physics and Institute of Theoretical Physics, Nanjing Normal University, Nanjing, 210023, China}
\affiliation{Key Laboratory of Dark Matter and Space Astronomy, Purple Mountain Observatory, Chinese Academy of Sciences, Nanjing 210033, China}

\author{Yue-Lin Sming Tsai}
\email{smingtsai@pmo.ac.cn}
\affiliation{Key Laboratory of Dark Matter and Space Astronomy, Purple Mountain Observatory, Chinese Academy of Sciences, Nanjing 210033, China}
\affiliation{School of Astronomy and Space Science, University of Science and Technology of China, Hefei, Anhui 230026, China}

\author{Lei Wu}
\email{leiwu@njnu.edu.cn}
\affiliation{Department of Physics and Institute of Theoretical Physics, Nanjing Normal University, Nanjing, 210023, China}
\affiliation{Nanjing Key Laboratory of Particle Physics and Astrophysics, Nanjing, 210023, China}

\author{Zi-Qing Xia}
\email{xiazq@pmo.ac.cn}
\affiliation{Key Laboratory of Dark Matter and Space Astronomy, Purple Mountain Observatory, Chinese Academy of Sciences, Nanjing 210033, China}

\date{\today}

\begin{abstract}

Axionlike particles (ALPs), hypothetical particles beyond the Standard Model (SM), are considered as promising dark matter candidates.
ALPs can convert into photons and vice versa in a magnetic field via the Primakoff effect, potentially generating detectable oscillation in $\gamma$-ray spectra.
This study analyzes 16.5 years of data from the Fermi Large Area Telescope (Fermi-LAT) on NGC 1275, the brightest galaxy in the Perseus cluster, to constrain the ALP parameter space.
Our results improve the previous 95\% exclusion limits of the photon-ALP coupling $g_{a\gamma}$ by a factor of 2 in the ALP mass range of $4\times 10^{-10}\,\mathrm{eV}\lesssim m_{a}\lesssim 5\times 10^{-9}\,\mathrm{eV}$.
Moreover, we investigate the projected sensitivity of the future Very Large Area $\gamma$-ray Space Telescope (VLAST) on searching for ALPs.
We find that (i) the expected sensitivity on the photon-ALP coupling $g_{a\gamma}$ can be stronger than that from the upcoming International Axion Observatory (IAXO) in the ALP mass range of $2\times 10^{-11}\,\mathrm{eV}\lesssim m_{a}\lesssim 1\times 10^{-7}\,\mathrm{eV}$, with the best sensitivity of $g_{a\gamma}\sim 7\times 10^{-13}\,\mathrm{GeV^{-1}}$ at $m_{a}\sim 2\times 10^{-10}\,\mathrm{eV}$; (ii) VLAST can extend the sensitivity of the ALP masses below $5\times 10^{-12}\,\mathrm{eV}$, where the photon-ALP coupling $g_{a\gamma}\gtrsim 1.5\times 10^{-11}\,\mathrm{GeV^{-1}}$ will be excluded; (iii) the entire parameter space of ALP accounting for TeV transparency can be fully tested.
These results demonstrate that VLAST will offer an excellent opportunity for ALPs searches.
\end{abstract}

\pacs{Valid PACS appear here}


\maketitle

\section{\label{sec:Introduction}Introduction}

The strong \textit{CP} problem in quantum chromodynamics (QCD) motivated the proposal of the axion, a hypothetical particle beyond the Standard Model (SM)~\cite{Peccei:1977hh,Weinberg:1977ma,Wilczek:1977pj,Peccei:2006as}, with the mass postulated to be directly proportional to its coupling with photons.
More general axionlike particles (ALPs) predicted in many string-theory-motivated extensions of the SM~\cite{Svrcek:2006yi,Arvanitaki:2009fg,Cicoli:2012sz}, share similarities with axions but are not necessarily tied to the strong \textit{CP} problem, allowing independent variation of mass and coupling.~\footnote{For convenience, we use ALPs to refer to the axions and ALPs in the following context.}
Besides, the ALPs have been proposed as one of the promising dark matter candidates, whose cosmological abundance can be naturally consistent with observational data via the nonthermal production mechanism~\cite{Preskill:1982cy,Abbott:1982af,Dine:1982ah,Hwang:2009js,Duffy:2009ig,Arias:2012az,Chadha-Day:2021szb}.
Additionally, ALPs offer a potential solution to various astrophysical anomalies, such as the apparent transparency of TeV photons from extragalactic sources~\cite{DeAngelis:2007dqd,Simet:2007sa,Horns:2012fx,Meyer:2013pny}.
These motivate the great efforts to detect ALPs through ground-based experiments and astronomical observations ~\cite{Carosi:2013rla,Graham:2015ouw,PhysRevD.93.045019,Irastorza:2018dyq,Majumdar:2018sbv,DiLuzio:2020wdo,Choi:2020rgn,Chen:2021lvo,PhysRevD.106.083006,2022JHEP...10..141A,Guo:2023hyp,Gong:2023ilg,Arza:2023rcs,Li:2024ivs,Song:2024rru,Lella:2024dmx,Yang:2024jtp,Zhu:2024kmu,FASER:2024bbl,Chen:2024ekh,Guo:2024oqo,Malyshev:2025iis,Pankratov:2025cby}.

The majority of ALP searches rely on the Primakoff effect, which utilizes the conversion between ALPs and photons in the presence of an external magnetic field.
The relevant interaction Lagrangian is given by
\begin{equation}
\mathcal{L}=\frac{1}{4}g_{a\gamma}aF^{\mu\nu}\tilde{F}_{\mu\nu}=g_{a\gamma}aE\cdot B,
\label{eq:1}
\end{equation}
where $g_{a\gamma}$ is the photon-ALP coupling, ${F}^{\mu\nu}$ is the electromagnetic field tensor, $\tilde{F}_{\mu\nu}$ is its dual tensor, $E$ is the electric field, $B$ is the magnetic field, and $a$ is the ALPs field.

Ground-based experiments can be categorized into three main types.
The first, light-shining-through-walls experiments, such as the any light particle search at Deutsches Elektronen Synchrotron~\cite{Ehret:2010mh,Ortiz:2020tgs}, aim to detect ALPs by shining light through a barrier and observing its conversion into axions.
The second type focuses on axion dark matter detection, exemplified by the Axion Dark Matter eXperiment (ADMX)~\cite{ADMX:2009iij,ADMX:2018gho,ADMX:2018ogs,ADMX:2019uok,Crisosto:2019fcj,ADMX:2021mio,ADMX:2021nhd,ADMX:2024xbv}, which uses a strong magnetic field to convert dark matter axions into detectable microwave photons.
The third type involves solar axion detection, represented by the Axion Solar Telescope (CAST) at CERN~\cite{CAST:2017uph} and the upcoming International Axion Observatory (IAXO)~\cite{IAXO:2019mpb}, which aims to greatly enhance sensitivity to solar axions.

Astronomical observations, especially with $\gamma$-ray telescopes, have been believed to be capable of advancing the search for ALPs and effectively narrowing down the viable parameter space for ALPs~\cite{Hooper:2007bq,DeAngelis:2007wiw,Meyer:2016wrm,Benabou:2025jcv}.
Using six years of the Fermi Large Area Telescope (Fermi-LAT) observation on NGC 1275, the couplings above $5\times 10^{-12}\,\mathrm{GeV^{-1}}$ have been excluded for ALP mass $5\times 10^{-10}\,\mathrm{eV}\lesssim m_{a}\lesssim 5\times 10^{-9}\,\mathrm{eV}$~\cite{Fermi-LAT:2016nkz}.
Similarly, the $\gamma$-ray observation for SN 1987A rules out the couplings above $2\times 10^{-13}\,\mathrm{GeV^{-1}}$ for ALP mass $m_{a}\lesssim 1\times 10^{-9}\,\mathrm{eV}$~\cite{2015JCAP...02..006P}.
Further constraints come from the $\gamma$-ray spectra of extragalactic core-collapse supernovae and bright Fermi flat-spectrum radio quasars (3C 454.3, 3C 279, and CTA 102)~\cite{Meyer:2020vzy,Davies:2022wvj}.
The High Energy Stereoscopic System (HESS) collaboration, using observation from PKS 2155$-$304, excludes the couplings above $2.1\times 10^{-11}\,\mathrm{GeV^{-1}}$ for ALP mass $1.5\times 10^{-8}\,\mathrm{eV}\lesssim m_{a}\lesssim 6\times 10^{-8}\,\mathrm{eV}$~\cite{HESS:2013udx}.
Notably, Ref.~\cite{Zhang:2018wpc} uses eight years of Fermi-LAT observations on PKS 2155$-$304 to exclude a previously allowed ``hole'' parameter region in Ref.~\cite{Fermi-LAT:2016nkz}.
Based on HESS observations, Ref.~\cite{Liang:2018mqm} analyzes the $\gamma$-ray spectra of TeV sources in the Galactic plane, excluding certain parameter space regions and further providing projected sensitivity for the next-generation Cherenkov Telescope Array.
Reference~\cite{Xia:2019yud} combines GeV (Fermi-LAT) and TeV (MAGIC, VERITAS, and HESS) data from three bright supernova remnants to search for ALPs.
Reference~\cite{Yuan:2020xui} performs high-resolution polarimetric measurements of radiation near the Galactic Center’s supermassive black hole using a subset of the Event Horizon Telescope array, excluding the couplings above $1\times 10^{-12}\,\mathrm{GeV^{-1}}$ for ALP mass $1\times 10^{-19}\,\mathrm{eV}\lesssim m_{a}\lesssim 1\times 10^{-18}\,\mathrm{eV}$.

In this work, we employ $\gamma$-ray observations of NGC 1275 to explore the potential for detecting ALPs through the Primakoff effect.
This is because that NGC 1275 is exceptionally bright in the $\gamma$-ray band, with a Fermi-LAT detection significance surpassing 100$\sigma$~\cite{Fermi-LAT:2015bdd}.
Besides, it has a well-understood broadband spectrum that follows a synchrotron self-Compton emission model~\cite{MAGIC:2013mvu,Tavecchio:2014oja}.
Additionally, its location within the Perseus cluster allows for well-constrained magnetic fields through rotation measures~\cite{Taylor:2006ta}.
By utilizing nearly 16.5 years of Fermi-LAT $\gamma$-ray data on NGC 1275, we derive more stringent constraints on the ALP parameter space than that previously reported in Ref.~\cite{Fermi-LAT:2016nkz}.
Moreover, we investigate the detection potential for ALPs using the Very Large Area $\gamma$-ray Space Telescope (VLAST)~\cite{2022AcASn..63...27F}.
As a next-generation $\gamma$-ray observatory with broad energy coverage from MeV to TeV, VLAST is specifically designed to achieve high sensitivity in the GeV$–$TeV range while maintaining observational capabilities in the MeV$–$GeV band.
Its scientific objectives include probing dark matter, studying high-energy time-domain phenomena, investigating cosmic-ray physics, and exploring cosmological questions.
The VLAST Collaboration has established a comprehensive simulation framework based on \texttt{GEANT4}~\cite{GEANT4:2002zbu}.
This framework enables precise evaluation of critical instrument performance parameters including acceptance, effective area, angular resolution, and energy resolution.
Simulation results demonstrate that VLAST achieves a maximum effective acceptance of approximately $12\,\mathrm{m^{2}\,sr}$ with an effective area of about $4\,\mathrm{m^{2}}$ at normal incidence.
This represents a fivefold improvement over Fermi-LAT for energies above 1 GeV. In the MeV energy range, the effective area reaches $0.5\,\mathrm{m^{2}}$. For high-energy events involving pair conversion (above 10 MeV), the instrument achieves an angular resolution ranging from $0.03^{\circ}$ to $20^{\circ}$ and an energy resolution between 1\% and 30\%. For low-energy Compton events (below 10 MeV), the corresponding values are $3^{\circ}$–$6^{\circ}$ for angular resolution and 8\%–20\% for energy resolution~\cite{2022AcASn..63...27F}.
Our analysis of simulated five-year observations of NGC 1275 indicates that VLAST will provide significantly enhanced sensitivity for ALP detection compared to current facilities.

The paper is structured as follows.
Section~\ref{sec:Photon-ALP Oscillation} details the calculation of the photon survival probability in the photon-ALP conversion process and the magnetic field model.
Section~\ref{sec:Data Analysis} systematically describes the Fermi-LAT data processing pipeline, $\chi^{2}$ analysis, and Monte Carlo simulations.
Section~\ref{sec:Result} presents the Fermi-LAT observational results and the projected sensitivity of VLAST.
Section~\ref{sec:Discussion} discusses potential key factors that may influence our result.
Section~\ref{sec:Summary} provides a comprehensive summary of our main results.

\section{\label{sec:Photon-ALP Oscillation}Photon-ALP Oscillation}

\begin{table*}
\caption{\label{tab:table}Fundamental model parameters, including those for the double $\beta$ model and the magnetic field model.}
\begin{ruledtabular}
\begin{tabular}{ccc}
Parameter                 & Description                                    & Value                                  \\
\colrule
$n_{0_{\mathrm{int}}}$    & the internal central electron density          & $3.9\times 10^{-2}\,\mathrm{cm^{-3}}$  \\
$n_{0_{\mathrm{ext}}}$    & the external central electron density          & $4.05\times 10^{-3}\,\mathrm{cm^{-3}}$ \\
$r_{\mathrm{core_{int}}}$ & the internal cluster core radius               & 80 kpc                                 \\
$r_{\mathrm{core_{ext}}}$ & the external cluster core radius               & 280 kpc                                \\
$\beta_{\mathrm{int}}$    & the internal electron density slope parameter  & 1.2                                    \\
$\beta_{\mathrm{ext}}$    & the external electron density slope parameter  & 0.58                                   \\
$r_{\mathrm{max}}$        & the fiducial cluster radius                    & 500 kpc                                \\
$\Lambda_{\mathrm{min}}$  & the minimum scale of fluctuation,
                            $\Lambda_{\mathrm{min}}=2\pi/k_{\mathrm{max}}$ & 0.7 kpc                                \\
$\Lambda_{\mathrm{max}}$  & the maximum scale of fluctuation,
                            $\Lambda_{\mathrm{max}}=2\pi/k_{\mathrm{min}}$ & 35 kpc                                 \\
$q$                       & the power spectrum index                       & -2.80                                  \\
$B_{0}$                   & the average magnetic field strength
                            at the center of the cluster                   & $10\,\mathrm{\mu G}$                   \\
$\eta$                    & the radial slope                               & 0.5                                    \\
\end{tabular}
\end{ruledtabular}
\end{table*}

\begin{figure*}
\includegraphics[scale=0.4]{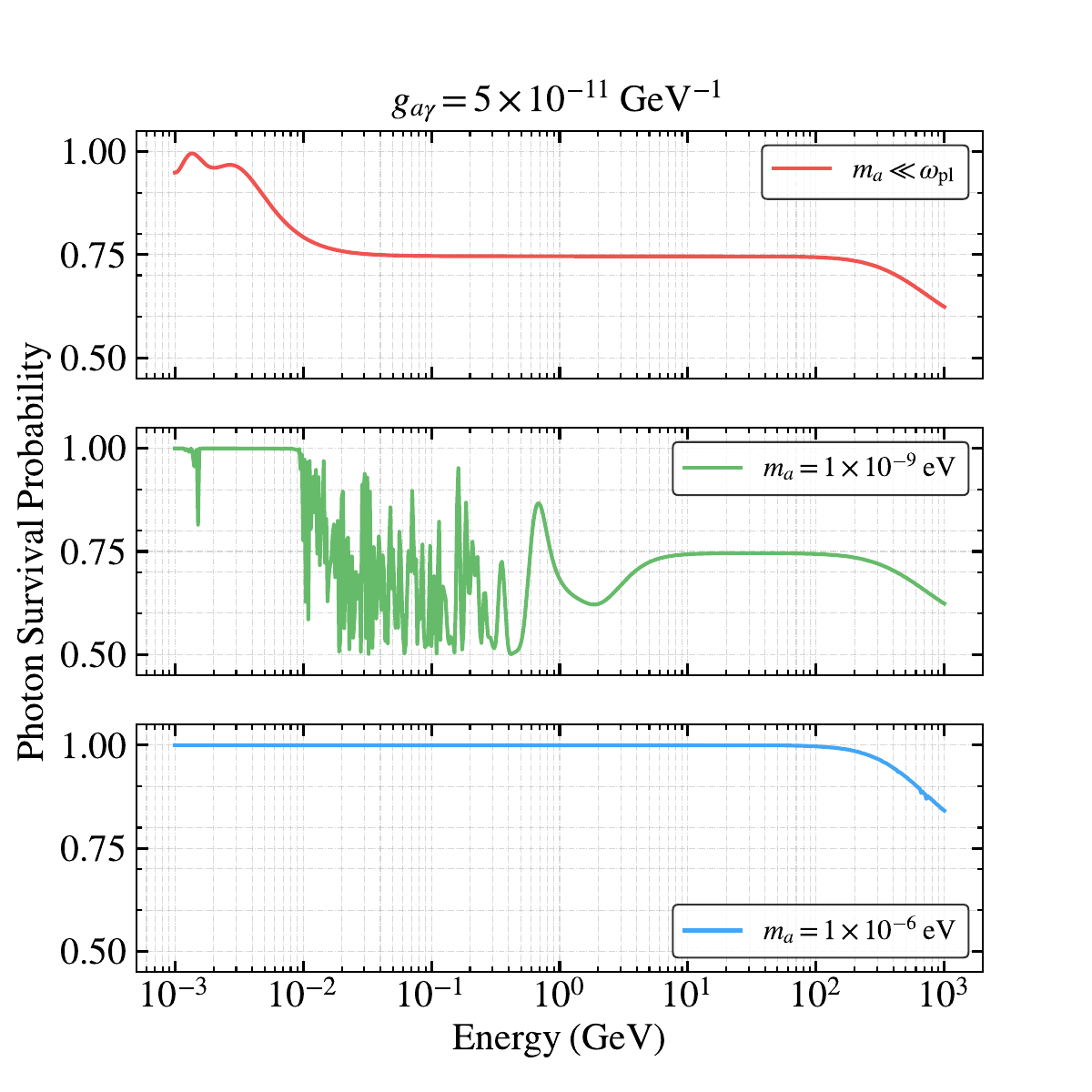}
\includegraphics[scale=0.4]{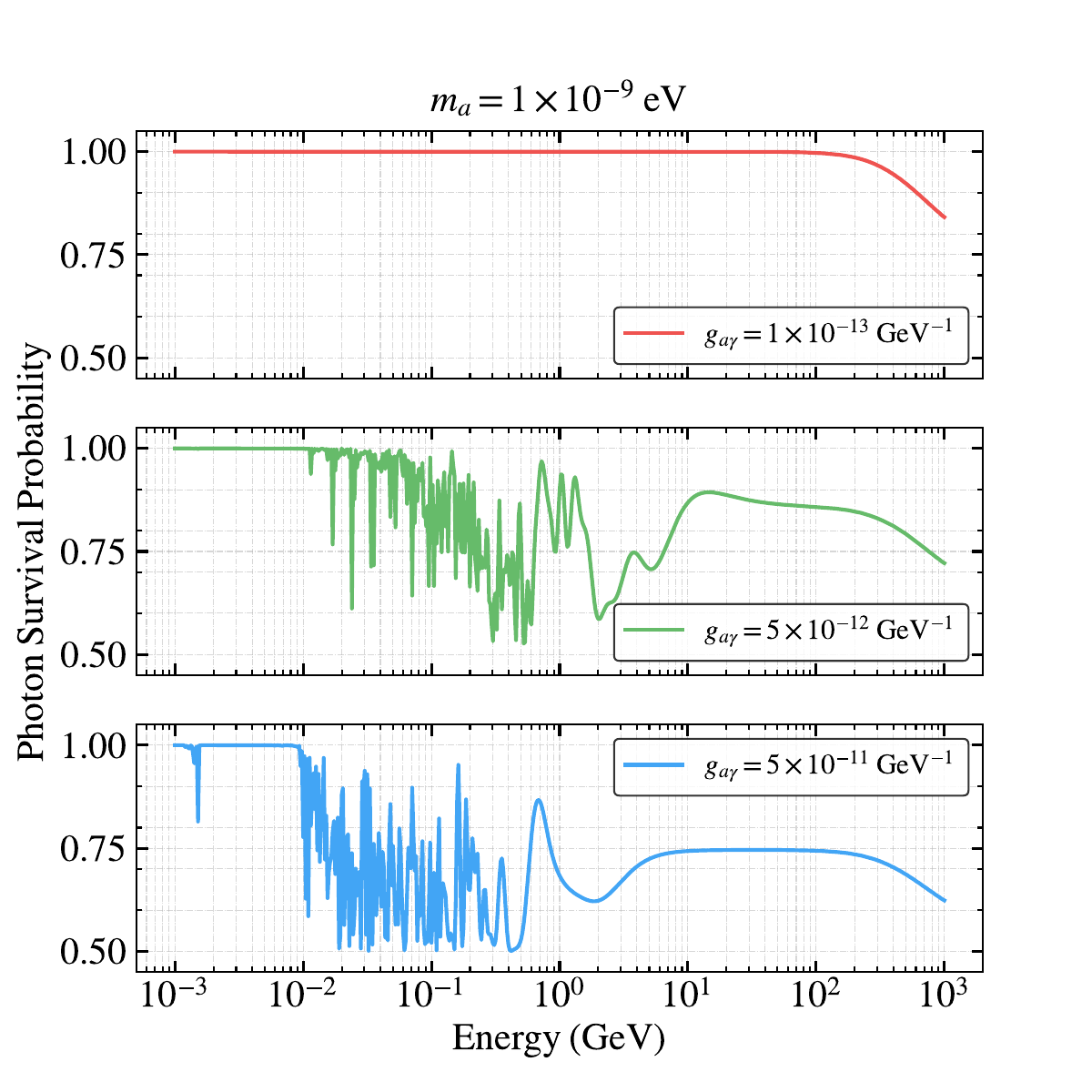}
\caption{Survival probability of photons from NGC 1275 under different ALP parameters based on one realization of the Gaussian turbulent magnetic field, the ``\texttt{Dominguez}'' EBL model and the ``\texttt{Jansson12}'' GMF model. In the left panel, with the fixed coupling ($g_{a\gamma}=5\times 10^{-11}\,\mathrm{GeV^{-1}}$), the photon survival probability is presented for different ALP masses (red solid line: $m_{a}\ll\omega_{\mathrm{pl}}$; green solid line: $m_{a}=1\times 10^{-9}\,\mathrm{eV}$; blue solid line: $m_{a}=1\times 10^{-6}\,\mathrm{eV}$). When the ALP mass is significantly smaller than the plasma frequency, its effect on the photon survival probability becomes negligible. In the right panel, with the ALP mass fixed ($m_{a}=1\times 10^{-9}\,\mathrm{eV}$), the photon survival probability is presented for different couplings (red solid line: $g_{a\gamma}=1\times 10^{-13}\,\mathrm{GeV^{-1}}$; green solid line: $g_{a\gamma}=5\times 10^{-12}\,\mathrm{GeV^{-1}}$; blue solid line: $g_{a\gamma}=5\times 10^{-11}\,\mathrm{GeV^{-1}}$).}
\label{fig:p}
\end{figure*}

Observational evidence indicates the ubiquitous presence of magnetic fields in Perseus clusters (with the typical strength of $B\sim\mathrm{10\,\mu G}$) and the Milky Way ($B\sim\mathrm{1\,\mu G}$).
The $\gamma$-ray photons emitted from NGC 1275 propagate through these magnetic fields and undergo mutual conversions with ALPs via the Primakoff effect.
This process, known as photon-ALP oscillation, can introduce irregular oscillatory features into the observed $\gamma$-ray energy spectra of the sources.
Such features that are difficult to produce through known astrophysical processes, serve as a smoking-gun signature for the existence of ALPs.
Assuming a region of size $\ell$ with a uniform magnetic field, the probability of a photon with energy $E_{\gamma}$ converting into an ALP is given by $P_{\gamma\to a}$~\cite{Raffelt:1987im,Grossman:2002by,Csaki:2003ef,DeAngelis:2007dqd,Hooper:2007bq,Mirizzi:2006zy,Mirizzi:2009aj,DeAngelis:2011id,Meyer:2014epa,Majumdar:2018sbv,Choi:2018mvk}, and the photon survival probability $P\left(g_{a\gamma},m_{a},E_\gamma\right)$ is
\begin{equation}
1-P_{\gamma\to a}=1-\frac{1}{1+\left(E_{c}/E_{\gamma}\right)^{2}}\sin^{2}\left[\frac{g_{a\gamma}B_{\mathrm{T}}l}{2}\sqrt{1+\left(\frac{E_{c}}{E_{\gamma}}\right)^{2}}\right],
\label{eq:2}
\end{equation}
where $B_{\mathrm{T}}$ denotes the magnetic field component along the photon polarization vector, assumed constant within the region.
The characteristic energy $E_{c}$ is defined as
\begin{equation}
E_c=\frac{\left|m_{a}^{2}-\omega_{\mathrm{pl}}^{2}\right|}{2g_{a\gamma}B_{\mathrm{T}}},
\label{eq:3}
\end{equation}
where $\omega_{\mathrm{pl}}$ is the plasma frequency,
\begin{equation}
\omega_{\mathrm{pl}}=\sqrt{4\pi\alpha n_{e}/m_{e}},
\label{eq:4}
\end{equation}
with the electron density $n_{e}$ and the electron mass $m_{e}$.

This conversion process exhibits strong energy dependence, as detailed in Fig.~\ref{fig:p}, 

(i) $\mathbf{E_{\gamma} \ll E_{c}:}$ the photon-ALP oscillation is weak, with a low conversion probability, and the photon survival probability is close to 1.

(ii) $\mathbf{E_{\gamma} \sim E_{c}:}$ photons and ALPs undergo strong mutual conversion, leading to obvious oscillation.

(iii) $\mathbf{E_{\gamma} \gg E_{c}:}$ the conversion probability is still large, however, the photon survival probability in Eq.~\eqref{eq:2} mildly varies with $E_{\gamma}$ and becomes gradual.

To distinguish from known astrophysical processes, we focus on searching for the oscillation features with $E_{\gamma}\sim E_{c}$.
For typical astrophysical and ALP parameters ($B_{\mathrm{T}}=1\,\mu G,n_{e}=0.01\,\mathrm{cm^{-3}}$ for the galaxy clusters or $n_{e}=0.1\,\mathrm{cm^{-3}}$ for the Milky Way, $m_{a}=1\times 10^{-9}\,\mathrm{eV},g_{a\gamma}=1\times 10^{-10}\,\mathrm{GeV^{-1}}$), the resulting critical energy $E_{c}\approx 250\,\mathrm{MeV}$ falls within the optimal detectable energy ranges of both Fermi-LAT and VLAST, rendering them highly suitable for ALPs searches.

To accurately calculate the photon survival probability, it is essential to precisely reconstruct the spatial distribution characteristics of the magnetic field within galaxy clusters.
This study adopts a systematic magnetic field modeling framework, referencing~\cite{Fermi-LAT:2016nkz} and its cited series of works~\cite{Taylor:2006ta,Meyer:2014epa,Dolag:2008ks,Dubois:2008mz,Feretti:2012vk,Churazov:2003hr,MAGIC:2011vay,Kuchar:2009jj,Vacca:2012up}.
First, a double $\beta$ model is employed to describe the spatial distribution of electron density, expressed as 
\begin{equation}
n_{e}\left(r\right)=n_{0_{\mathrm{int}}}\left(1+\frac{r^{2}}{r_{c_{\mathrm{int}}}^{2}}\right)^{-\frac{3}{2}\beta _{\mathrm{int}}}+n_{0_{\mathrm{ext}}}\left(1+\frac{r^{2}}{r_{c_{\mathrm{ext}}}^{2}}\right)^{-\frac{3}{2}\beta _{\mathrm{ext}}},
\label{eq:5}
\end{equation}
where $r$ represents the radial distance from the X-ray centroid of the galaxy cluster.
Second, based on turbulent magnetic field theory, a power-law power spectrum model 
\begin{equation}
M\left(k\right)\propto k^{q}
\label{eq:6}
\end{equation}
is adopted to characterize the magnetic field fluctuations, which is valid within the wave number range $k_{\mathrm{min}}$ to $k_{\mathrm{max}}$, with the power spectrum set to zero outside this interval.
Finally, a correlation model between magnetic field strength and electron density is established as
\begin{equation}
B\left(r\right)=B_{0}\left[n_{e}\left(r\right)/n_{e}\left(r=0\right)\right]^{\eta}.
\label{eq:7}
\end{equation}
Through this series of physical models, the spatial distribution characteristics of the magnetic field in galaxy clusters are comprehensively described, laying the theoretical foundation for the subsequent precise calculation of photon survival probability.

The \texttt{gammaALPs} software package~\footnote{\url{https://github.com/me-manu/gammaALPs}} is employed to calculate the photon survival probability, incorporating: Gaussian turbulent cluster magnetic field models, the ``\texttt{Dominguez}'’ extragalactic background light (EBL) model~\cite{Dominguez:2010bv} and the ``\texttt{Jansson12}'' galactic magnetic field (GMF) model~\cite{Jansson:2012pc}.
The complete set of model parameters is provided in Table~\ref{tab:table}~\cite{Churazov:2003hr,Fermi-LAT:2016nkz,Vacca:2012up}.

We investigate the survival probability of photons from the NGC 1275, under different ALP parameters, as illustrated in Fig.~\ref{fig:p}.
The left panels show the photon survival probability for different ALP masses, with a constant coupling ($g_{a\gamma}=5\times 10^{-11}\,\mathrm{GeV^{-1}}$).
From top to bottom, the ALP mass increases progressively, resulting in the shift of the oscillation energy range:
(i) in the top subplot ($m_{a}\ll\omega_{\mathrm{pl}}$), the oscillation range lies to the left of the observation window of Fermi-LAT (100 MeV$-$500 GeV), but still is in that of VLAST (1 MeV$-$1 TeV);
(ii) in the middle subplot ($m_{a}=1\times 10^{-9}\,\mathrm{eV}$), it falls within both observation windows;
(iii) in the bottom subplot ($m_{a}=1\times 10^{-6}\,\mathrm{eV}$), it shifts to the right, outside both observation windows.
When the ALP mass is much smaller than the plasma frequency (in the top subplot of the left panel), changes in the ALP mass do not affect the trend of photon survival probability according to Eq.~\eqref{eq:2} and Eq.~\eqref{eq:3}.
The right panel presents the photon survival probability for different couplings while keeping the ALP mass fixed ($m_{a}=1\times 10^{-9}\,\mathrm{eV}$).
From top to bottom, the coupling increases gradually.
Based on Eq.~\eqref{eq:2} and Eq.~\eqref{eq:3}, the characteristic energy decreases continuously.
Consequently, the oscillation frequency of the photon survival probability increases, and the oscillation range shifts progressively to the left.
These results reveal the distinct dependencies of the ALP conversion process on the ALP mass $m_{a}$ and the coupling $g_{a\gamma}$.

\section{\label{sec:Data Analysis}Data Analysis}

\subsection{\label{Fermi-LAT data}Fermi-LAT data}

\begin{figure}
\includegraphics[scale=0.45]{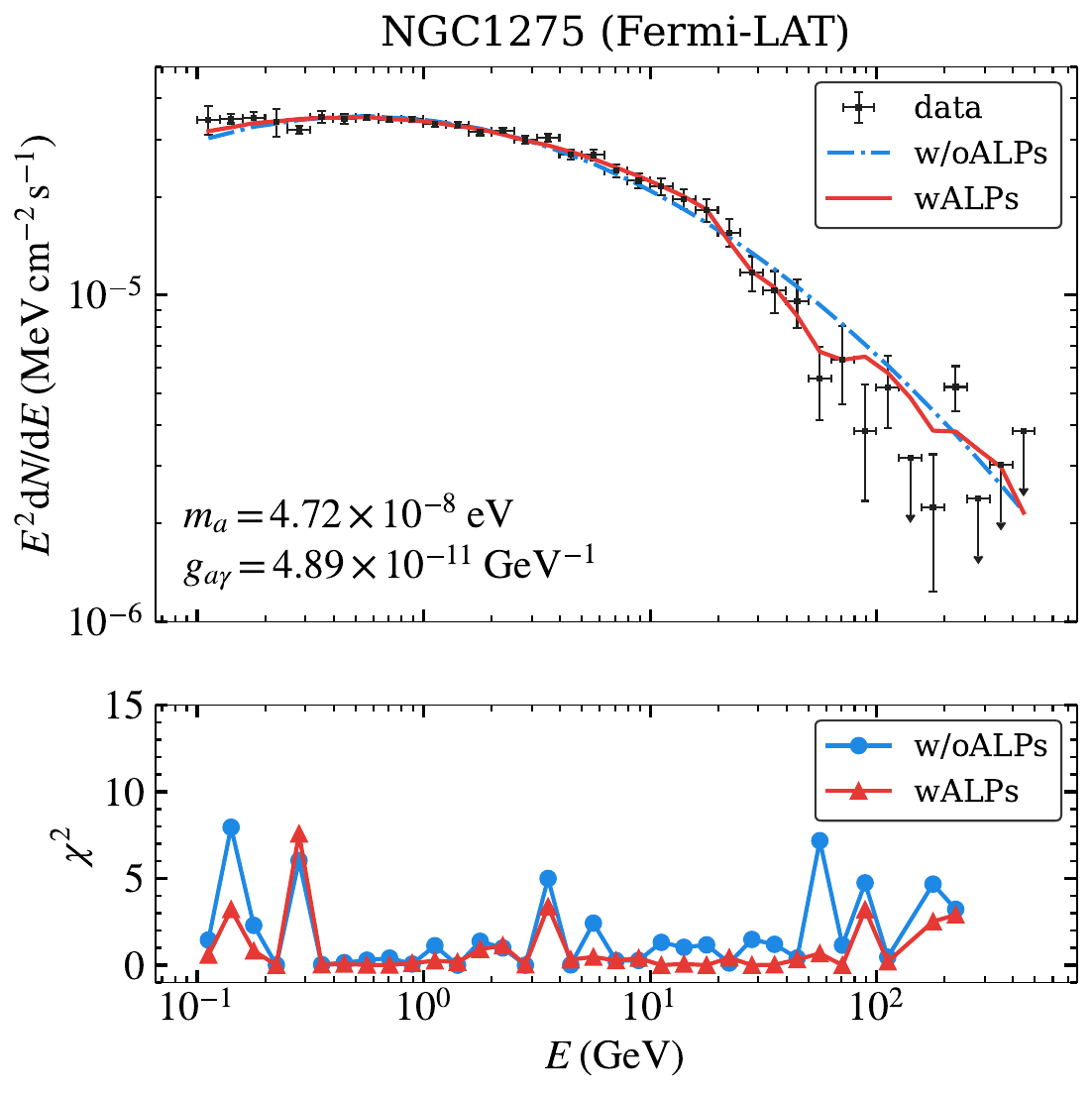}
\caption{\label{fig:SED}The SEDs analysis results for NGC 1275. Top panel: The SEDs of NGC 1275. Black data points show the Fermi-LAT observational results. The blue dashdot line represents the best-fit spectrum without an ALP model, while the red solid line shows the best-fit spectrum including an ALP model with $m_{a}=4.72\times 10^{-8}\,\mathrm{eV},g_{a\gamma}=4.89\times 10^{-11}\,\mathrm{GeV^{-1}}$. Bottom panel: $\chi^{2}$ values for each energy bin. Blue solid line with circles indicates $\chi^{2}$ values without the ALP model, and red solid line with triangles correspond to $\chi^{2}$ values including the ALP model with the above specified parameters.}
\end{figure}

In this work, we utilize nearly 16.5 years of Fermi-LAT P8R3 data~\cite{Fermi-LAT:2013jgq,fermi_data_access} collected from October 27, 2008 (MET=246823875), to March 10, 2025 (MET=763293464).
Data prior to October 27, 2008 are excluded due to notable background contamination issues at energies above ~30 GeV, ensuring the reliability of the analysis.~\footnote{\url{https://fermi.gsfc.nasa.gov/ssc/data/analysis/LAT_caveats.html}}

We select the photon events in a region of interest with a radius of $15^{\circ}$ centered on NGC 1275 and in the energy range from 100 MeV to 500 GeV.
The ``\texttt{SOURCE}'' event class and ``\texttt{FRONT+BACK}'' event type are used for data selection, with the maximum zenith angle set to $90^{\circ}$ to minimize contamination from Earth’s limb.
Additionally, the recommended \texttt{gtmktime} filter expression \texttt{(DATA\_QUAL\textgreater 0)\,\&\&\,(LAT\_CONFIG==1)} is applied to ensure that the satellite operates in standard data-taking mode and selects intervals with optimal data quality.
Here we use the standard binned likelihood analysis with the Fermi-LAT official software package \texttt{fermitools} (v11r5p3)~\footnote{\url{https://github.com/fermi-lat/Fermitools-conda/}} with the instrument response function \texttt{P8R3\_SOURCE\_V3}.

With the user-contributed script \texttt{make4FGLxml.py},~\footnote{\url{https://fermi.gsfc.nasa.gov/ssc/data/analysis/user/make4FGLxml.py}} the initial model is built as the combination of the Galactic diffuse emission model (\texttt{gll\_iem\_v07.fits}), the isotropic background spectral template (\texttt{iso\_P8R3\_SOURCE\_V3\_v1.txt}), and the incremental version of the fourth full catalog of Fermi-LAT $\gamma$-ray sources (4FGL-DR4)~\cite{Ballet:2023qzs}.
The Galactic and isotropic background models are available from the Fermi Science Support Center.~\footnote{\url{https://fermi.gsfc.nasa.gov/ssc/}}
Spectral parameters for sources within $7^{\circ}$ of NGC 1275 and the normalizations of background models are set as free.
We perform the likelihood fit in the energy range (100 MeV$-$500 GeV) with the \texttt{gtlike} tool to obtain the optimized model.

To derive the observed spectral energy distribution (SED) of NGC 1275, we divide the selected events into 37 logarithmically spaced uniform energy bins from 100 MeV to 500 GeV.
We fix the spectral indexes of all sources as the values optimized in the global fit, while permitting normalizations to vary freely during the likelihood fit.
The test statistic (TS) is defined as minus twice the log-likelihood ratio of the models with and without the target source.
Then we repeat the standard binned likelihood analysis in each energy bin to calculate corresponding flux and TS value.
For a given energy bin where the TS value is less than 9, we further calculate the 95\% confidence level upper limit on the flux using the \texttt{pyLikelihood UpperLimits} tool.
The SEDs of NGC 1275 we obtain are shown in Fig.~\ref{fig:SED} as black markers with error bars indicating uncertainties.

\subsection{\label{Chi2 analysis}$\chi^{2}$ analysis}

To investigate the signature of photon-ALP oscillation, we perform the $\chi^{2}$ analysis on the obtained SEDs of NGC 1275.
Based on the 4FGL-DR4~\cite{Ballet:2023qzs}, the intrinsic spectrum for the null hypothesis (labeled as w/oALPs) can be modeled as a log-parabola function,
\begin{equation}
\left(\frac{\mathrm{d}N}{\mathrm{d}E}\right)_{\mathrm{w/oALPs}}=N_{0}\left(\frac{E}{E_{b}}\right)^{-\left[\alpha+\beta\ln_{}{\left({E}/{E_{b}}\right)}\right]},
\label{eq:8}
\end{equation}
where $N_{0}$ is the normalization constant, $\alpha$ is the spectral index, $\beta$ is the curvature parameter, and $E_{b}$ is the scale parameter, typically fixed to the characteristic energy value near the low-energy region of the fitted spectrum.
In this study, it is set to 0.9578 GeV.

Incorporating the photon-ALP oscillation effect, the intrinsic spectrum is multiplied by the photon survival probability to obtain the oscillatory spectrum for the alternative hypothesis (labeled as wALPs),
\begin{equation}
\left(\frac{\mathrm{d}N}{\mathrm{d}E}\right)_{\mathrm{wALPs}}=P\left(g_{a\gamma},m_{a},E\right)\left(\frac{\mathrm{d}N}{\mathrm{d}E}\right)_{\mathrm{w/oALPs}}.
\label{eq:9}
\end{equation}

Due to the limited energy resolution of Fermi-LAT Pass 8 data at low energies ($\sim$ 10\% for energies below 1 GeV), energy dispersion is introduced across the studied energy range (100 MeV$–$500 GeV).
The energy dispersion function $D_{\mathrm{eff}}\left(E^{\prime},E\right)$ is convolved with the spectral model to yield the final observed spectrum,
\begin{equation}
\frac{\mathrm{d}N}{\mathrm{d}E^{\prime}}=D_{\mathrm{eff}}\left(E^{\prime},E\right)\otimes\frac{\mathrm{d}N}{\mathrm{d}E},
\label{eq:10}
\end{equation}
where $E^{\prime}$ and $E$ represent the observed and true photon energies, respectively.

Incorporating the energy dispersion of Fermi-LAT, we conduct the $\chi^{2}$ fits on the measured SEDs using the models with and without photon-ALP oscillation, respectively.
For the model with ALPs, we simulate 100 random realizations of Gaussian turbulent magnetic fields and repeat the fitting procedure on a grid of ALP masses and the couplings, covering the parameter spaces of ($m_{a}:1\times 10^{-11}-1\times 10^{-6}\,\mathrm{eV};\;g_{a\gamma}:1\times 10^{-13}-1\times 10^{-10}\,\mathrm{GeV^{-1}}$).
In each fit, the ALP mass $m_{a}$, coupling $g_{a\gamma}$, and scale parameter $E_{b}$ are fixed, while the normalization $N_{0}$, spectral index $\alpha$, and curvature parameter $\beta$ are left free.
To constrain the ALP parameter space, we calculate $\Delta\chi^{2}$ denoted as
\begin{equation}
\Delta\chi^{2}=\chi_{\mathrm{wALPs},5}^{2}-\chi_{\mathrm{w/oALPs}}^{2}.
\label{eq:11}
\end{equation}
Considering the uncertainty of the random turbulent magnetic field, we conservatively select $\chi_{\mathrm{wALPs},5}^{2}$, the fifth percentile of the best-fit $\chi_{\mathrm{wALPs}}^{2}$ values among the 100 magnetic field realization sorted in ascending order, to set constraints.
This means that our limiting results are applicable to at least 95\% of simulated magnetic field realizations~\cite{Fermi-LAT:2016nkz}.

\subsection{\label{sec:Monte Carlo Simulation}Monte Carlo simulation}

\begin{figure}
\includegraphics[scale=0.45]{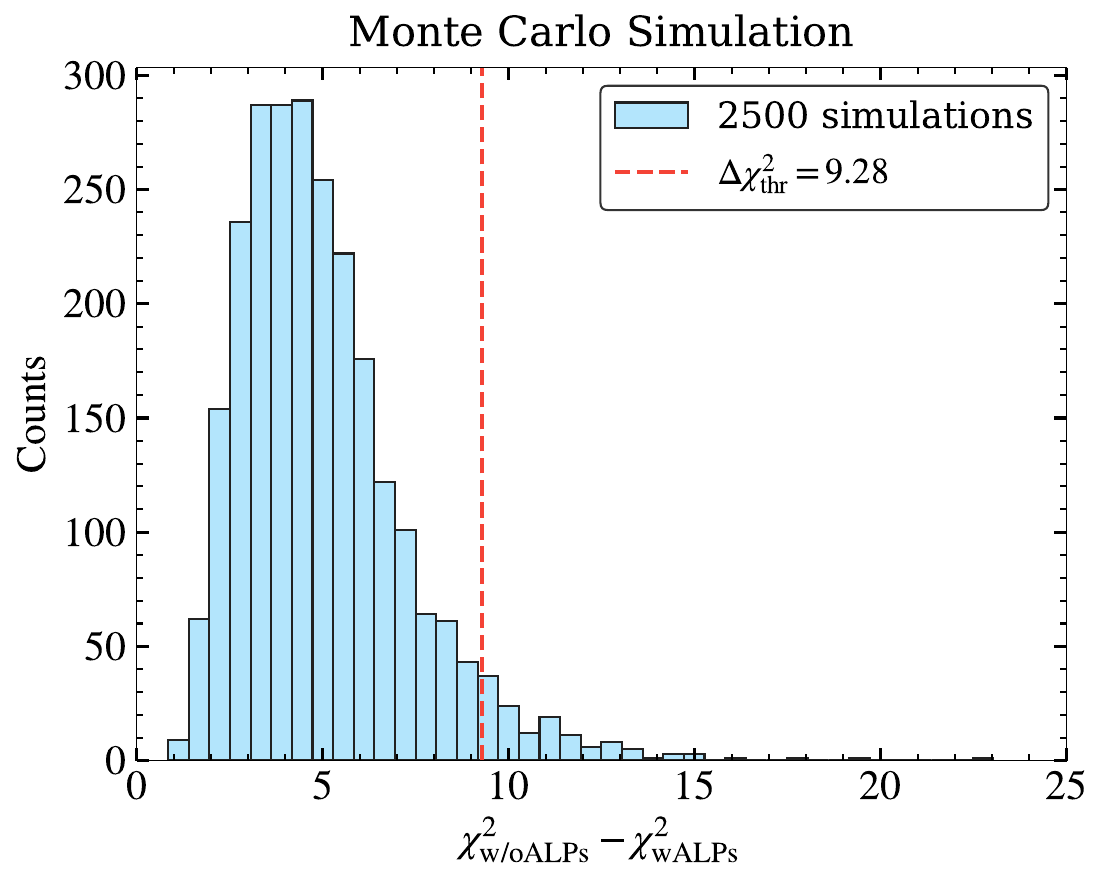}
\caption{\label{fig:Monte_Carlo_Simulation}Results of 2500 Monte Carlo simulations based on the null hypothesis spectral model. The red vertical dashed line indicates the statistical significance threshold $\Delta\chi^{2}_{\mathrm{thr}}=9.28$ corresponding to the 95\% confidence level.}
\end{figure}

The threshold $\Delta\chi^{2}_{\mathrm{thr}}$ corresponds to the 95\% confidence level, such that ALP parameters ($m_{a},g_{a\gamma}$) are excluded if $\Delta\chi^{2}>\Delta\chi^{2}_{\mathrm{thr}}$.
For two degrees of freedom and a significance level of 0.05, the critical $\chi^{2}$ value is 5.99.
However, due to the nonlinear relationship between the irregular spectral features and ALP parameters, theoretical thresholds deviate from actual values.
Following the method developed in Ref.~\cite{Fermi-LAT:2016nkz,Liang:2018mqm}, 
we perform Monte Carlo simulations to determine the threshold $\Delta\chi^{2}_{\mathrm{thr}}$ by computing the null distribution of $\chi^{2}$ differences between models with and without ALPs.

We simulate spectra of the null hypothesis by adopting the same energy bins as the observed SEDs.
For each energy bin, the simulated flux is randomly generated from a Gaussian distribution with the mean given by the best-fit intrinsic spectral model and the variance set to the observed flux error.
Using this method, 2500 spectral realizations are generated.
For each simulated spectrum, the aforementioned $\chi^{2}$ fitting procedure is repeated to obtain the best-fit $\chi^{2}$ values for both models (wALPs and w/oALPs).
Figure~\ref{fig:Monte_Carlo_Simulation} presents the $\chi_{\mathrm{w/oALPs}}^{2}-\chi_{\mathrm{wALPs}}^{2}$ distribution obtained from 2500 Monte Carlo simulations of the null hypothesis.
The exclusion threshold is determined to be $\Delta\chi^{2}_\mathrm{thr}=9.28$ (red dashed line) based on the 95\% quantile of this null distribution.
Subsequently, the ALP parameter space where $\Delta\chi^{2}$ exceeds this threshold value can be excluded at the 95\% confidence level.

\section{\label{sec:Result}Result}

\subsection{\label{sec:Fermi-LAT Result}Fermi-LAT}

\begin{figure}
\includegraphics[scale=0.325]{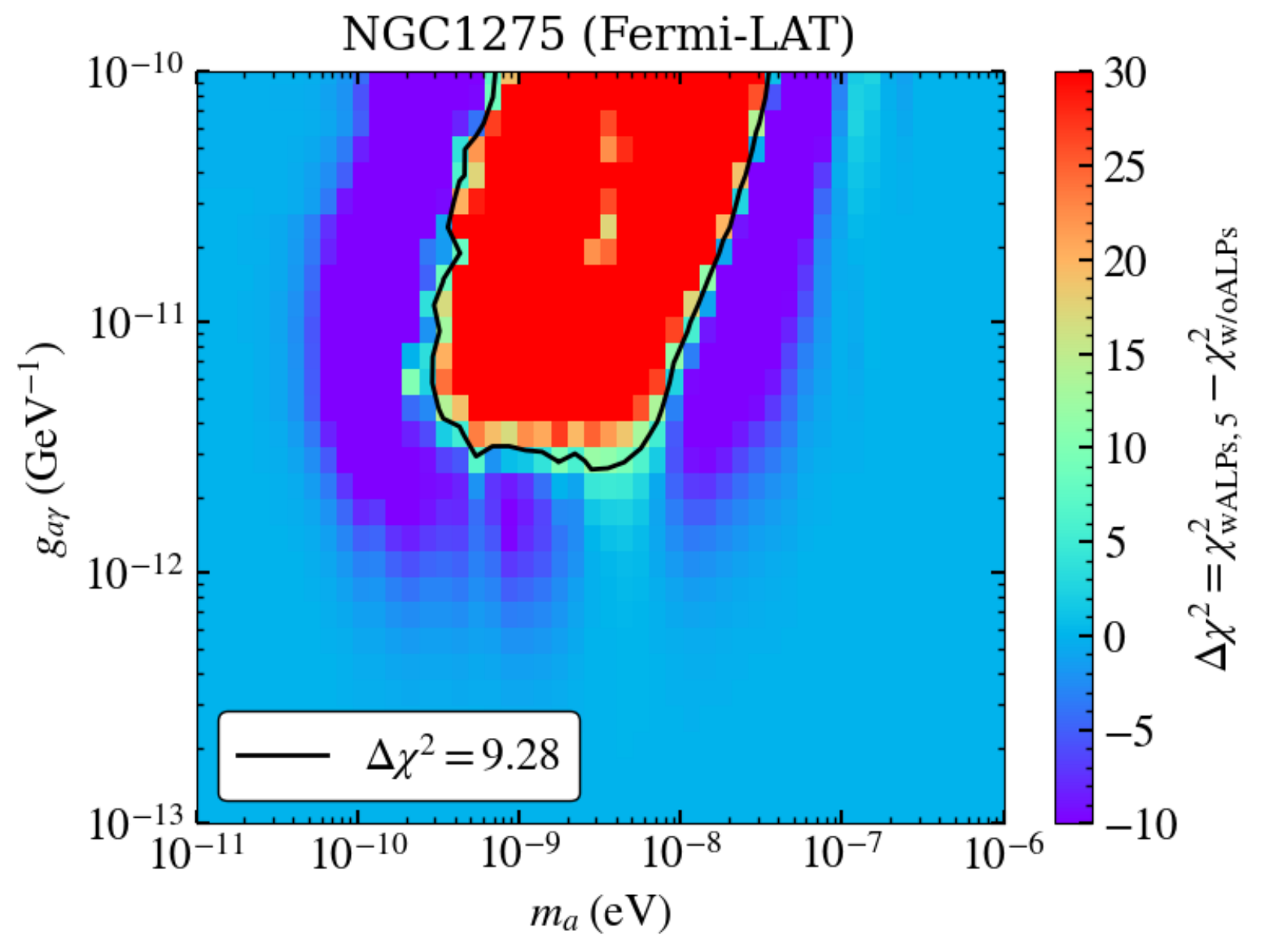}
\caption{\label{fig:chi2}Two-dimensional $\Delta\chi^{2}$ distribution in the parameter space of ALP masses versus photon-ALP couplings for NGC 1275. The solid black line indicates the 95\% confidence level threshold ($\Delta\chi^{2}$=9.28).}
\end{figure}

Following the method introduced in Sec.~\ref{sec:Data Analysis}, we obtain the SEDs (black) of NGC 1275 based on 16.5 years of Fermi-LAT observation and make the $\chi^{2}$ fits for the wALPs (red solid line) and w/oALPs (blue dashdot line) models, as shown in Fig.~\ref{fig:SED}.
The $\Delta\chi^{2}$ map is shown in Fig.~\ref{fig:chi2}, where the red region indicates parameter space disfavoring the signal model (wALPs).
The region with $\Delta\chi^{2}>\Delta\chi^{2}_{\mathrm{thr}} =9.28$, enclosed by the black solid line in Figure~\ref{fig:chi2}, has been excluded at the 95\% confidence level.
We find that with 16.5 years of Fermi-LAT observation, the ALP parameter space with the photon-ALP coupling of $g_{a\gamma}\gtrsim 3\times 10^{-12}\,\mathrm{GeV^{-1}}$ is excluded for the mass range $4\times 10^{-10}\,\mathrm{eV}\lesssim m_{a} \lesssim 5\times 10^{-9}\,\mathrm{eV}$, which improves previous limits (dark green region) for six-year data in Ref.~\cite{Fermi-LAT:2016nkz} by a factor of $\sim$ 2.
Additionally, as shown in Fig.~\ref{fig:limits}, our result (dark orange region) excludes the previously reported ``hole'' region  survived in Ref.~\cite{Fermi-LAT:2016nkz} (dark green region).
To check our method, we further repeat the above analysis with the same photon data selected in Ref.~\cite{Fermi-LAT:2016nkz} and almost the same limits with this survived ``hole'' region are obtained.
Hence, the stronger ALP limits we set are primarily due to the additional ten years of Fermi-LAT photon data included in our study.

\subsection{\label{sec:VLAST}VLAST}

\begin{figure}
\includegraphics[scale=0.45]{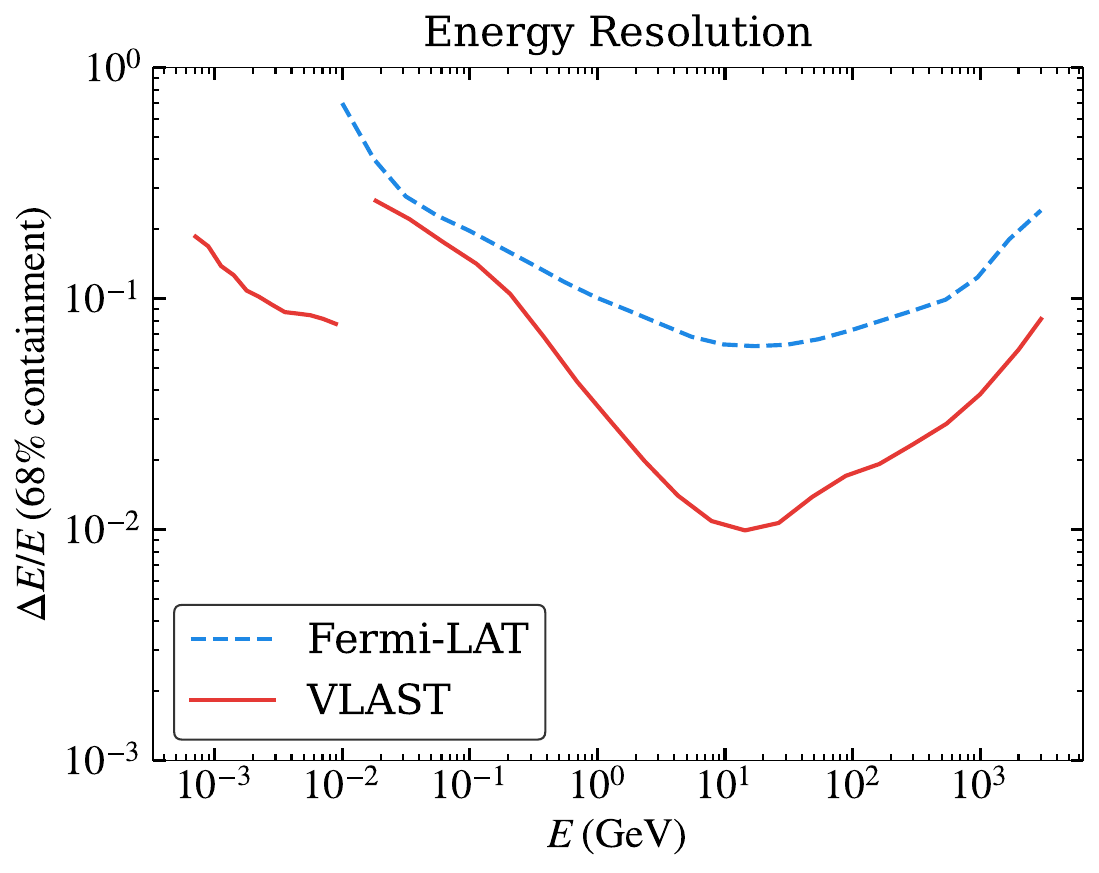}
\caption{\label{fig:Energy_Resolution}Sixty-eight percent energy resolution as a function of incident energy. The blue dashed line represents the energy resolution of Fermi-LAT (data version: \texttt{P8R3\_SOURCE\_V3}), while the red solid line shows the preliminary simulated energy resolution for VLAST. The curve in the low-energy regime corresponds to Compton scattering events, whereas the high-energy portion reflects pair-production events.}
\end{figure}

\begin{figure*}
\includegraphics[scale=0.56]{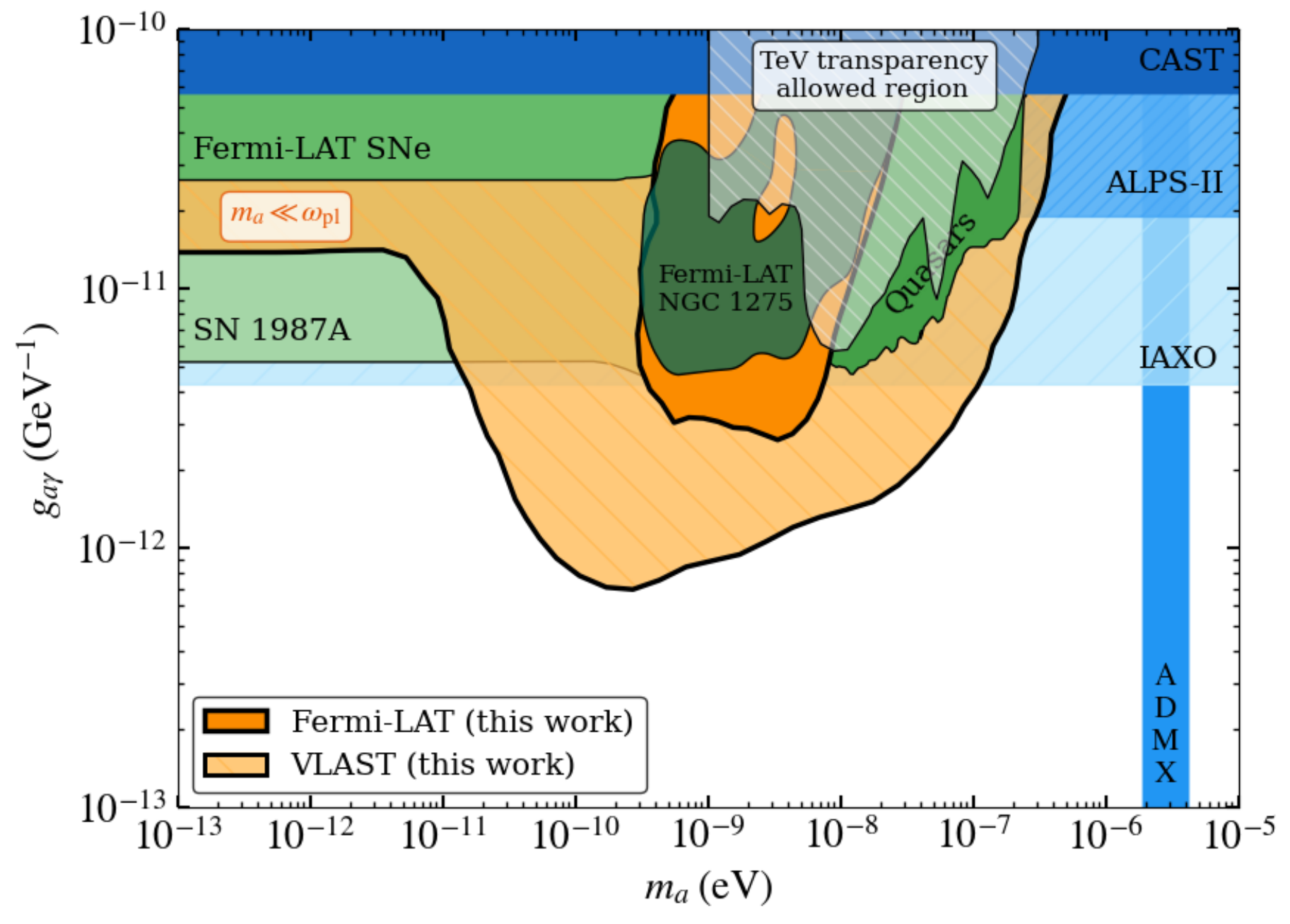}
\caption{\label{fig:limits}Comparison of current experimental constraints on the parameter space of ALP. 
The blue-shaded regions represent exclusion limits from ground-based experiments, with darker to lighter shades corresponding to the CAST experiment~\cite{CAST:2017uph}, the ADMX experiment~\cite{ADMX:2009iij,ADMX:2018gho,ADMX:2018ogs,ADMX:2019uok,Crisosto:2019fcj,ADMX:2021mio,ADMX:2021nhd,ADMX:2024xbv}, and the projected sensitivities of ALP-II~\cite{Ortiz:2020tgs} and IAXO~\cite{IAXO:2019mpb}, respectively. The green-shaded regions show exclusion bounds derived from astronomical observations, with darker to lighter shades representing constraints from Fermi-LAT observations of NGC 1275~\cite{Fermi-LAT:2016nkz}, quasars~\cite{Davies:2022wvj}, supernovae~\cite{Meyer:2020vzy}, and SN 1987A~\cite{2015JCAP...02..006P}, respectively. The white region denotes the parameter space where ALPs could explain the opacity of the cosmic $\gamma$-ray background~\cite{Meyer:2013pny}. In this study, exclusion bounds from observations of NGC 1275 using 16.5 years of Fermi-LAT data and five years of simulated VLAST data are displayed in dark and light orange, respectively.}
\end{figure*}

The next-generation $\gamma$-ray telescopes, such as the VLAST~\cite{2022AcASn..63...27F}, are expected to expand the detectable the ALP parameter space by extending the energy range, improving energy resolution, and increasing the effective detection area.
Compared to Fermi-LAT, VLAST extends the detection energy range down to below MeV and represents a significant improvement in the instrument performance for the GeV$-$TeV energy range.
For above 1 GeV, VLAST can achieve an effective acceptance and effective area approximately five times those of Fermi-LAT.
This significantly improves the detection efficiency for high-energy photons.
More crucially, VLAST demonstrates remarkable improvements in the energy resolution.
It reaches 8\% - 20\% for low-energy Compton scattering events ($<$ 10 MeV) and achieves 1\% - 30\% for high-energy pair-conversion events (10 MeV$-$1 TeV)~\cite{2022AcASn..63...27F}, representing substantial enhancement over Fermi-LAT's 6\% - 40\% resolution, as seen in Fig.~\ref{fig:Energy_Resolution}.
These breakthrough advancements make VLAST particularly well suited for detecting subtle spectral features that may arise from new physics phenomena such as ALPs.

We investigate the ALP detection sensitivity of the VLAST observations of NGC 1275 over a span of five years.
Assuming the spectrum below 100 MeV still follows the log-parabola model, the photon counts for each energy bin are simulated using the best-fit intrinsic spectral model and VLAST's exposure of five years.
In each energy bin, the simulated counts are randomly generated from a Poisson distribution.
Then we can obtain the simulated flux by ($\mathrm{count}/\mathrm{exposure}$) with flux error of ($\sqrt{\mathrm{count}}/\mathrm{exposure}$).
Using this method, 100 spectral realizations are generated, and the $\chi^{2}$ fitting procedure is repeated for each.

As shown in Fig.~\ref{fig:limits}, the light orange shaded region represents projected sensitivity of VLAST.
Notably, the sensitivity of VLAST in the ALP mass range of $2\times 10^{-11}\,\mathrm{eV}\lesssim m_{a}\lesssim 1\times 10^{-7}\,\mathrm{eV}$ is expected to be stronger than that of the upcoming Axion Solar Telescope IAXO (light sky-blue region in Fig.~\ref{fig:limits}).
Especially for $m_{a}\sim 2\times 10^{-10}\,\mathrm{eV}$, VLAST can probe the coupling down to $g_{a\gamma}\gtrsim 7\times 10^{-13}\,\mathrm{GeV^{-1}}$.
In the lower mass range $m_{a}\lesssim 5\times 10^{-12}\,\mathrm{eV}$, VLAST demonstrates sensitivity to the coupling $g_{a\gamma}\gtrsim 1.5\times 10^{-11}\,\mathrm{GeV^{-1}}$.
A distinct ``plateau'' feature appears in the parameter space for $m_{a}\lesssim 5\times 10^{-12}\,\mathrm{eV}$.
As shown in Fig.~\ref{fig:p} and Eq.~\eqref{eq:1}, when $m_{a}\ll\omega _{\mathrm{pl}}$, the photon survival probability does not change with the ALP mass, leading to this ``plateau''.
Furthermore, VLAST detectable parameter space fully encompasses the region predicted by the TeV transparency phenomenon~\cite{Meyer:2013pny}, indicating that future observations may provide recheck on the constraints of these theoretical predictions.
The projected sensitivity of the future VLAST experiment is expected to significantly expand the detectable ALP parameter space.

\section{\label{sec:Discussion}Discussion}

\begin{figure}[t]
\includegraphics[width=0.4\textwidth]{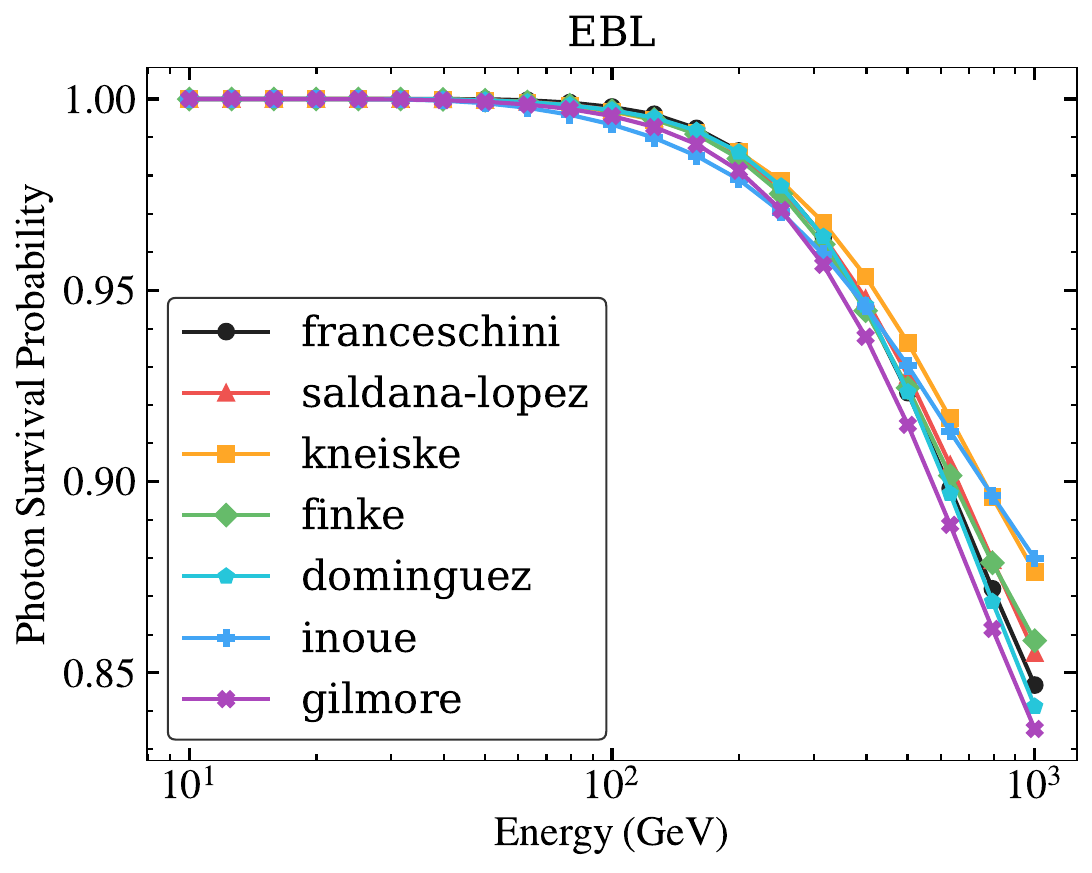}
\caption{\label{fig:EBL}Photon survival probability predicted by different EBL models, with each model represented by distinct line colors and markers: black solid line with circles (``\texttt{Franceschini}'' model), red solid line with triangles (``\texttt{Saldaña-López}'' model), yellow solid line with squares (``\texttt{Kneiske}'' model), green solid line with diamonds (``\texttt{Finke}''model), cyan solid line with pentagons (``\texttt{Domínguez}'' model, the reference model used in the baseline analysis), blue solid line with crosses (``\texttt{Inoue}'' model), and purple solid line with x markers (``\texttt{Gilmore}'' model).}
\end{figure}

\begin{figure*}
\includegraphics[scale=0.4]{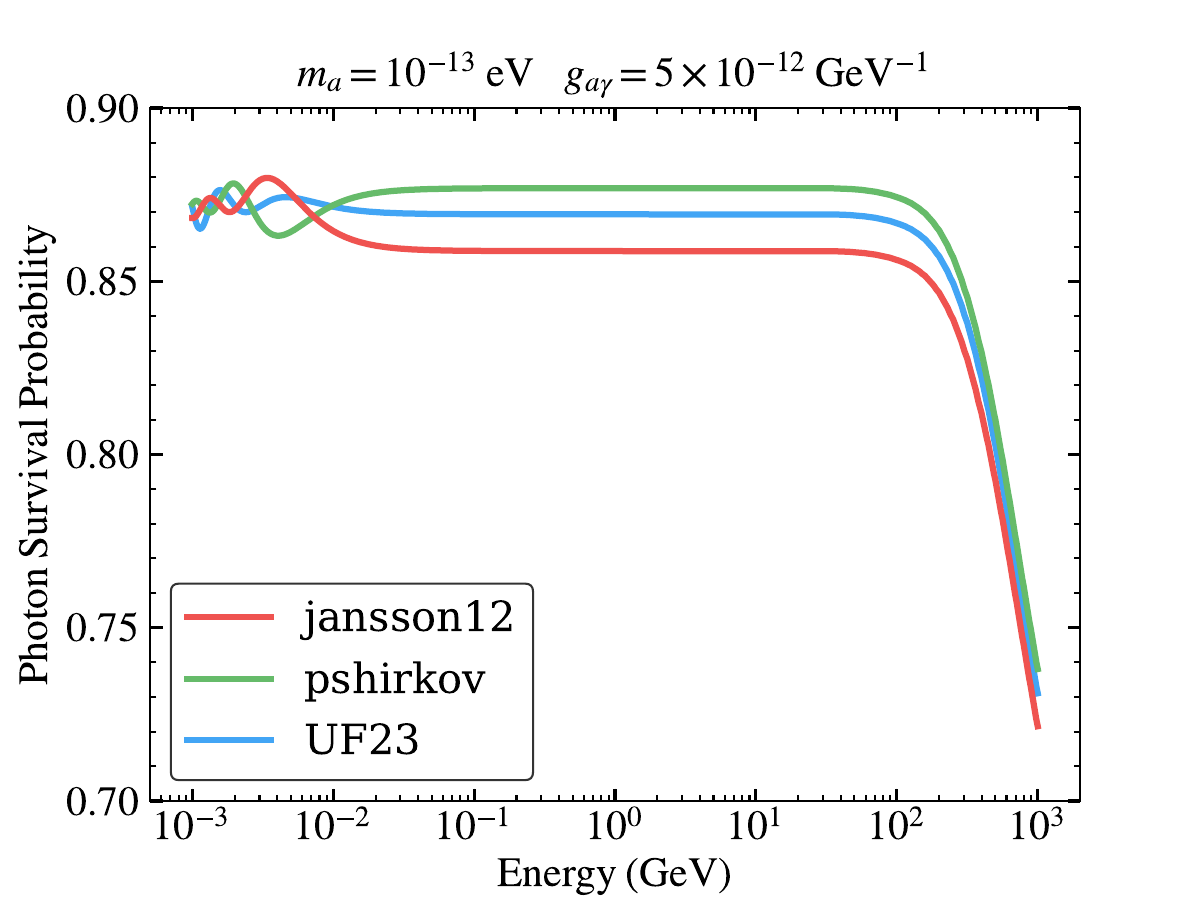}
\includegraphics[scale=0.4]{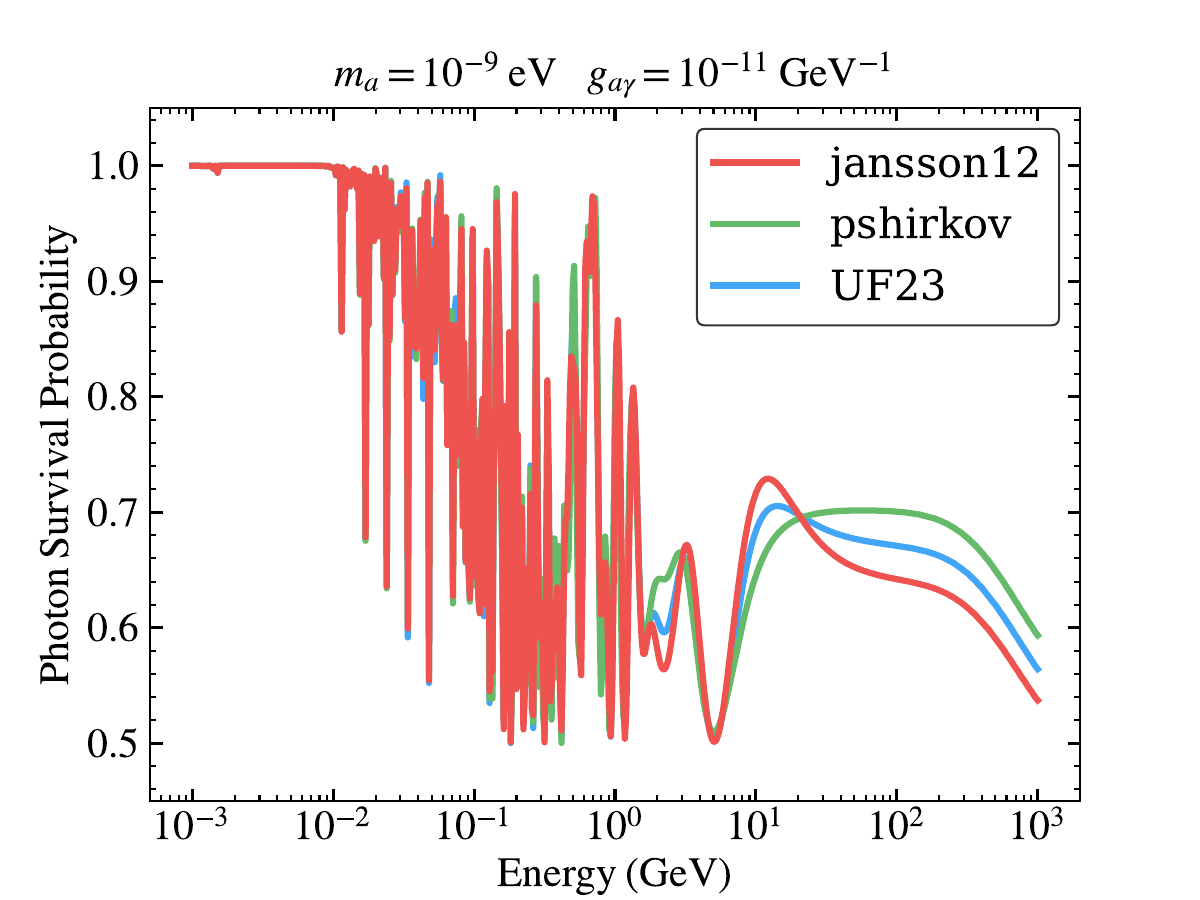}
\caption{\label{fig:GMF}Photon survival probability under different GMF models. Left panel: Results for ALP mass significantly below the plasma frequency, showing three characteristic curves: the red solid line represents the ``\texttt{Jansson12}'' model (adopted in the baseline analysis), the green solid line corresponds to the ``\texttt{Pshirkov}'' model, and the blue solid line indicates the ``\texttt{UF23}'' model. Right panel: Results for ALP mass reaching or exceeding the plasma frequency.}
\end{figure*}

\begin{figure}
\includegraphics[scale=0.4]{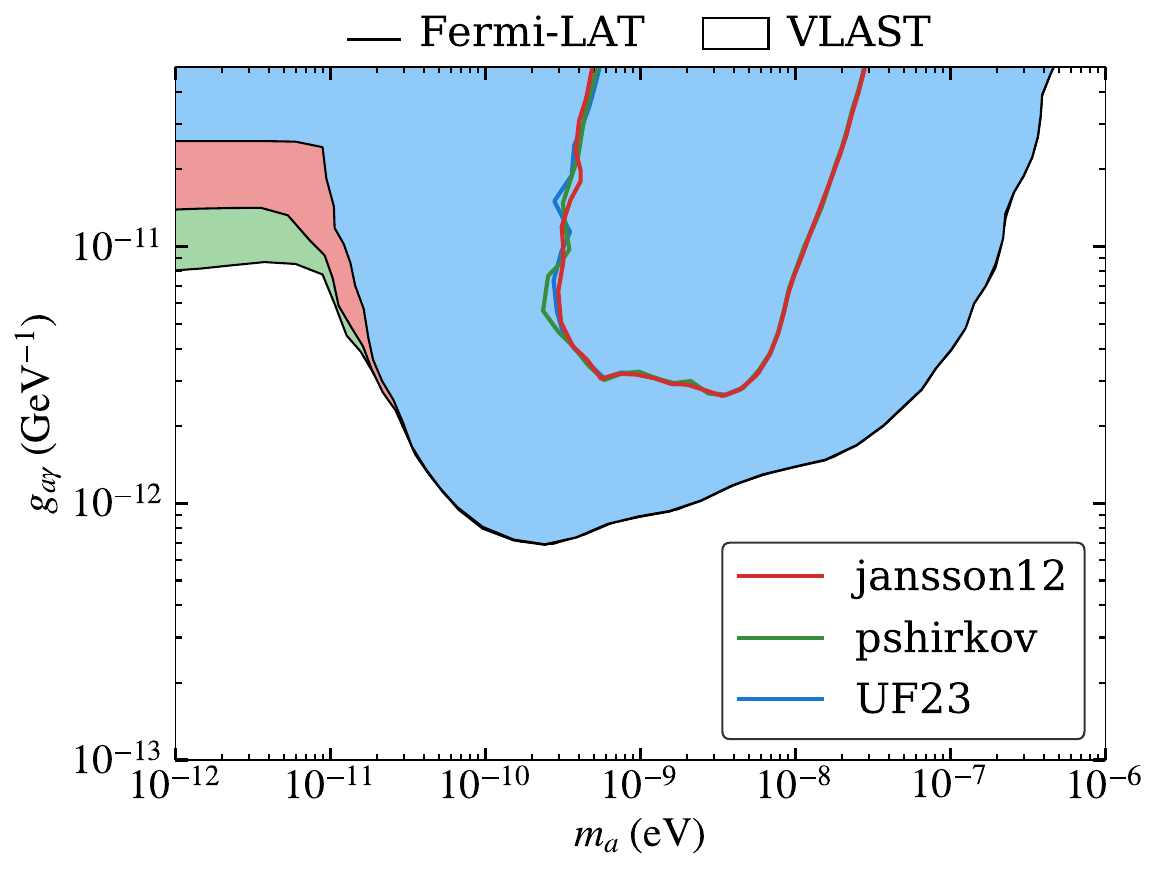}
\caption{\label{fig:GMF_limits}Exclusion regions in the parameter space of ALP under different GMF models. The solid lines represent the exclusion results from Fermi-LAT observations, while the colored regions indicate the detectable range of the future VLAST. Three colors are used to distinguish between different GMF models: red corresponds to the ``\texttt{Jansson12}'' model adopted in this study, green represents the ``\texttt{Pshirkov}'' model, and blue denotes the ``\texttt{UF23}'' model.}
\end{figure}

The EBL in the intergalactic medium and the GMF in the Milky Way, can influence the propagation of photons from extra-galactic sources to Earth.
In our baseline analysis, the ``\texttt{Dominguez}'' EBL model~\cite{Dominguez:2010bv} and ``\texttt{Jansson12}'' GMF model ~\cite{Jansson:2012pc} are adopted.
Here we conduct the comparative analysis on the potential systematic uncertainty of EBL models and GMF models.

The EBL, as the remnant radiation field from galaxy formation and evolution, has been conclusively demonstrated to significantly absorb $\gamma$-ray photons through the pair-production process~\cite{Dominguez:2010bv,Kneiske:2010pt}.
Consequently, the influence of the EBL is specifically considered when calculating the photon survival probability.
To evaluate the uncertainty of the model, comparisons are made with other EBL models~\cite{Kneiske:2010pt,Gilmore:2011ks,Inoue:2012bk,Franceschini:2017iwq,Saldana-Lopez:2020qzx,Finke:2022uvv}.
The differences in photon survival probability between models primarily occur at energies above 100 GeV, with a maximum discrepancy not exceeding 5\% (see Fig.~\ref{fig:EBL}).
This indicates that model selection does not substantially affect our conclusions.

The GMF is consisting of a large-scale regular field and a small-scale turbulent field~\cite{Planck:2016gdp,Jansson:2012pc} and plays an important role in the photon-ALP conversion process within the Milky Way.
To assess uncertainties, two other GMF models, the ``\texttt{Pshirkov}'' model~\cite{Pshirkov:2011um} and the ``\texttt{UF23}'' model~\cite{Unger:2023lob}, are also considered.
When the ALP mass is substantially smaller than the plasma frequency, as illustrated in the left panel of Fig.~\ref{fig:GMF}, the photon survival probability for different GMF models exhibits markedly different oscillation structures within the photon energy below 100 MeV, which leads to distinct VLAST sensitivities in the low ALP mass range ($m_{a}\lesssim 1\times 10^{-11}\,\mathrm{eV}$) in Fig.~\ref{fig:GMF_limits}.
When the ALP mass approaches or exceeds the plasma frequency in the right panel of Fig.~\ref{fig:GMF}, the photon survival probability below 10 GeV varies only slightly among different GMF models.
Hence, as seen in Fig.~\ref{fig:GMF_limits}, the uncertainties of GMF models have a negligible effect on the Fermi-LAT limits and the VLAST sensitivity in the high-mass range ($m_{a}\gtrsim 5\times 10^{-11}\,\mathrm{eV}$), but for the low-mass range ($m_{a}\lesssim 1\times 10^{-11}\,\mathrm{eV}$) the VLAST sensitivity will be influenced.

\section{\label{sec:Summary}Summary}

Based on the $\gamma$-ray observations of NGC 1275, we investigate the photon-ALP oscillation structures in the Fermi-LAT measured SEDs to constrain the ALP properties and evaluate the expected detection sensitivity of VLAST over a five-year observation period.
Using 16.5 years of the Fermi-LAT observations, we exclude the parameter space with photon-ALP coupling $g_{a\gamma}\gtrsim 3\times 10^{-12}\,\mathrm{GeV^{-1}}$ within the ALP mass range of $4\times 10^{-10}\,\mathrm{eV}\lesssim m_{a}\lesssim 5\times 10^{-9}\,\mathrm{eV}$.
Our results not only can exclude the previously existing ``hole'' survived in the six-year observations~\cite{Fermi-LAT:2016nkz}, but also improve on previous limits by a factor of 2.

By simulating the five-year observation of NGC 1275 with VLAST, we also show that the expected sensitivity of VLAST may exceed that of the next generation Axion Solar Telescope IAXO for an ALP with the mass of $2\times 10^{-11}\,\mathrm{eV}\lesssim m_{a}\lesssim 1\times 10^{-7}\,\mathrm{eV}$.
We obtain the best sensitivity of $g_{a\gamma}\sim 7\times 10^{-13}\,\mathrm{GeV^{-1}}$ at $m_{a}\sim 2\times 10^{-10}\,\mathrm{eV}$.
Thanks to its detectable energy range extending below 1 MeV, VLAST will achieve sensitivity to ALP masses below $5\times 10^{-12}\,\mathrm{eV}$ with a sensitivity level of $g_{a\gamma}\gtrsim 1.5\times 10^{-11}\,\mathrm{GeV^{-1}}$ (affected by the uncertainty of the GMF model).
The ALP sensitivities in this low-mass range exhibits a characteristic ``plateau'' feature, due to the insensitivity of photon-ALP conversion probability to the ALP mass when $m_{a}\ll\omega _{\mathrm{pl}}$.
In the near future, the upcoming VLAST will extensively probe the ALP parameter space and potentially lead to important discoveries.

\begin{acknowledgments}
We would like to thank Kai-Kai Duan for insightful discussions and providing the VLAST instrument performance.
This work is supported in part by the National Key Research and Development Program of China (No. 2022YFF0503304), the Project for Young Scientists in Basic Research of CAS (No. YSBR-092) and the Strategic Priority Research Program of the Chinese Academy of Sciences (No. XDB0550400).
L.W. is supported by the National Natural Science Foundation of China (NSFC) under Grants No. 12275134 and No. 12335005, and by the State Key Laboratory of Dark Matter Physics.
\end{acknowledgments}

\section*{Data Availability}

The data that support the findings of this article are openly available~\cite{fermi_data_access}.

\nocite{*}

\bibliography{reference}

\providecommand{\noopsort}[1]{}
\begin{thebibliography}{101}%
\makeatletter
\providecommand \@ifxundefined [1]{%
 \@ifx{#1\undefined}
}%
\providecommand \@ifnum [1]{%
 \ifnum #1\expandafter \@firstoftwo
 \else \expandafter \@secondoftwo
 \fi
}%
\providecommand \@ifx [1]{%
 \ifx #1\expandafter \@firstoftwo
 \else \expandafter \@secondoftwo
 \fi
}%
\providecommand \natexlab [1]{#1}%
\providecommand \enquote  [1]{``#1''}%
\providecommand \bibnamefont  [1]{#1}%
\providecommand \bibfnamefont [1]{#1}%
\providecommand \citenamefont [1]{#1}%
\providecommand \href@noop [0]{\@secondoftwo}%
\providecommand \href [0]{\begingroup \@sanitize@url \@href}%
\providecommand \@href[1]{\@@startlink{#1}\@@href}%
\providecommand \@@href[1]{\endgroup#1\@@endlink}%
\providecommand \@sanitize@url [0]{\catcode `\\12\catcode `\$12\catcode `\&12\catcode `\#12\catcode `\^12\catcode `\_12\catcode `\%12\relax}%
\providecommand \@@startlink[1]{}%
\providecommand \@@endlink[0]{}%
\providecommand \url  [0]{\begingroup\@sanitize@url \@url }%
\providecommand \@url [1]{\endgroup\@href {#1}{\urlprefix }}%
\providecommand \urlprefix  [0]{URL }%
\providecommand \Eprint [0]{\href }%
\providecommand \doibase [0]{https://doi.org/}%
\providecommand \selectlanguage [0]{\@gobble}%
\providecommand \bibinfo  [0]{\@secondoftwo}%
\providecommand \bibfield  [0]{\@secondoftwo}%
\providecommand \translation [1]{[#1]}%
\providecommand \BibitemOpen [0]{}%
\providecommand \bibitemStop [0]{}%
\providecommand \bibitemNoStop [0]{.\EOS\space}%
\providecommand \EOS [0]{\spacefactor3000\relax}%
\providecommand \BibitemShut  [1]{\csname bibitem#1\endcsname}%
\let\auto@bib@innerbib\@empty
\bibitem [{\citenamefont {Peccei}\ and\ \citenamefont {Quinn}(1977)}]{Peccei:1977hh}%
  \BibitemOpen
  \bibfield  {author} {\bibinfo {author} {\bibfnamefont {R.~D.}\ \bibnamefont {Peccei}}\ and\ \bibinfo {author} {\bibfnamefont {H.~R.}\ \bibnamefont {Quinn}},\ }\bibfield  {title} {\bibinfo {title} {{CP Conservation in the Presence of Instantons}},\ }\href {https://doi.org/10.1103/PhysRevLett.38.1440} {\bibfield  {journal} {\bibinfo  {journal} {Phys. Rev. Lett.}\ }\textbf {\bibinfo {volume} {38}},\ \bibinfo {pages} {1440} (\bibinfo {year} {1977})}\BibitemShut {NoStop}%
\bibitem [{\citenamefont {Weinberg}(1978)}]{Weinberg:1977ma}%
  \BibitemOpen
  \bibfield  {author} {\bibinfo {author} {\bibfnamefont {S.}~\bibnamefont {Weinberg}},\ }\bibfield  {title} {\bibinfo {title} {{A New Light Boson?}},\ }\href {https://doi.org/10.1103/PhysRevLett.40.223} {\bibfield  {journal} {\bibinfo  {journal} {Phys. Rev. Lett.}\ }\textbf {\bibinfo {volume} {40}},\ \bibinfo {pages} {223} (\bibinfo {year} {1978})}\BibitemShut {NoStop}%
\bibitem [{\citenamefont {Wilczek}(1978)}]{Wilczek:1977pj}%
  \BibitemOpen
  \bibfield  {author} {\bibinfo {author} {\bibfnamefont {F.}~\bibnamefont {Wilczek}},\ }\bibfield  {title} {\bibinfo {title} {{Problem of Strong $P$ and $T$ Invariance in the Presence of Instantons}},\ }\href {https://doi.org/10.1103/PhysRevLett.40.279} {\bibfield  {journal} {\bibinfo  {journal} {Phys. Rev. Lett.}\ }\textbf {\bibinfo {volume} {40}},\ \bibinfo {pages} {279} (\bibinfo {year} {1978})}\BibitemShut {NoStop}%
\bibitem [{\citenamefont {Peccei}(2008)}]{Peccei:2006as}%
  \BibitemOpen
  \bibfield  {author} {\bibinfo {author} {\bibfnamefont {R.~D.}\ \bibnamefont {Peccei}},\ }\bibfield  {title} {\bibinfo {title} {{The Strong CP problem and axions}},\ }\href {https://doi.org/10.1007/978-3-540-73518-2_1} {\bibfield  {journal} {\bibinfo  {journal} {Lect. Notes Phys.}\ }\textbf {\bibinfo {volume} {741}},\ \bibinfo {pages} {3} (\bibinfo {year} {2008})},\ \Eprint {https://arxiv.org/abs/hep-ph/0607268} {arXiv:hep-ph/0607268} \BibitemShut {NoStop}%
\bibitem [{\citenamefont {Svrcek}\ and\ \citenamefont {Witten}(2006)}]{Svrcek:2006yi}%
  \BibitemOpen
  \bibfield  {author} {\bibinfo {author} {\bibfnamefont {P.}~\bibnamefont {Svrcek}}\ and\ \bibinfo {author} {\bibfnamefont {E.}~\bibnamefont {Witten}},\ }\bibfield  {title} {\bibinfo {title} {{Axions In String Theory}},\ }\href {https://doi.org/10.1088/1126-6708/2006/06/051} {\bibfield  {journal} {\bibinfo  {journal} {JHEP}\ }\textbf {\bibinfo {volume} {06}},\ \bibinfo {pages} {051}},\ \Eprint {https://arxiv.org/abs/hep-th/0605206} {arXiv:hep-th/0605206} \BibitemShut {NoStop}%
\bibitem [{\citenamefont {Arvanitaki}\ \emph {et~al.}(2010)\citenamefont {Arvanitaki}, \citenamefont {Dimopoulos}, \citenamefont {Dubovsky}, \citenamefont {Kaloper},\ and\ \citenamefont {March-Russell}}]{Arvanitaki:2009fg}%
  \BibitemOpen
  \bibfield  {author} {\bibinfo {author} {\bibfnamefont {A.}~\bibnamefont {Arvanitaki}}, \bibinfo {author} {\bibfnamefont {S.}~\bibnamefont {Dimopoulos}}, \bibinfo {author} {\bibfnamefont {S.}~\bibnamefont {Dubovsky}}, \bibinfo {author} {\bibfnamefont {N.}~\bibnamefont {Kaloper}},\ and\ \bibinfo {author} {\bibfnamefont {J.}~\bibnamefont {March-Russell}},\ }\bibfield  {title} {\bibinfo {title} {{String Axiverse}},\ }\href {https://doi.org/10.1103/PhysRevD.81.123530} {\bibfield  {journal} {\bibinfo  {journal} {Phys. Rev. D}\ }\textbf {\bibinfo {volume} {81}},\ \bibinfo {pages} {123530} (\bibinfo {year} {2010})},\ \Eprint {https://arxiv.org/abs/0905.4720} {arXiv:0905.4720 [hep-th]} \BibitemShut {NoStop}%
\bibitem [{\citenamefont {Cicoli}\ \emph {et~al.}(2012)\citenamefont {Cicoli}, \citenamefont {Goodsell},\ and\ \citenamefont {Ringwald}}]{Cicoli:2012sz}%
  \BibitemOpen
  \bibfield  {author} {\bibinfo {author} {\bibfnamefont {M.}~\bibnamefont {Cicoli}}, \bibinfo {author} {\bibfnamefont {M.}~\bibnamefont {Goodsell}},\ and\ \bibinfo {author} {\bibfnamefont {A.}~\bibnamefont {Ringwald}},\ }\bibfield  {title} {\bibinfo {title} {{The type IIB string axiverse and its low-energy phenomenology}},\ }\href {https://doi.org/10.1007/JHEP10(2012)146} {\bibfield  {journal} {\bibinfo  {journal} {JHEP}\ }\textbf {\bibinfo {volume} {10}},\ \bibinfo {pages} {146}},\ \Eprint {https://arxiv.org/abs/1206.0819} {arXiv:1206.0819 [hep-th]} \BibitemShut {NoStop}%
\bibitem [{\citenamefont {Preskill}\ \emph {et~al.}(1983)\citenamefont {Preskill}, \citenamefont {Wise},\ and\ \citenamefont {Wilczek}}]{Preskill:1982cy}%
  \BibitemOpen
  \bibfield  {author} {\bibinfo {author} {\bibfnamefont {J.}~\bibnamefont {Preskill}}, \bibinfo {author} {\bibfnamefont {M.~B.}\ \bibnamefont {Wise}},\ and\ \bibinfo {author} {\bibfnamefont {F.}~\bibnamefont {Wilczek}},\ }\bibfield  {title} {\bibinfo {title} {{Cosmology of the Invisible Axion}},\ }\href {https://doi.org/10.1016/0370-2693(83)90637-8} {\bibfield  {journal} {\bibinfo  {journal} {Phys. Lett. B}\ }\textbf {\bibinfo {volume} {120}},\ \bibinfo {pages} {127} (\bibinfo {year} {1983})}\BibitemShut {NoStop}%
\bibitem [{\citenamefont {Abbott}\ and\ \citenamefont {Sikivie}(1983)}]{Abbott:1982af}%
  \BibitemOpen
  \bibfield  {author} {\bibinfo {author} {\bibfnamefont {L.~F.}\ \bibnamefont {Abbott}}\ and\ \bibinfo {author} {\bibfnamefont {P.}~\bibnamefont {Sikivie}},\ }\bibfield  {title} {\bibinfo {title} {{A Cosmological Bound on the Invisible Axion}},\ }\href {https://doi.org/10.1016/0370-2693(83)90638-X} {\bibfield  {journal} {\bibinfo  {journal} {Phys. Lett. B}\ }\textbf {\bibinfo {volume} {120}},\ \bibinfo {pages} {133} (\bibinfo {year} {1983})}\BibitemShut {NoStop}%
\bibitem [{\citenamefont {Dine}\ and\ \citenamefont {Fischler}(1983)}]{Dine:1982ah}%
  \BibitemOpen
  \bibfield  {author} {\bibinfo {author} {\bibfnamefont {M.}~\bibnamefont {Dine}}\ and\ \bibinfo {author} {\bibfnamefont {W.}~\bibnamefont {Fischler}},\ }\bibfield  {title} {\bibinfo {title} {{The Not So Harmless Axion}},\ }\href {https://doi.org/10.1016/0370-2693(83)90639-1} {\bibfield  {journal} {\bibinfo  {journal} {Phys. Lett. B}\ }\textbf {\bibinfo {volume} {120}},\ \bibinfo {pages} {137} (\bibinfo {year} {1983})}\BibitemShut {NoStop}%
\bibitem [{\citenamefont {Hwang}\ and\ \citenamefont {Noh}(2009)}]{Hwang:2009js}%
  \BibitemOpen
  \bibfield  {author} {\bibinfo {author} {\bibfnamefont {J.-c.}\ \bibnamefont {Hwang}}\ and\ \bibinfo {author} {\bibfnamefont {H.}~\bibnamefont {Noh}},\ }\bibfield  {title} {\bibinfo {title} {{Axion as a Cold Dark Matter candidate}},\ }\href {https://doi.org/10.1016/j.physletb.2009.08.031} {\bibfield  {journal} {\bibinfo  {journal} {Phys. Lett. B}\ }\textbf {\bibinfo {volume} {680}},\ \bibinfo {pages} {1} (\bibinfo {year} {2009})},\ \Eprint {https://arxiv.org/abs/0902.4738} {arXiv:0902.4738 [astro-ph.CO]} \BibitemShut {NoStop}%
\bibitem [{\citenamefont {Duffy}\ and\ \citenamefont {van Bibber}(2009)}]{Duffy:2009ig}%
  \BibitemOpen
  \bibfield  {author} {\bibinfo {author} {\bibfnamefont {L.~D.}\ \bibnamefont {Duffy}}\ and\ \bibinfo {author} {\bibfnamefont {K.}~\bibnamefont {van Bibber}},\ }\bibfield  {title} {\bibinfo {title} {{Axions as Dark Matter Particles}},\ }\href {https://doi.org/10.1088/1367-2630/11/10/105008} {\bibfield  {journal} {\bibinfo  {journal} {New J. Phys.}\ }\textbf {\bibinfo {volume} {11}},\ \bibinfo {pages} {105008} (\bibinfo {year} {2009})},\ \Eprint {https://arxiv.org/abs/0904.3346} {arXiv:0904.3346 [hep-ph]} \BibitemShut {NoStop}%
\bibitem [{\citenamefont {Arias}\ \emph {et~al.}(2012)\citenamefont {Arias}, \citenamefont {Cadamuro}, \citenamefont {Goodsell}, \citenamefont {Jaeckel}, \citenamefont {Redondo},\ and\ \citenamefont {Ringwald}}]{Arias:2012az}%
  \BibitemOpen
  \bibfield  {author} {\bibinfo {author} {\bibfnamefont {P.}~\bibnamefont {Arias}}, \bibinfo {author} {\bibfnamefont {D.}~\bibnamefont {Cadamuro}}, \bibinfo {author} {\bibfnamefont {M.}~\bibnamefont {Goodsell}}, \bibinfo {author} {\bibfnamefont {J.}~\bibnamefont {Jaeckel}}, \bibinfo {author} {\bibfnamefont {J.}~\bibnamefont {Redondo}},\ and\ \bibinfo {author} {\bibfnamefont {A.}~\bibnamefont {Ringwald}},\ }\bibfield  {title} {\bibinfo {title} {{WISPy Cold Dark Matter}},\ }\href {https://doi.org/10.1088/1475-7516/2012/06/013} {\bibfield  {journal} {\bibinfo  {journal} {JCAP}\ }\textbf {\bibinfo {volume} {06}},\ \bibinfo {pages} {013}},\ \Eprint {https://arxiv.org/abs/1201.5902} {arXiv:1201.5902 [hep-ph]} \BibitemShut {NoStop}%
\bibitem [{\citenamefont {Chadha-Day}\ \emph {et~al.}(2022)\citenamefont {Chadha-Day}, \citenamefont {Ellis},\ and\ \citenamefont {Marsh}}]{Chadha-Day:2021szb}%
  \BibitemOpen
  \bibfield  {author} {\bibinfo {author} {\bibfnamefont {F.}~\bibnamefont {Chadha-Day}}, \bibinfo {author} {\bibfnamefont {J.}~\bibnamefont {Ellis}},\ and\ \bibinfo {author} {\bibfnamefont {D.~J.~E.}\ \bibnamefont {Marsh}},\ }\bibfield  {title} {\bibinfo {title} {{Axion dark matter: What is it and why now?}},\ }\href {https://doi.org/10.1126/sciadv.abj3618} {\bibfield  {journal} {\bibinfo  {journal} {Sci. Adv.}\ }\textbf {\bibinfo {volume} {8}},\ \bibinfo {pages} {abj3618} (\bibinfo {year} {2022})},\ \Eprint {https://arxiv.org/abs/2105.01406} {arXiv:2105.01406 [hep-ph]} \BibitemShut {NoStop}%
\bibitem [{\citenamefont {De~Angelis}\ \emph {et~al.}(2007)\citenamefont {De~Angelis}, \citenamefont {Roncadelli},\ and\ \citenamefont {Mansutti}}]{DeAngelis:2007dqd}%
  \BibitemOpen
  \bibfield  {author} {\bibinfo {author} {\bibfnamefont {A.}~\bibnamefont {De~Angelis}}, \bibinfo {author} {\bibfnamefont {M.}~\bibnamefont {Roncadelli}},\ and\ \bibinfo {author} {\bibfnamefont {O.}~\bibnamefont {Mansutti}},\ }\bibfield  {title} {\bibinfo {title} {{Evidence for a new light spin-zero boson from cosmological gamma-ray propagation?}},\ }\href {https://doi.org/10.1103/PhysRevD.76.121301} {\bibfield  {journal} {\bibinfo  {journal} {Phys. Rev. D}\ }\textbf {\bibinfo {volume} {76}},\ \bibinfo {pages} {121301} (\bibinfo {year} {2007})},\ \Eprint {https://arxiv.org/abs/0707.4312} {arXiv:0707.4312 [astro-ph]} \BibitemShut {NoStop}%
\bibitem [{\citenamefont {Simet}\ \emph {et~al.}(2008)\citenamefont {Simet}, \citenamefont {Hooper},\ and\ \citenamefont {Serpico}}]{Simet:2007sa}%
  \BibitemOpen
  \bibfield  {author} {\bibinfo {author} {\bibfnamefont {M.}~\bibnamefont {Simet}}, \bibinfo {author} {\bibfnamefont {D.}~\bibnamefont {Hooper}},\ and\ \bibinfo {author} {\bibfnamefont {P.~D.}\ \bibnamefont {Serpico}},\ }\bibfield  {title} {\bibinfo {title} {{The Milky Way as a Kiloparsec-Scale Axionscope}},\ }\href {https://doi.org/10.1103/PhysRevD.77.063001} {\bibfield  {journal} {\bibinfo  {journal} {Phys. Rev. D}\ }\textbf {\bibinfo {volume} {77}},\ \bibinfo {pages} {063001} (\bibinfo {year} {2008})},\ \Eprint {https://arxiv.org/abs/0712.2825} {arXiv:0712.2825 [astro-ph]} \BibitemShut {NoStop}%
\bibitem [{\citenamefont {Horns}\ and\ \citenamefont {Meyer}(2012)}]{Horns:2012fx}%
  \BibitemOpen
  \bibfield  {author} {\bibinfo {author} {\bibfnamefont {D.}~\bibnamefont {Horns}}\ and\ \bibinfo {author} {\bibfnamefont {M.}~\bibnamefont {Meyer}},\ }\bibfield  {title} {\bibinfo {title} {{Indications for a pair-production anomaly from the propagation of VHE gamma-rays}},\ }\href {https://doi.org/10.1088/1475-7516/2012/02/033} {\bibfield  {journal} {\bibinfo  {journal} {JCAP}\ }\textbf {\bibinfo {volume} {02}},\ \bibinfo {pages} {033}},\ \Eprint {https://arxiv.org/abs/1201.4711} {arXiv:1201.4711 [astro-ph.CO]} \BibitemShut {NoStop}%
\bibitem [{\citenamefont {Meyer}\ \emph {et~al.}(2013)\citenamefont {Meyer}, \citenamefont {Horns},\ and\ \citenamefont {Raue}}]{Meyer:2013pny}%
  \BibitemOpen
  \bibfield  {author} {\bibinfo {author} {\bibfnamefont {M.}~\bibnamefont {Meyer}}, \bibinfo {author} {\bibfnamefont {D.}~\bibnamefont {Horns}},\ and\ \bibinfo {author} {\bibfnamefont {M.}~\bibnamefont {Raue}},\ }\bibfield  {title} {\bibinfo {title} {{First lower limits on the photon-axion-like particle coupling from very high energy gamma-ray observations}},\ }\href {https://doi.org/10.1103/PhysRevD.87.035027} {\bibfield  {journal} {\bibinfo  {journal} {Phys. Rev. D}\ }\textbf {\bibinfo {volume} {87}},\ \bibinfo {pages} {035027} (\bibinfo {year} {2013})},\ \Eprint {https://arxiv.org/abs/1302.1208} {arXiv:1302.1208 [astro-ph.HE]} \BibitemShut {NoStop}%
\bibitem [{\citenamefont {Carosi}\ \emph {et~al.}(2013)\citenamefont {Carosi}, \citenamefont {Friedland}, \citenamefont {Giannotti}, \citenamefont {Pivovaroff}, \citenamefont {Ruz},\ and\ \citenamefont {Vogel}}]{Carosi:2013rla}%
  \BibitemOpen
  \bibfield  {author} {\bibinfo {author} {\bibfnamefont {G.}~\bibnamefont {Carosi}}, \bibinfo {author} {\bibfnamefont {A.}~\bibnamefont {Friedland}}, \bibinfo {author} {\bibfnamefont {M.}~\bibnamefont {Giannotti}}, \bibinfo {author} {\bibfnamefont {M.~J.}\ \bibnamefont {Pivovaroff}}, \bibinfo {author} {\bibfnamefont {J.}~\bibnamefont {Ruz}},\ and\ \bibinfo {author} {\bibfnamefont {J.~K.}\ \bibnamefont {Vogel}},\ }\bibfield  {title} {\bibinfo {title} {{Probing the axion-photon coupling: phenomenological and experimental perspectives. A snowmass white paper}},\ }in\ \href@noop {} {\emph {\bibinfo {booktitle} {{Snowmass 2013}: {Snowmass on the Mississippi}}}}\ (\bibinfo {year} {2013})\ \Eprint {https://arxiv.org/abs/1309.7035} {arXiv:1309.7035 [hep-ph]} \BibitemShut {NoStop}%
\bibitem [{\citenamefont {Graham}\ \emph {et~al.}(2015)\citenamefont {Graham}, \citenamefont {Irastorza}, \citenamefont {Lamoreaux}, \citenamefont {Lindner},\ and\ \citenamefont {van Bibber}}]{Graham:2015ouw}%
  \BibitemOpen
  \bibfield  {author} {\bibinfo {author} {\bibfnamefont {P.~W.}\ \bibnamefont {Graham}}, \bibinfo {author} {\bibfnamefont {I.~G.}\ \bibnamefont {Irastorza}}, \bibinfo {author} {\bibfnamefont {S.~K.}\ \bibnamefont {Lamoreaux}}, \bibinfo {author} {\bibfnamefont {A.}~\bibnamefont {Lindner}},\ and\ \bibinfo {author} {\bibfnamefont {K.~A.}\ \bibnamefont {van Bibber}},\ }\bibfield  {title} {\bibinfo {title} {{Experimental Searches for the Axion and Axion-Like Particles}},\ }\href {https://doi.org/10.1146/annurev-nucl-102014-022120} {\bibfield  {journal} {\bibinfo  {journal} {Ann. Rev. Nucl. Part. Sci.}\ }\textbf {\bibinfo {volume} {65}},\ \bibinfo {pages} {485} (\bibinfo {year} {2015})},\ \Eprint {https://arxiv.org/abs/1602.00039} {arXiv:1602.00039 [hep-ex]} \BibitemShut {NoStop}%
\bibitem [{\citenamefont {Berenji}\ \emph {et~al.}(2016)\citenamefont {Berenji}, \citenamefont {Gaskins},\ and\ \citenamefont {Meyer}}]{PhysRevD.93.045019}%
  \BibitemOpen
  \bibfield  {author} {\bibinfo {author} {\bibfnamefont {B.}~\bibnamefont {Berenji}}, \bibinfo {author} {\bibfnamefont {J.}~\bibnamefont {Gaskins}},\ and\ \bibinfo {author} {\bibfnamefont {M.}~\bibnamefont {Meyer}},\ }\bibfield  {title} {\bibinfo {title} {Constraints on axions and axionlike particles from fermi large area telescope observations of neutron stars},\ }\href {https://doi.org/10.1103/PhysRevD.93.045019} {\bibfield  {journal} {\bibinfo  {journal} {Phys. Rev. D}\ }\textbf {\bibinfo {volume} {93}},\ \bibinfo {pages} {045019} (\bibinfo {year} {2016})}\BibitemShut {NoStop}%
\bibitem [{\citenamefont {Irastorza}\ and\ \citenamefont {Redondo}(2018)}]{Irastorza:2018dyq}%
  \BibitemOpen
  \bibfield  {author} {\bibinfo {author} {\bibfnamefont {I.~G.}\ \bibnamefont {Irastorza}}\ and\ \bibinfo {author} {\bibfnamefont {J.}~\bibnamefont {Redondo}},\ }\bibfield  {title} {\bibinfo {title} {{New experimental approaches in the search for axion-like particles}},\ }\href {https://doi.org/10.1016/j.ppnp.2018.05.003} {\bibfield  {journal} {\bibinfo  {journal} {Prog. Part. Nucl. Phys.}\ }\textbf {\bibinfo {volume} {102}},\ \bibinfo {pages} {89} (\bibinfo {year} {2018})},\ \Eprint {https://arxiv.org/abs/1801.08127} {arXiv:1801.08127 [hep-ph]} \BibitemShut {NoStop}%
\bibitem [{\citenamefont {Majumdar}\ \emph {et~al.}(2018)\citenamefont {Majumdar}, \citenamefont {Calore},\ and\ \citenamefont {Horns}}]{Majumdar:2018sbv}%
  \BibitemOpen
  \bibfield  {author} {\bibinfo {author} {\bibfnamefont {J.}~\bibnamefont {Majumdar}}, \bibinfo {author} {\bibfnamefont {F.}~\bibnamefont {Calore}},\ and\ \bibinfo {author} {\bibfnamefont {D.}~\bibnamefont {Horns}},\ }\bibfield  {title} {\bibinfo {title} {{Search for gamma-ray spectral modulations in Galactic pulsars}},\ }\href {https://doi.org/10.1088/1475-7516/2018/04/048} {\bibfield  {journal} {\bibinfo  {journal} {JCAP}\ }\textbf {\bibinfo {volume} {04}},\ \bibinfo {pages} {048}},\ \Eprint {https://arxiv.org/abs/1801.08813} {arXiv:1801.08813 [hep-ph]} \BibitemShut {NoStop}%
\bibitem [{\citenamefont {Di~Luzio}\ \emph {et~al.}(2020)\citenamefont {Di~Luzio}, \citenamefont {Giannotti}, \citenamefont {Nardi},\ and\ \citenamefont {Visinelli}}]{DiLuzio:2020wdo}%
  \BibitemOpen
  \bibfield  {author} {\bibinfo {author} {\bibfnamefont {L.}~\bibnamefont {Di~Luzio}}, \bibinfo {author} {\bibfnamefont {M.}~\bibnamefont {Giannotti}}, \bibinfo {author} {\bibfnamefont {E.}~\bibnamefont {Nardi}},\ and\ \bibinfo {author} {\bibfnamefont {L.}~\bibnamefont {Visinelli}},\ }\bibfield  {title} {\bibinfo {title} {{The landscape of QCD axion models}},\ }\href {https://doi.org/10.1016/j.physrep.2020.06.002} {\bibfield  {journal} {\bibinfo  {journal} {Phys. Rept.}\ }\textbf {\bibinfo {volume} {870}},\ \bibinfo {pages} {1} (\bibinfo {year} {2020})},\ \Eprint {https://arxiv.org/abs/2003.01100} {arXiv:2003.01100 [hep-ph]} \BibitemShut {NoStop}%
\bibitem [{\citenamefont {Choi}\ \emph {et~al.}(2021)\citenamefont {Choi}, \citenamefont {Im},\ and\ \citenamefont {Sub~Shin}}]{Choi:2020rgn}%
  \BibitemOpen
  \bibfield  {author} {\bibinfo {author} {\bibfnamefont {K.}~\bibnamefont {Choi}}, \bibinfo {author} {\bibfnamefont {S.~H.}\ \bibnamefont {Im}},\ and\ \bibinfo {author} {\bibfnamefont {C.}~\bibnamefont {Sub~Shin}},\ }\bibfield  {title} {\bibinfo {title} {{Recent Progress in the Physics of Axions and Axion-Like Particles}},\ }\href {https://doi.org/10.1146/annurev-nucl-120720-031147} {\bibfield  {journal} {\bibinfo  {journal} {Ann. Rev. Nucl. Part. Sci.}\ }\textbf {\bibinfo {volume} {71}},\ \bibinfo {pages} {225} (\bibinfo {year} {2021})},\ \Eprint {https://arxiv.org/abs/2012.05029} {arXiv:2012.05029 [hep-ph]} \BibitemShut {NoStop}%
\bibitem [{\citenamefont {Chen}\ \emph {et~al.}(2022)\citenamefont {Chen}, \citenamefont {Liu}, \citenamefont {Lu}, \citenamefont {Mizuno}, \citenamefont {Shu}, \citenamefont {Xue}, \citenamefont {Yuan},\ and\ \citenamefont {Zhao}}]{Chen:2021lvo}%
  \BibitemOpen
  \bibfield  {author} {\bibinfo {author} {\bibfnamefont {Y.}~\bibnamefont {Chen}}, \bibinfo {author} {\bibfnamefont {Y.}~\bibnamefont {Liu}}, \bibinfo {author} {\bibfnamefont {R.-S.}\ \bibnamefont {Lu}}, \bibinfo {author} {\bibfnamefont {Y.}~\bibnamefont {Mizuno}}, \bibinfo {author} {\bibfnamefont {J.}~\bibnamefont {Shu}}, \bibinfo {author} {\bibfnamefont {X.}~\bibnamefont {Xue}}, \bibinfo {author} {\bibfnamefont {Q.}~\bibnamefont {Yuan}},\ and\ \bibinfo {author} {\bibfnamefont {Y.}~\bibnamefont {Zhao}},\ }\bibfield  {title} {\bibinfo {title} {{Stringent axion constraints with Event Horizon Telescope polarimetric measurements of M87*}},\ }\href {https://doi.org/10.1038/s41550-022-01620-3} {\bibfield  {journal} {\bibinfo  {journal} {Nature Astron.}\ }\textbf {\bibinfo {volume} {6}},\ \bibinfo {pages} {592} (\bibinfo {year} {2022})},\ \Eprint {https://arxiv.org/abs/2105.04572} {arXiv:2105.04572 [hep-ph]} \BibitemShut {NoStop}%
\bibitem [{\citenamefont {Zhou}\ \emph {et~al.}(2022)\citenamefont {Zhou}, \citenamefont {Houston}, \citenamefont {J\'ozsa}, \citenamefont {Chen}, \citenamefont {Ma}, \citenamefont {Yuan}, \citenamefont {An}, \citenamefont {Chandola}, \citenamefont {Ding}, \citenamefont {Du}, \citenamefont {Guo}, \citenamefont {Huang}, \citenamefont {Li},\ and\ \citenamefont {Sengupta}}]{PhysRevD.106.083006}%
  \BibitemOpen
  \bibfield  {author} {\bibinfo {author} {\bibfnamefont {Y.-F.}\ \bibnamefont {Zhou}}, \bibinfo {author} {\bibfnamefont {N.}~\bibnamefont {Houston}}, \bibinfo {author} {\bibfnamefont {G.~I.~G.}\ \bibnamefont {J\'ozsa}}, \bibinfo {author} {\bibfnamefont {H.}~\bibnamefont {Chen}}, \bibinfo {author} {\bibfnamefont {Y.-Z.}\ \bibnamefont {Ma}}, \bibinfo {author} {\bibfnamefont {Q.}~\bibnamefont {Yuan}}, \bibinfo {author} {\bibfnamefont {T.}~\bibnamefont {An}}, \bibinfo {author} {\bibfnamefont {Y.}~\bibnamefont {Chandola}}, \bibinfo {author} {\bibfnamefont {R.}~\bibnamefont {Ding}}, \bibinfo {author} {\bibfnamefont {F.}~\bibnamefont {Du}}, \bibinfo {author} {\bibfnamefont {S.-G.}\ \bibnamefont {Guo}}, \bibinfo {author} {\bibfnamefont {X.}~\bibnamefont {Huang}}, \bibinfo {author} {\bibfnamefont {M.}~\bibnamefont {Li}},\ and\ \bibinfo {author} {\bibfnamefont {C.}~\bibnamefont {Sengupta}},\ }\bibfield  {title} {\bibinfo {title} {Searching for axion dark matter with the meerkat radio telescope},\ }\href
  {https://doi.org/10.1103/PhysRevD.106.083006} {\bibfield  {journal} {\bibinfo  {journal} {Phys. Rev. D}\ }\textbf {\bibinfo {volume} {106}},\ \bibinfo {pages} {083006} (\bibinfo {year} {2022})}\BibitemShut {NoStop}%
\bibitem [{\citenamefont {{Agrawal}}\ \emph {et~al.}(2022)\citenamefont {{Agrawal}}, \citenamefont {{Nee}},\ and\ \citenamefont {{Reig}}}]{2022JHEP...10..141A}%
  \BibitemOpen
  \bibfield  {author} {\bibinfo {author} {\bibfnamefont {P.}~\bibnamefont {{Agrawal}}}, \bibinfo {author} {\bibfnamefont {M.}~\bibnamefont {{Nee}}},\ and\ \bibinfo {author} {\bibfnamefont {M.}~\bibnamefont {{Reig}}},\ }\bibfield  {title} {\bibinfo {title} {{Axion couplings in grand unified theories}},\ }\href {https://doi.org/10.1007/JHEP10(2022)141} {\bibfield  {journal} {\bibinfo  {journal} {Journal of High Energy Physics}\ }\textbf {\bibinfo {volume} {2022}},\ \bibinfo {eid} {141} (\bibinfo {year} {2022})},\ \Eprint {https://arxiv.org/abs/2206.07053} {arXiv:2206.07053 [hep-ph]} \BibitemShut {NoStop}%
\bibitem [{\citenamefont {Guo}\ \emph {et~al.}(2024{\natexlab{a}})\citenamefont {Guo}, \citenamefont {Khlopov}, \citenamefont {Liu}, \citenamefont {Wu}, \citenamefont {Wu},\ and\ \citenamefont {Zhu}}]{Guo:2023hyp}%
  \BibitemOpen
  \bibfield  {author} {\bibinfo {author} {\bibfnamefont {S.-Y.}\ \bibnamefont {Guo}}, \bibinfo {author} {\bibfnamefont {M.}~\bibnamefont {Khlopov}}, \bibinfo {author} {\bibfnamefont {X.}~\bibnamefont {Liu}}, \bibinfo {author} {\bibfnamefont {L.}~\bibnamefont {Wu}}, \bibinfo {author} {\bibfnamefont {Y.}~\bibnamefont {Wu}},\ and\ \bibinfo {author} {\bibfnamefont {B.}~\bibnamefont {Zhu}},\ }\bibfield  {title} {\bibinfo {title} {{Footprints of axion-like particle in pulsar timing array data and James Webb Space Telescope observations}},\ }\href {https://doi.org/10.1007/s11433-024-2445-1} {\bibfield  {journal} {\bibinfo  {journal} {Sci. China Phys. Mech. Astron.}\ }\textbf {\bibinfo {volume} {67}},\ \bibinfo {pages} {111011} (\bibinfo {year} {2024}{\natexlab{a}})},\ \Eprint {https://arxiv.org/abs/2306.17022} {arXiv:2306.17022 [hep-ph]} \BibitemShut {NoStop}%
\bibitem [{\citenamefont {Gong}\ \emph {et~al.}(2024)\citenamefont {Gong}, \citenamefont {Liu}, \citenamefont {Wu}, \citenamefont {Yang},\ and\ \citenamefont {Zhu}}]{Gong:2023ilg}%
  \BibitemOpen
  \bibfield  {author} {\bibinfo {author} {\bibfnamefont {Y.}~\bibnamefont {Gong}}, \bibinfo {author} {\bibfnamefont {X.}~\bibnamefont {Liu}}, \bibinfo {author} {\bibfnamefont {L.}~\bibnamefont {Wu}}, \bibinfo {author} {\bibfnamefont {Q.}~\bibnamefont {Yang}},\ and\ \bibinfo {author} {\bibfnamefont {B.}~\bibnamefont {Zhu}},\ }\bibfield  {title} {\bibinfo {title} {{Detecting quadratically coupled ultralight dark matter with stimulated annihilation}},\ }\href {https://doi.org/10.1103/PhysRevD.109.055026} {\bibfield  {journal} {\bibinfo  {journal} {Phys. Rev. D}\ }\textbf {\bibinfo {volume} {109}},\ \bibinfo {pages} {055026} (\bibinfo {year} {2024})},\ \Eprint {https://arxiv.org/abs/2308.08477} {arXiv:2308.08477 [hep-ph]} \BibitemShut {NoStop}%
\bibitem [{\citenamefont {Arza}\ \emph {et~al.}(2024)\citenamefont {Arza}, \citenamefont {Guo}, \citenamefont {Wu}, \citenamefont {Yang}, \citenamefont {Yang}, \citenamefont {Yuan},\ and\ \citenamefont {Zhu}}]{Arza:2023rcs}%
  \BibitemOpen
  \bibfield  {author} {\bibinfo {author} {\bibfnamefont {A.}~\bibnamefont {Arza}}, \bibinfo {author} {\bibfnamefont {Q.}~\bibnamefont {Guo}}, \bibinfo {author} {\bibfnamefont {L.}~\bibnamefont {Wu}}, \bibinfo {author} {\bibfnamefont {Q.}~\bibnamefont {Yang}}, \bibinfo {author} {\bibfnamefont {X.}~\bibnamefont {Yang}}, \bibinfo {author} {\bibfnamefont {Q.}~\bibnamefont {Yuan}},\ and\ \bibinfo {author} {\bibfnamefont {B.}~\bibnamefont {Zhu}},\ }\bibfield  {title} {\bibinfo {title} {{Listening for echo from the stimulated axion decay with the 21 centimeter array}},\ }\href {https://doi.org/10.1016/j.scib.2024.08.003} {\bibfield  {journal} {\bibinfo  {journal} {Sci. Bull.}\ }\textbf {\bibinfo {volume} {69}},\ \bibinfo {pages} {2971} (\bibinfo {year} {2024})},\ \Eprint {https://arxiv.org/abs/2309.06857} {arXiv:2309.06857 [hep-ph]} \BibitemShut {NoStop}%
\bibitem [{\citenamefont {Li}\ \emph {et~al.}(2024)\citenamefont {Li}, \citenamefont {Bi}, \citenamefont {Gao}, \citenamefont {Huang}, \citenamefont {Yao},\ and\ \citenamefont {Yin}}]{Li:2024ivs}%
  \BibitemOpen
  \bibfield  {author} {\bibinfo {author} {\bibfnamefont {J.}~\bibnamefont {Li}}, \bibinfo {author} {\bibfnamefont {X.-J.}\ \bibnamefont {Bi}}, \bibinfo {author} {\bibfnamefont {L.-Q.}\ \bibnamefont {Gao}}, \bibinfo {author} {\bibfnamefont {X.}~\bibnamefont {Huang}}, \bibinfo {author} {\bibfnamefont {R.-M.}\ \bibnamefont {Yao}},\ and\ \bibinfo {author} {\bibfnamefont {P.-F.}\ \bibnamefont {Yin}},\ }\bibfield  {title} {\bibinfo {title} {{Constraints on axion-like particles from the observation of Galactic sources by the LHAASO*}},\ }\href {https://doi.org/10.1088/1674-1137/ad361e} {\bibfield  {journal} {\bibinfo  {journal} {Chin. Phys. C}\ }\textbf {\bibinfo {volume} {48}},\ \bibinfo {pages} {065107} (\bibinfo {year} {2024})},\ \Eprint {https://arxiv.org/abs/2401.01829} {arXiv:2401.01829 [astro-ph.HE]} \BibitemShut {NoStop}%
\bibitem [{\citenamefont {Song}\ \emph {et~al.}(2025)\citenamefont {Song}, \citenamefont {Su},\ and\ \citenamefont {Wu}}]{Song:2024rru}%
  \BibitemOpen
  \bibfield  {author} {\bibinfo {author} {\bibfnamefont {N.}~\bibnamefont {Song}}, \bibinfo {author} {\bibfnamefont {L.}~\bibnamefont {Su}},\ and\ \bibinfo {author} {\bibfnamefont {L.}~\bibnamefont {Wu}},\ }\bibfield  {title} {\bibinfo {title} {{Polarization signals from axion-photon resonant conversion in a neutron star magnetosphere}},\ }\href {https://doi.org/10.1103/PhysRevD.111.043025} {\bibfield  {journal} {\bibinfo  {journal} {Phys. Rev. D}\ }\textbf {\bibinfo {volume} {111}},\ \bibinfo {pages} {043025} (\bibinfo {year} {2025})},\ \Eprint {https://arxiv.org/abs/2402.15144} {arXiv:2402.15144 [hep-ph]} \BibitemShut {NoStop}%
\bibitem [{\citenamefont {Lella}\ \emph {et~al.}(2024)\citenamefont {Lella}, \citenamefont {Ravensburg}, \citenamefont {Carenza},\ and\ \citenamefont {Marsh}}]{Lella:2024dmx}%
  \BibitemOpen
  \bibfield  {author} {\bibinfo {author} {\bibfnamefont {A.}~\bibnamefont {Lella}}, \bibinfo {author} {\bibfnamefont {E.}~\bibnamefont {Ravensburg}}, \bibinfo {author} {\bibfnamefont {P.}~\bibnamefont {Carenza}},\ and\ \bibinfo {author} {\bibfnamefont {M.~C.~D.}\ \bibnamefont {Marsh}},\ }\bibfield  {title} {\bibinfo {title} {{Supernova limits on QCD axionlike particles}},\ }\href {https://doi.org/10.1103/PhysRevD.110.043019} {\bibfield  {journal} {\bibinfo  {journal} {Phys. Rev. D}\ }\textbf {\bibinfo {volume} {110}},\ \bibinfo {pages} {043019} (\bibinfo {year} {2024})},\ \Eprint {https://arxiv.org/abs/2405.00153} {arXiv:2405.00153 [hep-ph]} \BibitemShut {NoStop}%
\bibitem [{\citenamefont {Yang}\ \emph {et~al.}(2024)\citenamefont {Yang}, \citenamefont {Guo}, \citenamefont {Luo}, \citenamefont {Shen}, \citenamefont {Xia}, \citenamefont {Lu}, \citenamefont {Tsai},\ and\ \citenamefont {Fan}}]{Yang:2024jtp}%
  \BibitemOpen
  \bibfield  {author} {\bibinfo {author} {\bibfnamefont {M.}~\bibnamefont {Yang}}, \bibinfo {author} {\bibfnamefont {Z.-Q.}\ \bibnamefont {Guo}}, \bibinfo {author} {\bibfnamefont {X.-Y.}\ \bibnamefont {Luo}}, \bibinfo {author} {\bibfnamefont {Z.-Q.}\ \bibnamefont {Shen}}, \bibinfo {author} {\bibfnamefont {Z.-Q.}\ \bibnamefont {Xia}}, \bibinfo {author} {\bibfnamefont {C.-T.}\ \bibnamefont {Lu}}, \bibinfo {author} {\bibfnamefont {Y.-L.~S.}\ \bibnamefont {Tsai}},\ and\ \bibinfo {author} {\bibfnamefont {Y.-Z.}\ \bibnamefont {Fan}},\ }\bibfield  {title} {\bibinfo {title} {{Searching accretion-enhanced dark matter annihilation signals in the Galactic Centre}},\ }\href {https://doi.org/10.1007/JHEP10(2024)094} {\bibfield  {journal} {\bibinfo  {journal} {JHEP}\ }\textbf {\bibinfo {volume} {10}},\ \bibinfo {pages} {094}},\ \Eprint {https://arxiv.org/abs/2407.06815} {arXiv:2407.06815 [hep-ph]} \BibitemShut {NoStop}%
\bibitem [{\citenamefont {Zhu}\ \emph {et~al.}(2025)\citenamefont {Zhu}, \citenamefont {Huang},\ and\ \citenamefont {Yin}}]{Zhu:2024kmu}%
  \BibitemOpen
  \bibfield  {author} {\bibinfo {author} {\bibfnamefont {B.-Y.}\ \bibnamefont {Zhu}}, \bibinfo {author} {\bibfnamefont {X.}~\bibnamefont {Huang}},\ and\ \bibinfo {author} {\bibfnamefont {P.-F.}\ \bibnamefont {Yin}},\ }\bibfield  {title} {\bibinfo {title} {{Constraints on axion-like particles from the gamma-ray observation of the Galactic Center}},\ }\href {https://doi.org/10.1088/1475-7516/2025/01/030} {\bibfield  {journal} {\bibinfo  {journal} {JCAP}\ }\textbf {\bibinfo {volume} {01}},\ \bibinfo {pages} {030}},\ \Eprint {https://arxiv.org/abs/2408.12234} {arXiv:2408.12234 [astro-ph.HE]} \BibitemShut {NoStop}%
\bibitem [{\citenamefont {Mammen~Abraham}\ \emph {et~al.}(2025)\citenamefont {Mammen~Abraham} \emph {et~al.}}]{FASER:2024bbl}%
  \BibitemOpen
  \bibfield  {author} {\bibinfo {author} {\bibfnamefont {R.}~\bibnamefont {Mammen~Abraham}} \emph {et~al.} (\bibinfo {collaboration} {FASER}),\ }\bibfield  {title} {\bibinfo {title} {{Shining light on the dark sector: search for axion-like particles and other new physics in photonic final states with FASER}},\ }\href {https://doi.org/10.1007/JHEP01(2025)199} {\bibfield  {journal} {\bibinfo  {journal} {JHEP}\ }\textbf {\bibinfo {volume} {01}},\ \bibinfo {pages} {199}},\ \Eprint {https://arxiv.org/abs/2410.10363} {arXiv:2410.10363 [hep-ex]} \BibitemShut {NoStop}%
\bibitem [{\citenamefont {Chen}\ \emph {et~al.}(2025)\citenamefont {Chen}, \citenamefont {Lei}, \citenamefont {Xia}, \citenamefont {Wang}, \citenamefont {Tsai},\ and\ \citenamefont {Fan}}]{Chen:2024ekh}%
  \BibitemOpen
  \bibfield  {author} {\bibinfo {author} {\bibfnamefont {Y.-X.}\ \bibnamefont {Chen}}, \bibinfo {author} {\bibfnamefont {L.}~\bibnamefont {Lei}}, \bibinfo {author} {\bibfnamefont {Z.-Q.}\ \bibnamefont {Xia}}, \bibinfo {author} {\bibfnamefont {Z.}~\bibnamefont {Wang}}, \bibinfo {author} {\bibfnamefont {Y.-L.~S.}\ \bibnamefont {Tsai}},\ and\ \bibinfo {author} {\bibfnamefont {Y.-Z.}\ \bibnamefont {Fan}},\ }\bibfield  {title} {\bibinfo {title} {{Searching for Axionlike Particles with X-Ray Observations of Alpha Centauri}},\ }\href {https://doi.org/10.1103/wy1x-1lh7} {\bibfield  {journal} {\bibinfo  {journal} {Phys. Rev. Lett.}\ }\textbf {\bibinfo {volume} {134}},\ \bibinfo {pages} {241001} (\bibinfo {year} {2025})},\ \Eprint {https://arxiv.org/abs/2410.16065} {arXiv:2410.16065 [astro-ph.HE]} \BibitemShut {NoStop}%
\bibitem [{\citenamefont {Guo}\ \emph {et~al.}(2024{\natexlab{b}})\citenamefont {Guo}, \citenamefont {Xia},\ and\ \citenamefont {Huang}}]{Guo:2024oqo}%
  \BibitemOpen
  \bibfield  {author} {\bibinfo {author} {\bibfnamefont {W.-Q.}\ \bibnamefont {Guo}}, \bibinfo {author} {\bibfnamefont {Z.-Q.}\ \bibnamefont {Xia}},\ and\ \bibinfo {author} {\bibfnamefont {X.}~\bibnamefont {Huang}},\ }\bibfield  {title} {\bibinfo {title} {{Constraining axion-like particles dark matter in Coma Berenices with FAST}},\ }\href {https://doi.org/10.1016/j.physletb.2024.138631} {\bibfield  {journal} {\bibinfo  {journal} {Phys. Lett. B}\ }\textbf {\bibinfo {volume} {852}},\ \bibinfo {pages} {138631} (\bibinfo {year} {2024}{\natexlab{b}})}\BibitemShut {NoStop}%
\bibitem [{\citenamefont {Malyshev}\ \emph {et~al.}(2025)\citenamefont {Malyshev}, \citenamefont {Zadorozhna}, \citenamefont {Bidasyuk}, \citenamefont {Santangelo},\ and\ \citenamefont {Ruchayskiy}}]{Malyshev:2025iis}%
  \BibitemOpen
  \bibfield  {author} {\bibinfo {author} {\bibfnamefont {D.}~\bibnamefont {Malyshev}}, \bibinfo {author} {\bibfnamefont {L.}~\bibnamefont {Zadorozhna}}, \bibinfo {author} {\bibfnamefont {Y.}~\bibnamefont {Bidasyuk}}, \bibinfo {author} {\bibfnamefont {A.}~\bibnamefont {Santangelo}},\ and\ \bibinfo {author} {\bibfnamefont {O.}~\bibnamefont {Ruchayskiy}},\ }\bibfield  {title} {\bibinfo {title} {{Active galactic nuclei through the prism of galaxy clusters: bounds on axion-like particles}}\ }\href {https://doi.org/10.1038/s41550-025-02621-8} {10.1038/s41550-025-02621-8} (\bibinfo {year} {2025}),\ \Eprint {https://arxiv.org/abs/2506.02848} {arXiv:2506.02848 [astro-ph.HE]} \BibitemShut {NoStop}%
\bibitem [{\citenamefont {{Pankratov}}\ \emph {et~al.}(2025)\citenamefont {{Pankratov}}, \citenamefont {{Belov}}, \citenamefont {{Boos}}, \citenamefont {{Chepurnov}}, \citenamefont {{Chiginev}}, \citenamefont {{Derbin}}, \citenamefont {{Drachnev}}, \citenamefont {{Dudko}}, \citenamefont {{Gorbunov}}, \citenamefont {{Gorlach}}, \citenamefont {{Ivanov}}, \citenamefont {{Kravchuk}}, \citenamefont {{Libanov}}, \citenamefont {{Merkin}}, \citenamefont {{Muratova}}, \citenamefont {{Pukhov}}, \citenamefont {{Salnikov}}, \citenamefont {{Satunin}}, \citenamefont {{Semenov}}, \citenamefont {{Sergeev}}, \citenamefont {{Starostin}}, \citenamefont {{Tkachev}}, \citenamefont {{Troitsky}}, \citenamefont {{Trushin}}, \citenamefont {{Unzhakov}}, \citenamefont {{Vyalkov}},\ and\ \citenamefont {{Yukhimchuk}}}]{Pankratov:2025cby}%
  \BibitemOpen
  \bibfield  {author} {\bibinfo {author} {\bibfnamefont {A.~L.}\ \bibnamefont {{Pankratov}}}, \bibinfo {author} {\bibfnamefont {P.~A.}\ \bibnamefont {{Belov}}}, \bibinfo {author} {\bibfnamefont {E.~E.}\ \bibnamefont {{Boos}}}, \bibinfo {author} {\bibfnamefont {A.~S.}\ \bibnamefont {{Chepurnov}}}, \bibinfo {author} {\bibfnamefont {A.~V.}\ \bibnamefont {{Chiginev}}}, \bibinfo {author} {\bibfnamefont {A.~V.}\ \bibnamefont {{Derbin}}}, \bibinfo {author} {\bibfnamefont {I.~S.}\ \bibnamefont {{Drachnev}}}, \bibinfo {author} {\bibfnamefont {L.~V.}\ \bibnamefont {{Dudko}}}, \bibinfo {author} {\bibfnamefont {D.~S.}\ \bibnamefont {{Gorbunov}}}, \bibinfo {author} {\bibfnamefont {M.~A.}\ \bibnamefont {{Gorlach}}}, \bibinfo {author} {\bibfnamefont {V.~V.}\ \bibnamefont {{Ivanov}}}, \bibinfo {author} {\bibfnamefont {L.~V.}\ \bibnamefont {{Kravchuk}}}, \bibinfo {author} {\bibfnamefont {M.~V.}\ \bibnamefont {{Libanov}}}, \bibinfo {author} {\bibfnamefont {M.~M.}\ \bibnamefont {{Merkin}}}, \bibinfo {author} {\bibfnamefont
  {V.~N.}\ \bibnamefont {{Muratova}}}, \bibinfo {author} {\bibfnamefont {A.~E.}\ \bibnamefont {{Pukhov}}}, \bibinfo {author} {\bibfnamefont {D.~V.}\ \bibnamefont {{Salnikov}}}, \bibinfo {author} {\bibfnamefont {P.~S.}\ \bibnamefont {{Satunin}}}, \bibinfo {author} {\bibfnamefont {D.~A.}\ \bibnamefont {{Semenov}}}, \bibinfo {author} {\bibfnamefont {A.~M.}\ \bibnamefont {{Sergeev}}}, \bibinfo {author} {\bibfnamefont {M.~I.}\ \bibnamefont {{Starostin}}}, \bibinfo {author} {\bibfnamefont {I.~I.}\ \bibnamefont {{Tkachev}}}, \bibinfo {author} {\bibfnamefont {S.~V.}\ \bibnamefont {{Troitsky}}}, \bibinfo {author} {\bibfnamefont {M.~V.}\ \bibnamefont {{Trushin}}}, \bibinfo {author} {\bibfnamefont {E.~V.}\ \bibnamefont {{Unzhakov}}}, \bibinfo {author} {\bibfnamefont {M.~M.}\ \bibnamefont {{Vyalkov}}},\ and\ \bibinfo {author} {\bibfnamefont {A.~A.}\ \bibnamefont {{Yukhimchuk}}},\ }\bibfield  {title} {\bibinfo {title} {{Search for dark-matter axions beyond the quantum limit: the Cosmological Axion Sarov Haloscope (CASH)
  proposal}},\ }\href {https://doi.org/10.48550/arXiv.2506.18595} {\bibfield  {journal} {\bibinfo  {journal} {arXiv e-prints}\ ,\ \bibinfo {eid} {arXiv:2506.18595}} (\bibinfo {year} {2025})},\ \Eprint {https://arxiv.org/abs/2506.18595} {arXiv:2506.18595 [hep-ph]} \BibitemShut {NoStop}%
\bibitem [{\citenamefont {Ehret}\ \emph {et~al.}(2010)\citenamefont {Ehret} \emph {et~al.}}]{Ehret:2010mh}%
  \BibitemOpen
  \bibfield  {author} {\bibinfo {author} {\bibfnamefont {K.}~\bibnamefont {Ehret}} \emph {et~al.},\ }\bibfield  {title} {\bibinfo {title} {{New ALPS Results on Hidden-Sector Lightweights}},\ }\href {https://doi.org/10.1016/j.physletb.2010.04.066} {\bibfield  {journal} {\bibinfo  {journal} {Phys. Lett. B}\ }\textbf {\bibinfo {volume} {689}},\ \bibinfo {pages} {149} (\bibinfo {year} {2010})},\ \Eprint {https://arxiv.org/abs/1004.1313} {arXiv:1004.1313 [hep-ex]} \BibitemShut {NoStop}%
\bibitem [{\citenamefont {Ortiz}\ \emph {et~al.}(2022)\citenamefont {Ortiz} \emph {et~al.}}]{Ortiz:2020tgs}%
  \BibitemOpen
  \bibfield  {author} {\bibinfo {author} {\bibfnamefont {M.~D.}\ \bibnamefont {Ortiz}} \emph {et~al.},\ }\bibfield  {title} {\bibinfo {title} {{Design of the ALPS II optical system}},\ }\href {https://doi.org/10.1016/j.dark.2022.100968} {\bibfield  {journal} {\bibinfo  {journal} {Phys. Dark Univ.}\ }\textbf {\bibinfo {volume} {35}},\ \bibinfo {pages} {100968} (\bibinfo {year} {2022})},\ \Eprint {https://arxiv.org/abs/2009.14294} {arXiv:2009.14294 [physics.optics]} \BibitemShut {NoStop}%
\bibitem [{\citenamefont {Asztalos}\ \emph {et~al.}(2010)\citenamefont {Asztalos} \emph {et~al.}}]{ADMX:2009iij}%
  \BibitemOpen
  \bibfield  {author} {\bibinfo {author} {\bibfnamefont {S.~J.}\ \bibnamefont {Asztalos}} \emph {et~al.} (\bibinfo {collaboration} {ADMX}),\ }\bibfield  {title} {\bibinfo {title} {{A SQUID-based microwave cavity search for dark-matter axions}},\ }\href {https://doi.org/10.1103/PhysRevLett.104.041301} {\bibfield  {journal} {\bibinfo  {journal} {Phys. Rev. Lett.}\ }\textbf {\bibinfo {volume} {104}},\ \bibinfo {pages} {041301} (\bibinfo {year} {2010})},\ \Eprint {https://arxiv.org/abs/0910.5914} {arXiv:0910.5914 [astro-ph.CO]} \BibitemShut {NoStop}%
\bibitem [{\citenamefont {Du}\ \emph {et~al.}(2018)\citenamefont {Du} \emph {et~al.}}]{ADMX:2018gho}%
  \BibitemOpen
  \bibfield  {author} {\bibinfo {author} {\bibfnamefont {N.}~\bibnamefont {Du}} \emph {et~al.} (\bibinfo {collaboration} {ADMX}),\ }\bibfield  {title} {\bibinfo {title} {{A Search for Invisible Axion Dark Matter with the Axion Dark Matter Experiment}},\ }\href {https://doi.org/10.1103/PhysRevLett.120.151301} {\bibfield  {journal} {\bibinfo  {journal} {Phys. Rev. Lett.}\ }\textbf {\bibinfo {volume} {120}},\ \bibinfo {pages} {151301} (\bibinfo {year} {2018})},\ \Eprint {https://arxiv.org/abs/1804.05750} {arXiv:1804.05750 [hep-ex]} \BibitemShut {NoStop}%
\bibitem [{\citenamefont {Boutan}\ \emph {et~al.}(2018)\citenamefont {Boutan} \emph {et~al.}}]{ADMX:2018ogs}%
  \BibitemOpen
  \bibfield  {author} {\bibinfo {author} {\bibfnamefont {C.}~\bibnamefont {Boutan}} \emph {et~al.} (\bibinfo {collaboration} {ADMX}),\ }\bibfield  {title} {\bibinfo {title} {{Piezoelectrically Tuned Multimode Cavity Search for Axion Dark Matter}},\ }\href {https://doi.org/10.1103/PhysRevLett.121.261302} {\bibfield  {journal} {\bibinfo  {journal} {Phys. Rev. Lett.}\ }\textbf {\bibinfo {volume} {121}},\ \bibinfo {pages} {261302} (\bibinfo {year} {2018})},\ \Eprint {https://arxiv.org/abs/1901.00920} {arXiv:1901.00920 [hep-ex]} \BibitemShut {NoStop}%
\bibitem [{\citenamefont {Braine}\ \emph {et~al.}(2020)\citenamefont {Braine} \emph {et~al.}}]{ADMX:2019uok}%
  \BibitemOpen
  \bibfield  {author} {\bibinfo {author} {\bibfnamefont {T.}~\bibnamefont {Braine}} \emph {et~al.} (\bibinfo {collaboration} {ADMX}),\ }\bibfield  {title} {\bibinfo {title} {{Extended Search for the Invisible Axion with the Axion Dark Matter Experiment}},\ }\href {https://doi.org/10.1103/PhysRevLett.124.101303} {\bibfield  {journal} {\bibinfo  {journal} {Phys. Rev. Lett.}\ }\textbf {\bibinfo {volume} {124}},\ \bibinfo {pages} {101303} (\bibinfo {year} {2020})},\ \Eprint {https://arxiv.org/abs/1910.08638} {arXiv:1910.08638 [hep-ex]} \BibitemShut {NoStop}%
\bibitem [{\citenamefont {Crisosto}\ \emph {et~al.}(2020)\citenamefont {Crisosto}, \citenamefont {Sikivie}, \citenamefont {Sullivan}, \citenamefont {Tanner}, \citenamefont {Yang},\ and\ \citenamefont {Rybka}}]{Crisosto:2019fcj}%
  \BibitemOpen
  \bibfield  {author} {\bibinfo {author} {\bibfnamefont {N.}~\bibnamefont {Crisosto}}, \bibinfo {author} {\bibfnamefont {P.}~\bibnamefont {Sikivie}}, \bibinfo {author} {\bibfnamefont {N.~S.}\ \bibnamefont {Sullivan}}, \bibinfo {author} {\bibfnamefont {D.~B.}\ \bibnamefont {Tanner}}, \bibinfo {author} {\bibfnamefont {J.}~\bibnamefont {Yang}},\ and\ \bibinfo {author} {\bibfnamefont {G.}~\bibnamefont {Rybka}},\ }\bibfield  {title} {\bibinfo {title} {{ADMX SLIC: Results from a Superconducting $LC$ Circuit Investigating Cold Axions}},\ }\href {https://doi.org/10.1103/PhysRevLett.124.241101} {\bibfield  {journal} {\bibinfo  {journal} {Phys. Rev. Lett.}\ }\textbf {\bibinfo {volume} {124}},\ \bibinfo {pages} {241101} (\bibinfo {year} {2020})},\ \Eprint {https://arxiv.org/abs/1911.05772} {arXiv:1911.05772 [astro-ph.CO]} \BibitemShut {NoStop}%
\bibitem [{\citenamefont {Bartram}\ \emph {et~al.}(2023)\citenamefont {Bartram} \emph {et~al.}}]{ADMX:2021mio}%
  \BibitemOpen
  \bibfield  {author} {\bibinfo {author} {\bibfnamefont {C.}~\bibnamefont {Bartram}} \emph {et~al.} (\bibinfo {collaboration} {ADMX}),\ }\bibfield  {title} {\bibinfo {title} {{Dark matter axion search using a Josephson Traveling wave parametric amplifier}},\ }\href {https://doi.org/10.1063/5.0122907} {\bibfield  {journal} {\bibinfo  {journal} {Rev. Sci. Instrum.}\ }\textbf {\bibinfo {volume} {94}},\ \bibinfo {pages} {044703} (\bibinfo {year} {2023})},\ \Eprint {https://arxiv.org/abs/2110.10262} {arXiv:2110.10262 [hep-ex]} \BibitemShut {NoStop}%
\bibitem [{\citenamefont {Bartram}\ \emph {et~al.}(2021)\citenamefont {Bartram} \emph {et~al.}}]{ADMX:2021nhd}%
  \BibitemOpen
  \bibfield  {author} {\bibinfo {author} {\bibfnamefont {C.}~\bibnamefont {Bartram}} \emph {et~al.} (\bibinfo {collaboration} {ADMX}),\ }\bibfield  {title} {\bibinfo {title} {{Search for Invisible Axion Dark Matter in the 3.3\textendash{}4.2\,\,\ensuremath{\mu}eV Mass Range}},\ }\href {https://doi.org/10.1103/PhysRevLett.127.261803} {\bibfield  {journal} {\bibinfo  {journal} {Phys. Rev. Lett.}\ }\textbf {\bibinfo {volume} {127}},\ \bibinfo {pages} {261803} (\bibinfo {year} {2021})},\ \Eprint {https://arxiv.org/abs/2110.06096} {arXiv:2110.06096 [hep-ex]} \BibitemShut {NoStop}%
\bibitem [{\citenamefont {Goodman}\ \emph {et~al.}(2025)\citenamefont {Goodman} \emph {et~al.}}]{ADMX:2024xbv}%
  \BibitemOpen
  \bibfield  {author} {\bibinfo {author} {\bibfnamefont {C.}~\bibnamefont {Goodman}} \emph {et~al.} (\bibinfo {collaboration} {ADMX}),\ }\bibfield  {title} {\bibinfo {title} {{ADMX Axion Dark Matter Bounds around 3.3\,\,\ensuremath{\mu}eV with Dine-Fischler-Srednicki-Zhitnitsky Discovery Ability}},\ }\href {https://doi.org/10.1103/PhysRevLett.134.111002} {\bibfield  {journal} {\bibinfo  {journal} {Phys. Rev. Lett.}\ }\textbf {\bibinfo {volume} {134}},\ \bibinfo {pages} {111002} (\bibinfo {year} {2025})},\ \Eprint {https://arxiv.org/abs/2408.15227} {arXiv:2408.15227 [hep-ex]} \BibitemShut {NoStop}%
\bibitem [{\citenamefont {Anastassopoulos}\ \emph {et~al.}(2017)\citenamefont {Anastassopoulos} \emph {et~al.}}]{CAST:2017uph}%
  \BibitemOpen
  \bibfield  {author} {\bibinfo {author} {\bibfnamefont {V.}~\bibnamefont {Anastassopoulos}} \emph {et~al.} (\bibinfo {collaboration} {CAST}),\ }\bibfield  {title} {\bibinfo {title} {{New CAST Limit on the Axion-Photon Interaction}},\ }\href {https://doi.org/10.1038/nphys4109} {\bibfield  {journal} {\bibinfo  {journal} {Nature Phys.}\ }\textbf {\bibinfo {volume} {13}},\ \bibinfo {pages} {584} (\bibinfo {year} {2017})},\ \Eprint {https://arxiv.org/abs/1705.02290} {arXiv:1705.02290 [hep-ex]} \BibitemShut {NoStop}%
\bibitem [{\citenamefont {Armengaud}\ \emph {et~al.}(2019)\citenamefont {Armengaud} \emph {et~al.}}]{IAXO:2019mpb}%
  \BibitemOpen
  \bibfield  {author} {\bibinfo {author} {\bibfnamefont {E.}~\bibnamefont {Armengaud}} \emph {et~al.} (\bibinfo {collaboration} {IAXO}),\ }\bibfield  {title} {\bibinfo {title} {{Physics potential of the International Axion Observatory (IAXO)}},\ }\href {https://doi.org/10.1088/1475-7516/2019/06/047} {\bibfield  {journal} {\bibinfo  {journal} {JCAP}\ }\textbf {\bibinfo {volume} {06}},\ \bibinfo {pages} {047}},\ \Eprint {https://arxiv.org/abs/1904.09155} {arXiv:1904.09155 [hep-ph]} \BibitemShut {NoStop}%
\bibitem [{\citenamefont {Hooper}\ and\ \citenamefont {Serpico}(2007)}]{Hooper:2007bq}%
  \BibitemOpen
  \bibfield  {author} {\bibinfo {author} {\bibfnamefont {D.}~\bibnamefont {Hooper}}\ and\ \bibinfo {author} {\bibfnamefont {P.~D.}\ \bibnamefont {Serpico}},\ }\bibfield  {title} {\bibinfo {title} {{Detecting Axion-Like Particles With Gamma Ray Telescopes}},\ }\href {https://doi.org/10.1103/PhysRevLett.99.231102} {\bibfield  {journal} {\bibinfo  {journal} {Phys. Rev. Lett.}\ }\textbf {\bibinfo {volume} {99}},\ \bibinfo {pages} {231102} (\bibinfo {year} {2007})},\ \Eprint {https://arxiv.org/abs/0706.3203} {arXiv:0706.3203 [hep-ph]} \BibitemShut {NoStop}%
\bibitem [{\citenamefont {De~Angelis}\ \emph {et~al.}(2008)\citenamefont {De~Angelis}, \citenamefont {Mansutti},\ and\ \citenamefont {Roncadelli}}]{DeAngelis:2007wiw}%
  \BibitemOpen
  \bibfield  {author} {\bibinfo {author} {\bibfnamefont {A.}~\bibnamefont {De~Angelis}}, \bibinfo {author} {\bibfnamefont {O.}~\bibnamefont {Mansutti}},\ and\ \bibinfo {author} {\bibfnamefont {M.}~\bibnamefont {Roncadelli}},\ }\bibfield  {title} {\bibinfo {title} {{Axion-Like Particles, Cosmic Magnetic Fields and Gamma-Ray Astrophysics}},\ }\href {https://doi.org/10.1016/j.physletb.2007.12.012} {\bibfield  {journal} {\bibinfo  {journal} {Phys. Lett. B}\ }\textbf {\bibinfo {volume} {659}},\ \bibinfo {pages} {847} (\bibinfo {year} {2008})},\ \Eprint {https://arxiv.org/abs/0707.2695} {arXiv:0707.2695 [astro-ph]} \BibitemShut {NoStop}%
\bibitem [{\citenamefont {Meyer}\ \emph {et~al.}(2017)\citenamefont {Meyer}, \citenamefont {Giannotti}, \citenamefont {Mirizzi}, \citenamefont {Conrad},\ and\ \citenamefont {S\'anchez-Conde}}]{Meyer:2016wrm}%
  \BibitemOpen
  \bibfield  {author} {\bibinfo {author} {\bibfnamefont {M.}~\bibnamefont {Meyer}}, \bibinfo {author} {\bibfnamefont {M.}~\bibnamefont {Giannotti}}, \bibinfo {author} {\bibfnamefont {A.}~\bibnamefont {Mirizzi}}, \bibinfo {author} {\bibfnamefont {J.}~\bibnamefont {Conrad}},\ and\ \bibinfo {author} {\bibfnamefont {M.~A.}\ \bibnamefont {S\'anchez-Conde}},\ }\bibfield  {title} {\bibinfo {title} {{Fermi Large Area Telescope as a Galactic Supernovae Axionscope}},\ }\href {https://doi.org/10.1103/PhysRevLett.118.011103} {\bibfield  {journal} {\bibinfo  {journal} {Phys. Rev. Lett.}\ }\textbf {\bibinfo {volume} {118}},\ \bibinfo {pages} {011103} (\bibinfo {year} {2017})},\ \Eprint {https://arxiv.org/abs/1609.02350} {arXiv:1609.02350 [astro-ph.HE]} \BibitemShut {NoStop}%
\bibitem [{\citenamefont {{Benabou}}\ \emph {et~al.}(2025)\citenamefont {{Benabou}}, \citenamefont {{Dessert}}, \citenamefont {{Patra}}, \citenamefont {{Brink}}, \citenamefont {{Zheng}}, \citenamefont {{Filippenko}},\ and\ \citenamefont {{Safdi}}}]{Benabou:2025jcv}%
  \BibitemOpen
  \bibfield  {author} {\bibinfo {author} {\bibfnamefont {J.~N.}\ \bibnamefont {{Benabou}}}, \bibinfo {author} {\bibfnamefont {C.}~\bibnamefont {{Dessert}}}, \bibinfo {author} {\bibfnamefont {K.~C.}\ \bibnamefont {{Patra}}}, \bibinfo {author} {\bibfnamefont {T.~G.}\ \bibnamefont {{Brink}}}, \bibinfo {author} {\bibfnamefont {W.}~\bibnamefont {{Zheng}}}, \bibinfo {author} {\bibfnamefont {A.~V.}\ \bibnamefont {{Filippenko}}},\ and\ \bibinfo {author} {\bibfnamefont {B.~R.}\ \bibnamefont {{Safdi}}},\ }\bibfield  {title} {\bibinfo {title} {{Search for Axions in Magnetic White Dwarf Polarization at Lick and Keck Observatories}},\ }\href {https://doi.org/10.48550/arXiv.2504.12377} {\bibfield  {journal} {\bibinfo  {journal} {arXiv e-prints}\ ,\ \bibinfo {eid} {arXiv:2504.12377}} (\bibinfo {year} {2025})},\ \Eprint {https://arxiv.org/abs/2504.12377} {arXiv:2504.12377 [hep-ph]} \BibitemShut {NoStop}%
\bibitem [{\citenamefont {Ajello}\ \emph {et~al.}(2016)\citenamefont {Ajello} \emph {et~al.}}]{Fermi-LAT:2016nkz}%
  \BibitemOpen
  \bibfield  {author} {\bibinfo {author} {\bibfnamefont {M.}~\bibnamefont {Ajello}} \emph {et~al.} (\bibinfo {collaboration} {Fermi-LAT}),\ }\bibfield  {title} {\bibinfo {title} {{Search for Spectral Irregularities due to Photon\textendash{}Axionlike-Particle Oscillations with the Fermi Large Area Telescope}},\ }\href {https://doi.org/10.1103/PhysRevLett.116.161101} {\bibfield  {journal} {\bibinfo  {journal} {Phys. Rev. Lett.}\ }\textbf {\bibinfo {volume} {116}},\ \bibinfo {pages} {161101} (\bibinfo {year} {2016})},\ \Eprint {https://arxiv.org/abs/1603.06978} {arXiv:1603.06978 [astro-ph.HE]} \BibitemShut {NoStop}%
\bibitem [{\citenamefont {{Payez}}\ \emph {et~al.}(2015)\citenamefont {{Payez}}, \citenamefont {{Evoli}}, \citenamefont {{Fischer}}, \citenamefont {{Giannotti}}, \citenamefont {{Mirizzi}},\ and\ \citenamefont {{Ringwald}}}]{2015JCAP...02..006P}%
  \BibitemOpen
  \bibfield  {author} {\bibinfo {author} {\bibfnamefont {A.}~\bibnamefont {{Payez}}}, \bibinfo {author} {\bibfnamefont {C.}~\bibnamefont {{Evoli}}}, \bibinfo {author} {\bibfnamefont {T.}~\bibnamefont {{Fischer}}}, \bibinfo {author} {\bibfnamefont {M.}~\bibnamefont {{Giannotti}}}, \bibinfo {author} {\bibfnamefont {A.}~\bibnamefont {{Mirizzi}}},\ and\ \bibinfo {author} {\bibfnamefont {A.}~\bibnamefont {{Ringwald}}},\ }\bibfield  {title} {\bibinfo {title} {{Revisiting the SN1987A gamma-ray limit on ultralight axion-like particles}},\ }\href {https://doi.org/10.1088/1475-7516/2015/02/006} {\bibfield  {journal} {\bibinfo  {journal} {Journal of Cosmology and Astroparticle Physics}\ }\textbf {\bibinfo {volume} {2015}}\bibfield  {number} {\bibinfo  {number} { (2)},\ \bibinfo {pages} {006}},\ }\Eprint {https://arxiv.org/abs/1410.3747} {arXiv:1410.3747 [astro-ph.HE]} \BibitemShut {NoStop}%
\bibitem [{\citenamefont {Meyer}\ and\ \citenamefont {Petrushevska}(2020)}]{Meyer:2020vzy}%
  \BibitemOpen
  \bibfield  {author} {\bibinfo {author} {\bibfnamefont {M.}~\bibnamefont {Meyer}}\ and\ \bibinfo {author} {\bibfnamefont {T.}~\bibnamefont {Petrushevska}},\ }\bibfield  {title} {\bibinfo {title} {{Search for Axionlike-Particle-Induced Prompt $\gamma$-Ray Emission from Extragalactic Core-Collapse Supernovae with the $Fermi$ Large Area Telescope}},\ }\href {https://doi.org/10.1103/PhysRevLett.124.231101} {\bibfield  {journal} {\bibinfo  {journal} {Phys. Rev. Lett.}\ }\textbf {\bibinfo {volume} {124}},\ \bibinfo {pages} {231101} (\bibinfo {year} {2020})},\ \bibinfo {note} {[Erratum: Phys.Rev.Lett. 125, 119901 (2020)]},\ \Eprint {https://arxiv.org/abs/2006.06722} {arXiv:2006.06722 [astro-ph.HE]} \BibitemShut {NoStop}%
\bibitem [{\citenamefont {Davies}\ \emph {et~al.}(2023)\citenamefont {Davies}, \citenamefont {Meyer},\ and\ \citenamefont {Cotter}}]{Davies:2022wvj}%
  \BibitemOpen
  \bibfield  {author} {\bibinfo {author} {\bibfnamefont {J.}~\bibnamefont {Davies}}, \bibinfo {author} {\bibfnamefont {M.}~\bibnamefont {Meyer}},\ and\ \bibinfo {author} {\bibfnamefont {G.}~\bibnamefont {Cotter}},\ }\bibfield  {title} {\bibinfo {title} {{Constraints on axionlike particles from a combined analysis of three flaring Fermi flat-spectrum radio quasars}},\ }\href {https://doi.org/10.1103/PhysRevD.107.083027} {\bibfield  {journal} {\bibinfo  {journal} {Phys. Rev. D}\ }\textbf {\bibinfo {volume} {107}},\ \bibinfo {pages} {083027} (\bibinfo {year} {2023})},\ \Eprint {https://arxiv.org/abs/2211.03414} {arXiv:2211.03414 [astro-ph.HE]} \BibitemShut {NoStop}%
\bibitem [{\citenamefont {Abramowski}\ \emph {et~al.}(2013)\citenamefont {Abramowski} \emph {et~al.}}]{HESS:2013udx}%
  \BibitemOpen
  \bibfield  {author} {\bibinfo {author} {\bibfnamefont {A.}~\bibnamefont {Abramowski}} \emph {et~al.} (\bibinfo {collaboration} {H.E.S.S.}),\ }\bibfield  {title} {\bibinfo {title} {{Constraints on axionlike particles with H.E.S.S. from the irregularity of the PKS 2155-304 energy spectrum}},\ }\href {https://doi.org/10.1103/PhysRevD.88.102003} {\bibfield  {journal} {\bibinfo  {journal} {Phys. Rev. D}\ }\textbf {\bibinfo {volume} {88}},\ \bibinfo {pages} {102003} (\bibinfo {year} {2013})},\ \Eprint {https://arxiv.org/abs/1311.3148} {arXiv:1311.3148 [astro-ph.HE]} \BibitemShut {NoStop}%
\bibitem [{\citenamefont {Zhang}\ \emph {et~al.}(2018)\citenamefont {Zhang}, \citenamefont {Liang}, \citenamefont {Li}, \citenamefont {Liao}, \citenamefont {Feng}, \citenamefont {Yuan}, \citenamefont {Fan},\ and\ \citenamefont {Ren}}]{Zhang:2018wpc}%
  \BibitemOpen
  \bibfield  {author} {\bibinfo {author} {\bibfnamefont {C.}~\bibnamefont {Zhang}}, \bibinfo {author} {\bibfnamefont {Y.-F.}\ \bibnamefont {Liang}}, \bibinfo {author} {\bibfnamefont {S.}~\bibnamefont {Li}}, \bibinfo {author} {\bibfnamefont {N.-H.}\ \bibnamefont {Liao}}, \bibinfo {author} {\bibfnamefont {L.}~\bibnamefont {Feng}}, \bibinfo {author} {\bibfnamefont {Q.}~\bibnamefont {Yuan}}, \bibinfo {author} {\bibfnamefont {Y.-Z.}\ \bibnamefont {Fan}},\ and\ \bibinfo {author} {\bibfnamefont {Z.-Z.}\ \bibnamefont {Ren}},\ }\bibfield  {title} {\bibinfo {title} {{New bounds on axionlike particles from the Fermi Large Area Telescope observation of PKS 2155-304}},\ }\href {https://doi.org/10.1103/PhysRevD.97.063009} {\bibfield  {journal} {\bibinfo  {journal} {Phys. Rev. D}\ }\textbf {\bibinfo {volume} {97}},\ \bibinfo {pages} {063009} (\bibinfo {year} {2018})},\ \Eprint {https://arxiv.org/abs/1802.08420} {arXiv:1802.08420 [hep-ph]} \BibitemShut {NoStop}%
\bibitem [{\citenamefont {Liang}\ \emph {et~al.}(2019)\citenamefont {Liang}, \citenamefont {Zhang}, \citenamefont {Xia}, \citenamefont {Feng}, \citenamefont {Yuan},\ and\ \citenamefont {Fan}}]{Liang:2018mqm}%
  \BibitemOpen
  \bibfield  {author} {\bibinfo {author} {\bibfnamefont {Y.-F.}\ \bibnamefont {Liang}}, \bibinfo {author} {\bibfnamefont {C.}~\bibnamefont {Zhang}}, \bibinfo {author} {\bibfnamefont {Z.-Q.}\ \bibnamefont {Xia}}, \bibinfo {author} {\bibfnamefont {L.}~\bibnamefont {Feng}}, \bibinfo {author} {\bibfnamefont {Q.}~\bibnamefont {Yuan}},\ and\ \bibinfo {author} {\bibfnamefont {Y.-Z.}\ \bibnamefont {Fan}},\ }\bibfield  {title} {\bibinfo {title} {{Constraints on axion-like particle properties with TeV gamma-ray observations of Galactic sources}},\ }\href {https://doi.org/10.1088/1475-7516/2019/06/042} {\bibfield  {journal} {\bibinfo  {journal} {JCAP}\ }\textbf {\bibinfo {volume} {06}},\ \bibinfo {pages} {042}},\ \Eprint {https://arxiv.org/abs/1804.07186} {arXiv:1804.07186 [hep-ph]} \BibitemShut {NoStop}%
\bibitem [{\citenamefont {Xia}\ \emph {et~al.}(2019)\citenamefont {Xia}, \citenamefont {Liang}, \citenamefont {Feng}, \citenamefont {Yuan}, \citenamefont {Fan},\ and\ \citenamefont {Wu}}]{Xia:2019yud}%
  \BibitemOpen
  \bibfield  {author} {\bibinfo {author} {\bibfnamefont {Z.-Q.}\ \bibnamefont {Xia}}, \bibinfo {author} {\bibfnamefont {Y.-F.}\ \bibnamefont {Liang}}, \bibinfo {author} {\bibfnamefont {L.}~\bibnamefont {Feng}}, \bibinfo {author} {\bibfnamefont {Q.}~\bibnamefont {Yuan}}, \bibinfo {author} {\bibfnamefont {Y.-Z.}\ \bibnamefont {Fan}},\ and\ \bibinfo {author} {\bibfnamefont {J.}~\bibnamefont {Wu}},\ }\bibfield  {title} {\bibinfo {title} {{Searching for the possible signal of the photon-axionlike particle oscillation in the combined GeV and TeV spectra of supernova remnants}},\ }\href {https://doi.org/10.1103/PhysRevD.100.123004} {\bibfield  {journal} {\bibinfo  {journal} {Phys. Rev. D}\ }\textbf {\bibinfo {volume} {100}},\ \bibinfo {pages} {123004} (\bibinfo {year} {2019})},\ \Eprint {https://arxiv.org/abs/1911.08096} {arXiv:1911.08096 [astro-ph.HE]} \BibitemShut {NoStop}%
\bibitem [{\citenamefont {Yuan}\ \emph {et~al.}(2021)\citenamefont {Yuan}, \citenamefont {Xia}, \citenamefont {Tang}, \citenamefont {Zhao}, \citenamefont {Cai}, \citenamefont {Chen}, \citenamefont {Shu},\ and\ \citenamefont {Yuan}}]{Yuan:2020xui}%
  \BibitemOpen
  \bibfield  {author} {\bibinfo {author} {\bibfnamefont {G.-W.}\ \bibnamefont {Yuan}}, \bibinfo {author} {\bibfnamefont {Z.-Q.}\ \bibnamefont {Xia}}, \bibinfo {author} {\bibfnamefont {C.}~\bibnamefont {Tang}}, \bibinfo {author} {\bibfnamefont {Y.}~\bibnamefont {Zhao}}, \bibinfo {author} {\bibfnamefont {Y.-F.}\ \bibnamefont {Cai}}, \bibinfo {author} {\bibfnamefont {Y.}~\bibnamefont {Chen}}, \bibinfo {author} {\bibfnamefont {J.}~\bibnamefont {Shu}},\ and\ \bibinfo {author} {\bibfnamefont {Q.}~\bibnamefont {Yuan}},\ }\bibfield  {title} {\bibinfo {title} {{Testing the ALP-photon coupling with polarization measurements of Sagittarius A$^*$}},\ }\href {https://doi.org/10.1088/1475-7516/2021/03/018} {\bibfield  {journal} {\bibinfo  {journal} {JCAP}\ }\textbf {\bibinfo {volume} {03}},\ \bibinfo {pages} {018}},\ \Eprint {https://arxiv.org/abs/2008.13662} {arXiv:2008.13662 [astro-ph.HE]} \BibitemShut {NoStop}%
\bibitem [{\citenamefont {Ackermann}\ \emph {et~al.}(2015)\citenamefont {Ackermann} \emph {et~al.}}]{Fermi-LAT:2015bdd}%
  \BibitemOpen
  \bibfield  {author} {\bibinfo {author} {\bibfnamefont {M.}~\bibnamefont {Ackermann}} \emph {et~al.} (\bibinfo {collaboration} {Fermi-LAT}),\ }\bibfield  {title} {\bibinfo {title} {{The Third Catalog of Active Galactic Nuclei Detected by the Fermi Large Area Telescope}},\ }\href {https://doi.org/10.1088/0004-637X/810/1/14} {\bibfield  {journal} {\bibinfo  {journal} {Astrophys. J.}\ }\textbf {\bibinfo {volume} {810}},\ \bibinfo {pages} {14} (\bibinfo {year} {2015})},\ \Eprint {https://arxiv.org/abs/1501.06054} {arXiv:1501.06054 [astro-ph.HE]} \BibitemShut {NoStop}%
\bibitem [{\citenamefont {Aleksi{\'c}}\ \emph {et~al.}(2014)\citenamefont {Aleksi{\'c}} \emph {et~al.}}]{MAGIC:2013mvu}%
  \BibitemOpen
  \bibfield  {author} {\bibinfo {author} {\bibfnamefont {J.}~\bibnamefont {Aleksi{\'c}}} \emph {et~al.} (\bibinfo {collaboration} {MAGIC}),\ }\bibfield  {title} {\bibinfo {title} {{Contemporaneous observations of the radio galaxy NGC 1275 from radio to very high energy $\gamma$-rays}},\ }\href {https://doi.org/10.1051/0004-6361/201322951} {\bibfield  {journal} {\bibinfo  {journal} {Astron. Astrophys.}\ }\textbf {\bibinfo {volume} {564}},\ \bibinfo {pages} {A5} (\bibinfo {year} {2014})},\ \Eprint {https://arxiv.org/abs/1310.8500} {arXiv:1310.8500 [astro-ph.HE]} \BibitemShut {NoStop}%
\bibitem [{\citenamefont {Tavecchio}\ and\ \citenamefont {Ghisellini}(2014)}]{Tavecchio:2014oja}%
  \BibitemOpen
  \bibfield  {author} {\bibinfo {author} {\bibfnamefont {F.}~\bibnamefont {Tavecchio}}\ and\ \bibinfo {author} {\bibfnamefont {G.}~\bibnamefont {Ghisellini}},\ }\bibfield  {title} {\bibinfo {title} {{On the spine-layer scenario for the very high-energy emission of NGC 1275}},\ }\href {https://doi.org/10.1093/mnras/stu1196} {\bibfield  {journal} {\bibinfo  {journal} {Mon. Not. Roy. Astron. Soc.}\ }\textbf {\bibinfo {volume} {443}},\ \bibinfo {pages} {1224} (\bibinfo {year} {2014})},\ \Eprint {https://arxiv.org/abs/1404.6894} {arXiv:1404.6894 [astro-ph.HE]} \BibitemShut {NoStop}%
\bibitem [{\citenamefont {Taylor}\ \emph {et~al.}(2006)\citenamefont {Taylor}, \citenamefont {Gugliucci}, \citenamefont {Fabian}, \citenamefont {Sanders}, \citenamefont {Gentile},\ and\ \citenamefont {Allen}}]{Taylor:2006ta}%
  \BibitemOpen
  \bibfield  {author} {\bibinfo {author} {\bibfnamefont {G.~B.}\ \bibnamefont {Taylor}}, \bibinfo {author} {\bibfnamefont {N.~E.}\ \bibnamefont {Gugliucci}}, \bibinfo {author} {\bibfnamefont {A.~C.}\ \bibnamefont {Fabian}}, \bibinfo {author} {\bibfnamefont {J.~S.}\ \bibnamefont {Sanders}}, \bibinfo {author} {\bibfnamefont {G.}~\bibnamefont {Gentile}},\ and\ \bibinfo {author} {\bibfnamefont {S.~W.}\ \bibnamefont {Allen}},\ }\bibfield  {title} {\bibinfo {title} {{Magnetic fields in the center of the perseus cluster}},\ }\href {https://doi.org/10.1111/j.1365-2966.2006.10244.x} {\bibfield  {journal} {\bibinfo  {journal} {Mon. Not. Roy. Astron. Soc.}\ }\textbf {\bibinfo {volume} {368}},\ \bibinfo {pages} {1500} (\bibinfo {year} {2006})},\ \Eprint {https://arxiv.org/abs/astro-ph/0602622} {arXiv:astro-ph/0602622} \BibitemShut {NoStop}%
\bibitem [{\citenamefont {{Fan}}\ \emph {et~al.}(2022)\citenamefont {{Fan}}, \citenamefont {{Chang}}, \citenamefont {{Guo}}, \citenamefont {{Yuan}}, \citenamefont {{Hu}}, \citenamefont {{Li}}, \citenamefont {{Yue}}, \citenamefont {{Huang}}, \citenamefont {{Liu}}, \citenamefont {{Feng}}, \citenamefont {{Zhang}}, \citenamefont {{Wei}}, \citenamefont {{Sun}}, \citenamefont {{Yu}}, \citenamefont {{Kong}}, \citenamefont {{Zhao}}, \citenamefont {{Zang}}, \citenamefont {{Jiang}}, \citenamefont {{Pan}}, \citenamefont {{Wei}}, \citenamefont {{Wang}}, \citenamefont {{Duan}}, \citenamefont {{Shen}}, \citenamefont {{Xia}}, \citenamefont {{Xu}}, \citenamefont {{Feng}}, \citenamefont {{Huang}}, \citenamefont {{TSAI}}, \citenamefont {{Wei}}, \citenamefont {{Zeng}}, \citenamefont {{He}}, \citenamefont {{Li}}, \citenamefont {{Yang}}, \citenamefont {{Yan}}, \citenamefont {{Zhang}}, \citenamefont {{Wu}},\ and\ \citenamefont {{Wei}}}]{2022AcASn..63...27F}%
  \BibitemOpen
  \bibfield  {author} {\bibinfo {author} {\bibfnamefont {Y.~Z.}\ \bibnamefont {{Fan}}}, \bibinfo {author} {\bibfnamefont {J.}~\bibnamefont {{Chang}}}, \bibinfo {author} {\bibfnamefont {J.~H.}\ \bibnamefont {{Guo}}}, \bibinfo {author} {\bibfnamefont {Q.}~\bibnamefont {{Yuan}}}, \bibinfo {author} {\bibfnamefont {Y.~M.}\ \bibnamefont {{Hu}}}, \bibinfo {author} {\bibfnamefont {X.}~\bibnamefont {{Li}}}, \bibinfo {author} {\bibfnamefont {C.}~\bibnamefont {{Yue}}}, \bibinfo {author} {\bibfnamefont {G.~S.}\ \bibnamefont {{Huang}}}, \bibinfo {author} {\bibfnamefont {S.~B.}\ \bibnamefont {{Liu}}}, \bibinfo {author} {\bibfnamefont {C.~Q.}\ \bibnamefont {{Feng}}}, \bibinfo {author} {\bibfnamefont {Y.~L.}\ \bibnamefont {{Zhang}}}, \bibinfo {author} {\bibfnamefont {Y.~F.}\ \bibnamefont {{Wei}}}, \bibinfo {author} {\bibfnamefont {Z.~Y.}\ \bibnamefont {{Sun}}}, \bibinfo {author} {\bibfnamefont {Y.~H.}\ \bibnamefont {{Yu}}}, \bibinfo {author} {\bibfnamefont {J.}~\bibnamefont {{Kong}}}, \bibinfo {author} {\bibfnamefont
  {C.~X.}\ \bibnamefont {{Zhao}}}, \bibinfo {author} {\bibfnamefont {J.~J.}\ \bibnamefont {{Zang}}}, \bibinfo {author} {\bibfnamefont {W.}~\bibnamefont {{Jiang}}}, \bibinfo {author} {\bibfnamefont {X.}~\bibnamefont {{Pan}}}, \bibinfo {author} {\bibfnamefont {J.~J.}\ \bibnamefont {{Wei}}}, \bibinfo {author} {\bibfnamefont {S.}~\bibnamefont {{Wang}}}, \bibinfo {author} {\bibfnamefont {K.~K.}\ \bibnamefont {{Duan}}}, \bibinfo {author} {\bibfnamefont {Z.~Q.}\ \bibnamefont {{Shen}}}, \bibinfo {author} {\bibfnamefont {Z.~Q.}\ \bibnamefont {{Xia}}}, \bibinfo {author} {\bibfnamefont {Z.~L.}\ \bibnamefont {{Xu}}}, \bibinfo {author} {\bibfnamefont {L.}~\bibnamefont {{Feng}}}, \bibinfo {author} {\bibfnamefont {X.~Y.}\ \bibnamefont {{Huang}}}, \bibinfo {author} {\bibfnamefont {Y.~L.}\ \bibnamefont {{TSAI}}}, \bibinfo {author} {\bibfnamefont {J.~J.}\ \bibnamefont {{Wei}}}, \bibinfo {author} {\bibfnamefont {H.~D.}\ \bibnamefont {{Zeng}}}, \bibinfo {author} {\bibfnamefont {H.~N.}\ \bibnamefont {{He}}}, \bibinfo {author}
  {\bibfnamefont {J.}~\bibnamefont {{Li}}}, \bibinfo {author} {\bibfnamefont {R.~Z.}\ \bibnamefont {{Yang}}}, \bibinfo {author} {\bibfnamefont {J.~Z.}\ \bibnamefont {{Yan}}}, \bibinfo {author} {\bibfnamefont {Y.}~\bibnamefont {{Zhang}}}, \bibinfo {author} {\bibfnamefont {X.~F.}\ \bibnamefont {{Wu}}},\ and\ \bibinfo {author} {\bibfnamefont {D.~M.}\ \bibnamefont {{Wei}}},\ }\bibfield  {title} {\bibinfo {title} {{Very Large Area Gamma-ray Space Telescope (VLAST)}},\ }\href@noop {} {\bibfield  {journal} {\bibinfo  {journal} {Acta Astronomica Sinica}\ }\textbf {\bibinfo {volume} {63}},\ \bibinfo {eid} {27} (\bibinfo {year} {2022})}\BibitemShut {NoStop}%
\bibitem [{\citenamefont {Agostinelli}\ \emph {et~al.}(2003)\citenamefont {Agostinelli} \emph {et~al.}}]{GEANT4:2002zbu}%
  \BibitemOpen
  \bibfield  {author} {\bibinfo {author} {\bibfnamefont {S.}~\bibnamefont {Agostinelli}} \emph {et~al.} (\bibinfo {collaboration} {GEANT4}),\ }\bibfield  {title} {\bibinfo {title} {{GEANT4 - A Simulation Toolkit}},\ }\href {https://doi.org/10.1016/S0168-9002(03)01368-8} {\bibfield  {journal} {\bibinfo  {journal} {Nucl. Instrum. Meth. A}\ }\textbf {\bibinfo {volume} {506}},\ \bibinfo {pages} {250} (\bibinfo {year} {2003})}\BibitemShut {NoStop}%
\bibitem [{\citenamefont {Raffelt}\ and\ \citenamefont {Stodolsky}(1988)}]{Raffelt:1987im}%
  \BibitemOpen
  \bibfield  {author} {\bibinfo {author} {\bibfnamefont {G.}~\bibnamefont {Raffelt}}\ and\ \bibinfo {author} {\bibfnamefont {L.}~\bibnamefont {Stodolsky}},\ }\bibfield  {title} {\bibinfo {title} {{Mixing of the Photon with Low Mass Particles}},\ }\href {https://doi.org/10.1103/PhysRevD.37.1237} {\bibfield  {journal} {\bibinfo  {journal} {Phys. Rev. D}\ }\textbf {\bibinfo {volume} {37}},\ \bibinfo {pages} {1237} (\bibinfo {year} {1988})}\BibitemShut {NoStop}%
\bibitem [{\citenamefont {Grossman}\ \emph {et~al.}(2002)\citenamefont {Grossman}, \citenamefont {Roy},\ and\ \citenamefont {Zupan}}]{Grossman:2002by}%
  \BibitemOpen
  \bibfield  {author} {\bibinfo {author} {\bibfnamefont {Y.}~\bibnamefont {Grossman}}, \bibinfo {author} {\bibfnamefont {S.}~\bibnamefont {Roy}},\ and\ \bibinfo {author} {\bibfnamefont {J.}~\bibnamefont {Zupan}},\ }\bibfield  {title} {\bibinfo {title} {{Effects of initial axion production and photon axion oscillation on type Ia supernova dimming}},\ }\href {https://doi.org/10.1016/S0370-2693(02)02448-6} {\bibfield  {journal} {\bibinfo  {journal} {Phys. Lett. B}\ }\textbf {\bibinfo {volume} {543}},\ \bibinfo {pages} {23} (\bibinfo {year} {2002})},\ \Eprint {https://arxiv.org/abs/hep-ph/0204216} {arXiv:hep-ph/0204216} \BibitemShut {NoStop}%
\bibitem [{\citenamefont {Csaki}\ \emph {et~al.}(2003)\citenamefont {Csaki}, \citenamefont {Kaloper}, \citenamefont {Peloso},\ and\ \citenamefont {Terning}}]{Csaki:2003ef}%
  \BibitemOpen
  \bibfield  {author} {\bibinfo {author} {\bibfnamefont {C.}~\bibnamefont {Csaki}}, \bibinfo {author} {\bibfnamefont {N.}~\bibnamefont {Kaloper}}, \bibinfo {author} {\bibfnamefont {M.}~\bibnamefont {Peloso}},\ and\ \bibinfo {author} {\bibfnamefont {J.}~\bibnamefont {Terning}},\ }\bibfield  {title} {\bibinfo {title} {{Super GZK photons from photon axion mixing}},\ }\href {https://doi.org/10.1088/1475-7516/2003/05/005} {\bibfield  {journal} {\bibinfo  {journal} {JCAP}\ }\textbf {\bibinfo {volume} {05}},\ \bibinfo {pages} {005}},\ \Eprint {https://arxiv.org/abs/hep-ph/0302030} {arXiv:hep-ph/0302030} \BibitemShut {NoStop}%
\bibitem [{\citenamefont {Mirizzi}\ \emph {et~al.}(2008)\citenamefont {Mirizzi}, \citenamefont {Raffelt},\ and\ \citenamefont {Serpico}}]{Mirizzi:2006zy}%
  \BibitemOpen
  \bibfield  {author} {\bibinfo {author} {\bibfnamefont {A.}~\bibnamefont {Mirizzi}}, \bibinfo {author} {\bibfnamefont {G.~G.}\ \bibnamefont {Raffelt}},\ and\ \bibinfo {author} {\bibfnamefont {P.~D.}\ \bibnamefont {Serpico}},\ }\bibfield  {title} {\bibinfo {title} {{Photon-axion conversion in intergalactic magnetic fields and cosmological consequences}},\ }\href {https://doi.org/10.1007/978-3-540-73518-2_7} {\bibfield  {journal} {\bibinfo  {journal} {Lect. Notes Phys.}\ }\textbf {\bibinfo {volume} {741}},\ \bibinfo {pages} {115} (\bibinfo {year} {2008})},\ \Eprint {https://arxiv.org/abs/astro-ph/0607415} {arXiv:astro-ph/0607415} \BibitemShut {NoStop}%
\bibitem [{\citenamefont {Mirizzi}\ and\ \citenamefont {Montanino}(2009)}]{Mirizzi:2009aj}%
  \BibitemOpen
  \bibfield  {author} {\bibinfo {author} {\bibfnamefont {A.}~\bibnamefont {Mirizzi}}\ and\ \bibinfo {author} {\bibfnamefont {D.}~\bibnamefont {Montanino}},\ }\bibfield  {title} {\bibinfo {title} {{Stochastic conversions of TeV photons into axion-like particles in extragalactic magnetic fields}},\ }\href {https://doi.org/10.1088/1475-7516/2009/12/004} {\bibfield  {journal} {\bibinfo  {journal} {JCAP}\ }\textbf {\bibinfo {volume} {12}},\ \bibinfo {pages} {004}},\ \Eprint {https://arxiv.org/abs/0911.0015} {arXiv:0911.0015 [astro-ph.HE]} \BibitemShut {NoStop}%
\bibitem [{\citenamefont {De~Angelis}\ \emph {et~al.}(2011)\citenamefont {De~Angelis}, \citenamefont {Galanti},\ and\ \citenamefont {Roncadelli}}]{DeAngelis:2011id}%
  \BibitemOpen
  \bibfield  {author} {\bibinfo {author} {\bibfnamefont {A.}~\bibnamefont {De~Angelis}}, \bibinfo {author} {\bibfnamefont {G.}~\bibnamefont {Galanti}},\ and\ \bibinfo {author} {\bibfnamefont {M.}~\bibnamefont {Roncadelli}},\ }\bibfield  {title} {\bibinfo {title} {{Relevance of axion-like particles for very-high-energy astrophysics}},\ }\href {https://doi.org/10.1103/PhysRevD.84.105030} {\bibfield  {journal} {\bibinfo  {journal} {Phys. Rev. D}\ }\textbf {\bibinfo {volume} {84}},\ \bibinfo {pages} {105030} (\bibinfo {year} {2011})},\ \bibinfo {note} {[Erratum: Phys.Rev.D 87, 109903 (2013)]},\ \Eprint {https://arxiv.org/abs/1106.1132} {arXiv:1106.1132 [astro-ph.HE]} \BibitemShut {NoStop}%
\bibitem [{\citenamefont {Meyer}\ \emph {et~al.}(2014)\citenamefont {Meyer}, \citenamefont {Montanino},\ and\ \citenamefont {Conrad}}]{Meyer:2014epa}%
  \BibitemOpen
  \bibfield  {author} {\bibinfo {author} {\bibfnamefont {M.}~\bibnamefont {Meyer}}, \bibinfo {author} {\bibfnamefont {D.}~\bibnamefont {Montanino}},\ and\ \bibinfo {author} {\bibfnamefont {J.}~\bibnamefont {Conrad}},\ }\bibfield  {title} {\bibinfo {title} {{On detecting oscillations of gamma rays into axion-like particles in turbulent and coherent magnetic fields}},\ }\href {https://doi.org/10.1088/1475-7516/2014/09/003} {\bibfield  {journal} {\bibinfo  {journal} {JCAP}\ }\textbf {\bibinfo {volume} {09}},\ \bibinfo {pages} {003}},\ \Eprint {https://arxiv.org/abs/1406.5972} {arXiv:1406.5972 [astro-ph.HE]} \BibitemShut {NoStop}%
\bibitem [{\citenamefont {Choi}\ \emph {et~al.}(2020)\citenamefont {Choi}, \citenamefont {Lee}, \citenamefont {Seong},\ and\ \citenamefont {Yun}}]{Choi:2018mvk}%
  \BibitemOpen
  \bibfield  {author} {\bibinfo {author} {\bibfnamefont {K.}~\bibnamefont {Choi}}, \bibinfo {author} {\bibfnamefont {S.}~\bibnamefont {Lee}}, \bibinfo {author} {\bibfnamefont {H.}~\bibnamefont {Seong}},\ and\ \bibinfo {author} {\bibfnamefont {S.}~\bibnamefont {Yun}},\ }\bibfield  {title} {\bibinfo {title} {{Gamma-ray spectral modulations induced by photon-ALP-dark photon oscillations}},\ }\href {https://doi.org/10.1103/PhysRevD.101.043007} {\bibfield  {journal} {\bibinfo  {journal} {Phys. Rev. D}\ }\textbf {\bibinfo {volume} {101}},\ \bibinfo {pages} {043007} (\bibinfo {year} {2020})},\ \Eprint {https://arxiv.org/abs/1806.09508} {arXiv:1806.09508 [hep-ph]} \BibitemShut {NoStop}%
\bibitem [{\citenamefont {Dolag}\ \emph {et~al.}(2008)\citenamefont {Dolag}, \citenamefont {Bykov},\ and\ \citenamefont {Diaferio}}]{Dolag:2008ks}%
  \BibitemOpen
  \bibfield  {author} {\bibinfo {author} {\bibfnamefont {K.}~\bibnamefont {Dolag}}, \bibinfo {author} {\bibfnamefont {A.~M.}\ \bibnamefont {Bykov}},\ and\ \bibinfo {author} {\bibfnamefont {A.}~\bibnamefont {Diaferio}},\ }\bibfield  {title} {\bibinfo {title} {{Non-thermal processes in cosmological simulations}},\ }\href {https://doi.org/10.1007/s11214-008-9319-2} {\bibfield  {journal} {\bibinfo  {journal} {Space Sci. Rev.}\ }\textbf {\bibinfo {volume} {134}},\ \bibinfo {pages} {311} (\bibinfo {year} {2008})},\ \Eprint {https://arxiv.org/abs/0801.1048} {arXiv:0801.1048 [astro-ph]} \BibitemShut {NoStop}%
\bibitem [{\citenamefont {Dubois}\ and\ \citenamefont {Teyssier}(2008)}]{Dubois:2008mz}%
  \BibitemOpen
  \bibfield  {author} {\bibinfo {author} {\bibfnamefont {Y.}~\bibnamefont {Dubois}}\ and\ \bibinfo {author} {\bibfnamefont {R.}~\bibnamefont {Teyssier}},\ }\bibfield  {title} {\bibinfo {title} {{Cosmological MHD simulation of a cooling flow cluster}},\ }\href {https://doi.org/10.1051/0004-6361:200809513} {\bibfield  {journal} {\bibinfo  {journal} {Astron. Astrophys.}\ }\textbf {\bibinfo {volume} {482}},\ \bibinfo {pages} {L13} (\bibinfo {year} {2008})},\ \Eprint {https://arxiv.org/abs/0802.0490} {arXiv:0802.0490 [astro-ph]} \BibitemShut {NoStop}%
\bibitem [{\citenamefont {Feretti}\ \emph {et~al.}(2012)\citenamefont {Feretti}, \citenamefont {Giovannini}, \citenamefont {Govoni},\ and\ \citenamefont {Murgia}}]{Feretti:2012vk}%
  \BibitemOpen
  \bibfield  {author} {\bibinfo {author} {\bibfnamefont {L.}~\bibnamefont {Feretti}}, \bibinfo {author} {\bibfnamefont {G.}~\bibnamefont {Giovannini}}, \bibinfo {author} {\bibfnamefont {F.}~\bibnamefont {Govoni}},\ and\ \bibinfo {author} {\bibfnamefont {M.}~\bibnamefont {Murgia}},\ }\bibfield  {title} {\bibinfo {title} {{Clusters of galaxies: observational properties of the diffuse radio emission}},\ }\href {https://doi.org/10.1007/s00159-012-0054-z} {\bibfield  {journal} {\bibinfo  {journal} {Astron. Astrophys. Rev.}\ }\textbf {\bibinfo {volume} {20}},\ \bibinfo {pages} {54} (\bibinfo {year} {2012})},\ \Eprint {https://arxiv.org/abs/1205.1919} {arXiv:1205.1919 [astro-ph.CO]} \BibitemShut {NoStop}%
\bibitem [{\citenamefont {Churazov}\ \emph {et~al.}(2003)\citenamefont {Churazov}, \citenamefont {Forman}, \citenamefont {Jones},\ and\ \citenamefont {Bohringer}}]{Churazov:2003hr}%
  \BibitemOpen
  \bibfield  {author} {\bibinfo {author} {\bibfnamefont {E.}~\bibnamefont {Churazov}}, \bibinfo {author} {\bibfnamefont {W.}~\bibnamefont {Forman}}, \bibinfo {author} {\bibfnamefont {C.}~\bibnamefont {Jones}},\ and\ \bibinfo {author} {\bibfnamefont {H.}~\bibnamefont {Bohringer}},\ }\bibfield  {title} {\bibinfo {title} {{Xmm-newton observations of the perseus cluster I: the temperature and surface brightness structure}},\ }\href {https://doi.org/10.1086/374923} {\bibfield  {journal} {\bibinfo  {journal} {Astrophys. J.}\ }\textbf {\bibinfo {volume} {590}},\ \bibinfo {pages} {225} (\bibinfo {year} {2003})},\ \Eprint {https://arxiv.org/abs/astro-ph/0301482} {arXiv:astro-ph/0301482} \BibitemShut {NoStop}%
\bibitem [{\citenamefont {Aleksic}\ \emph {et~al.}(2012)\citenamefont {Aleksic} \emph {et~al.}}]{MAGIC:2011vay}%
  \BibitemOpen
  \bibfield  {author} {\bibinfo {author} {\bibfnamefont {J.}~\bibnamefont {Aleksic}} \emph {et~al.} (\bibinfo {collaboration} {MAGIC}),\ }\bibfield  {title} {\bibinfo {title} {{Constraining Cosmic Rays and Magnetic Fields in the Perseus Galaxy Cluster with TeV observations by the MAGIC telescopes}},\ }\href {https://doi.org/10.1051/0004-6361/201118502} {\bibfield  {journal} {\bibinfo  {journal} {Astron. Astrophys.}\ }\textbf {\bibinfo {volume} {541}},\ \bibinfo {pages} {A99} (\bibinfo {year} {2012})},\ \Eprint {https://arxiv.org/abs/1111.5544} {arXiv:1111.5544 [astro-ph.HE]} \BibitemShut {NoStop}%
\bibitem [{\citenamefont {Kuchar}\ and\ \citenamefont {Ensslin}(2011)}]{Kuchar:2009jj}%
  \BibitemOpen
  \bibfield  {author} {\bibinfo {author} {\bibfnamefont {P.}~\bibnamefont {Kuchar}}\ and\ \bibinfo {author} {\bibfnamefont {T.~A.}\ \bibnamefont {Ensslin}},\ }\bibfield  {title} {\bibinfo {title} {{Magnetic power spectra from Faraday rotation maps - REALMAF and its use on Hydra A}},\ }\href {https://doi.org/10.1051/0004-6361/200913918} {\bibfield  {journal} {\bibinfo  {journal} {Astron. Astrophys.}\ }\textbf {\bibinfo {volume} {529}},\ \bibinfo {pages} {A13} (\bibinfo {year} {2011})},\ \Eprint {https://arxiv.org/abs/0912.3930} {arXiv:0912.3930 [astro-ph.CO]} \BibitemShut {NoStop}%
\bibitem [{\citenamefont {Vacca}\ \emph {et~al.}(2012)\citenamefont {Vacca}, \citenamefont {Murgia}, \citenamefont {Govoni}, \citenamefont {Feretti}, \citenamefont {Giovannini}, \citenamefont {Perley},\ and\ \citenamefont {Taylor}}]{Vacca:2012up}%
  \BibitemOpen
  \bibfield  {author} {\bibinfo {author} {\bibfnamefont {V.}~\bibnamefont {Vacca}}, \bibinfo {author} {\bibfnamefont {M.}~\bibnamefont {Murgia}}, \bibinfo {author} {\bibfnamefont {F.}~\bibnamefont {Govoni}}, \bibinfo {author} {\bibfnamefont {L.}~\bibnamefont {Feretti}}, \bibinfo {author} {\bibfnamefont {G.}~\bibnamefont {Giovannini}}, \bibinfo {author} {\bibfnamefont {R.~A.}\ \bibnamefont {Perley}},\ and\ \bibinfo {author} {\bibfnamefont {G.~B.}\ \bibnamefont {Taylor}},\ }\bibfield  {title} {\bibinfo {title} {{The intracluster magnetic field power spectrum in A2199}},\ }\href {https://doi.org/10.1051/0004-6361/201116622} {\bibfield  {journal} {\bibinfo  {journal} {Astron. Astrophys.}\ }\textbf {\bibinfo {volume} {540}},\ \bibinfo {pages} {A38} (\bibinfo {year} {2012})},\ \Eprint {https://arxiv.org/abs/1201.4119} {arXiv:1201.4119 [astro-ph.CO]} \BibitemShut {NoStop}%
\bibitem [{\citenamefont {Dominguez}\ \emph {et~al.}(2011)\citenamefont {Dominguez} \emph {et~al.}}]{Dominguez:2010bv}%
  \BibitemOpen
  \bibfield  {author} {\bibinfo {author} {\bibfnamefont {A.}~\bibnamefont {Dominguez}} \emph {et~al.},\ }\bibfield  {title} {\bibinfo {title} {{Extragalactic Background Light Inferred from AEGIS Galaxy SED-type Fractions}},\ }\href {https://doi.org/10.1111/j.1365-2966.2010.17631.x} {\bibfield  {journal} {\bibinfo  {journal} {Mon. Not. Roy. Astron. Soc.}\ }\textbf {\bibinfo {volume} {410}},\ \bibinfo {pages} {2556} (\bibinfo {year} {2011})},\ \Eprint {https://arxiv.org/abs/1007.1459} {arXiv:1007.1459 [astro-ph.CO]} \BibitemShut {NoStop}%
\bibitem [{\citenamefont {Jansson}\ and\ \citenamefont {Farrar}(2012)}]{Jansson:2012pc}%
  \BibitemOpen
  \bibfield  {author} {\bibinfo {author} {\bibfnamefont {R.}~\bibnamefont {Jansson}}\ and\ \bibinfo {author} {\bibfnamefont {G.~R.}\ \bibnamefont {Farrar}},\ }\bibfield  {title} {\bibinfo {title} {{A New Model of the Galactic Magnetic Field}},\ }\href {https://doi.org/10.1088/0004-637X/757/1/14} {\bibfield  {journal} {\bibinfo  {journal} {Astrophys. J.}\ }\textbf {\bibinfo {volume} {757}},\ \bibinfo {pages} {14} (\bibinfo {year} {2012})},\ \Eprint {https://arxiv.org/abs/1204.3662} {arXiv:1204.3662 [astro-ph.GA]} \BibitemShut {NoStop}%
\bibitem [{\citenamefont {Atwood}\ \emph {et~al.}(2013)\citenamefont {Atwood} \emph {et~al.}}]{Fermi-LAT:2013jgq}%
  \BibitemOpen
  \bibfield  {author} {\bibinfo {author} {\bibfnamefont {W.}~\bibnamefont {Atwood}} \emph {et~al.} (\bibinfo {collaboration} {Fermi-LAT}),\ }\bibfield  {title} {\bibinfo {title} {{Pass 8: Toward the Full Realization of the Fermi-LAT Scientific Potential}}\ }(\bibinfo {year} {2013})\ \Eprint {https://arxiv.org/abs/1303.3514} {arXiv:1303.3514 [astro-ph.IM]} \BibitemShut {NoStop}%
\bibitem [{fer()}]{fermi_data_access}%
  \BibitemOpen
  \href@noop {} {}\bibinfo {howpublished} {\url{https://fermi.gsfc.nasa.gov/ssc/data/access/}}\BibitemShut {NoStop}%
\bibitem [{\citenamefont {{Ballet}}\ \emph {et~al.}(2023)\citenamefont {{Ballet}}, \citenamefont {{Bruel}}, \citenamefont {{Burnett}}, \citenamefont {{Lott}},\ and\ \citenamefont {{The Fermi-LAT collaboration}}}]{Ballet:2023qzs}%
  \BibitemOpen
  \bibfield  {author} {\bibinfo {author} {\bibfnamefont {J.}~\bibnamefont {{Ballet}}}, \bibinfo {author} {\bibfnamefont {P.}~\bibnamefont {{Bruel}}}, \bibinfo {author} {\bibfnamefont {T.~H.}\ \bibnamefont {{Burnett}}}, \bibinfo {author} {\bibfnamefont {B.}~\bibnamefont {{Lott}}},\ and\ \bibinfo {author} {\bibnamefont {{The Fermi-LAT collaboration}}},\ }\bibfield  {title} {\bibinfo {title} {{Fermi Large Area Telescope Fourth Source Catalog Data Release 4 (4FGL-DR4)}},\ }\href {https://doi.org/10.48550/arXiv.2307.12546} {\bibfield  {journal} {\bibinfo  {journal} {arXiv e-prints}\ ,\ \bibinfo {eid} {arXiv:2307.12546}} (\bibinfo {year} {2023})},\ \Eprint {https://arxiv.org/abs/2307.12546} {arXiv:2307.12546 [astro-ph.HE]} \BibitemShut {NoStop}%
\bibitem [{\citenamefont {Kneiske}\ and\ \citenamefont {Dole}(2010)}]{Kneiske:2010pt}%
  \BibitemOpen
  \bibfield  {author} {\bibinfo {author} {\bibfnamefont {T.~M.}\ \bibnamefont {Kneiske}}\ and\ \bibinfo {author} {\bibfnamefont {H.}~\bibnamefont {Dole}},\ }\bibfield  {title} {\bibinfo {title} {{A Lower-Limit Flux for the Extragalactic Background Light}},\ }\href {https://doi.org/10.1051/0004-6361/200912000} {\bibfield  {journal} {\bibinfo  {journal} {Astron. Astrophys.}\ }\textbf {\bibinfo {volume} {515}},\ \bibinfo {pages} {A19} (\bibinfo {year} {2010})},\ \Eprint {https://arxiv.org/abs/1001.2132} {arXiv:1001.2132 [astro-ph.CO]} \BibitemShut {NoStop}%
\bibitem [{\citenamefont {Gilmore}\ \emph {et~al.}(2012)\citenamefont {Gilmore}, \citenamefont {Somerville}, \citenamefont {Primack},\ and\ \citenamefont {Dominguez}}]{Gilmore:2011ks}%
  \BibitemOpen
  \bibfield  {author} {\bibinfo {author} {\bibfnamefont {R.~C.}\ \bibnamefont {Gilmore}}, \bibinfo {author} {\bibfnamefont {R.~S.}\ \bibnamefont {Somerville}}, \bibinfo {author} {\bibfnamefont {J.~R.}\ \bibnamefont {Primack}},\ and\ \bibinfo {author} {\bibfnamefont {A.}~\bibnamefont {Dominguez}},\ }\bibfield  {title} {\bibinfo {title} {{Semi-analytic modeling of the EBL and consequences for extragalactic gamma-ray spectra}},\ }\href {https://doi.org/10.1111/j.1365-2966.2012.20841.x} {\bibfield  {journal} {\bibinfo  {journal} {Mon. Not. Roy. Astron. Soc.}\ }\textbf {\bibinfo {volume} {422}},\ \bibinfo {pages} {3189} (\bibinfo {year} {2012})},\ \Eprint {https://arxiv.org/abs/1104.0671} {arXiv:1104.0671 [astro-ph.CO]} \BibitemShut {NoStop}%
\bibitem [{\citenamefont {Inoue}\ \emph {et~al.}(2013)\citenamefont {Inoue}, \citenamefont {Inoue}, \citenamefont {Kobayashi}, \citenamefont {Makiya}, \citenamefont {Niino},\ and\ \citenamefont {Totani}}]{Inoue:2012bk}%
  \BibitemOpen
  \bibfield  {author} {\bibinfo {author} {\bibfnamefont {Y.}~\bibnamefont {Inoue}}, \bibinfo {author} {\bibfnamefont {S.}~\bibnamefont {Inoue}}, \bibinfo {author} {\bibfnamefont {M.~A.~R.}\ \bibnamefont {Kobayashi}}, \bibinfo {author} {\bibfnamefont {R.}~\bibnamefont {Makiya}}, \bibinfo {author} {\bibfnamefont {Y.}~\bibnamefont {Niino}},\ and\ \bibinfo {author} {\bibfnamefont {T.}~\bibnamefont {Totani}},\ }\bibfield  {title} {\bibinfo {title} {{Extragalactic Background Light from Hierarchical Galaxy Formation: Gamma-ray Attenuation up to the Epoch of Cosmic Reionization and the First Stars}},\ }\href {https://doi.org/10.1088/0004-637X/768/2/197} {\bibfield  {journal} {\bibinfo  {journal} {Astrophys. J.}\ }\textbf {\bibinfo {volume} {768}},\ \bibinfo {pages} {197} (\bibinfo {year} {2013})},\ \Eprint {https://arxiv.org/abs/1212.1683} {arXiv:1212.1683 [astro-ph.CO]} \BibitemShut {NoStop}%
\bibitem [{\citenamefont {Franceschini}\ and\ \citenamefont {Rodighiero}(2017)}]{Franceschini:2017iwq}%
  \BibitemOpen
  \bibfield  {author} {\bibinfo {author} {\bibfnamefont {A.}~\bibnamefont {Franceschini}}\ and\ \bibinfo {author} {\bibfnamefont {G.}~\bibnamefont {Rodighiero}},\ }\bibfield  {title} {\bibinfo {title} {{The extragalactic background light revisited and the cosmic photon-photon opacity}},\ }\href {https://doi.org/10.1051/0004-6361/201629684} {\bibfield  {journal} {\bibinfo  {journal} {Astron. Astrophys.}\ }\textbf {\bibinfo {volume} {603}},\ \bibinfo {pages} {A34} (\bibinfo {year} {2017})},\ \Eprint {https://arxiv.org/abs/1705.10256} {arXiv:1705.10256 [astro-ph.HE]} \BibitemShut {NoStop}%
\bibitem [{\citenamefont {Saldana-Lopez}\ \emph {et~al.}(2021)\citenamefont {Saldana-Lopez}, \citenamefont {Dom\'\i{}nguez}, \citenamefont {P\'erez-Gonz\'alez}, \citenamefont {Finke}, \citenamefont {Ajello}, \citenamefont {Primack}, \citenamefont {Paliya},\ and\ \citenamefont {Desai}}]{Saldana-Lopez:2020qzx}%
  \BibitemOpen
  \bibfield  {author} {\bibinfo {author} {\bibfnamefont {A.}~\bibnamefont {Saldana-Lopez}}, \bibinfo {author} {\bibfnamefont {A.}~\bibnamefont {Dom\'\i{}nguez}}, \bibinfo {author} {\bibfnamefont {P.~G.}\ \bibnamefont {P\'erez-Gonz\'alez}}, \bibinfo {author} {\bibfnamefont {J.}~\bibnamefont {Finke}}, \bibinfo {author} {\bibfnamefont {M.}~\bibnamefont {Ajello}}, \bibinfo {author} {\bibfnamefont {J.~R.}\ \bibnamefont {Primack}}, \bibinfo {author} {\bibfnamefont {V.~S.}\ \bibnamefont {Paliya}},\ and\ \bibinfo {author} {\bibfnamefont {A.}~\bibnamefont {Desai}},\ }\bibfield  {title} {\bibinfo {title} {{An observational determination of the evolving extragalactic background light from the multiwavelength HST/CANDELS survey in the Fermi and CTA era}},\ }\href {https://doi.org/10.1093/mnras/stab2393} {\bibfield  {journal} {\bibinfo  {journal} {Mon. Not. Roy. Astron. Soc.}\ }\textbf {\bibinfo {volume} {507}},\ \bibinfo {pages} {5144} (\bibinfo {year} {2021})},\ \Eprint {https://arxiv.org/abs/2012.03035}
  {arXiv:2012.03035 [astro-ph.CO]} \BibitemShut {NoStop}%
\bibitem [{\citenamefont {Finke}\ \emph {et~al.}(2022)\citenamefont {Finke}, \citenamefont {Ajello}, \citenamefont {Dominguez}, \citenamefont {Desai}, \citenamefont {Hartmann}, \citenamefont {Paliya},\ and\ \citenamefont {Saldana-Lopez}}]{Finke:2022uvv}%
  \BibitemOpen
  \bibfield  {author} {\bibinfo {author} {\bibfnamefont {J.~D.}\ \bibnamefont {Finke}}, \bibinfo {author} {\bibfnamefont {M.}~\bibnamefont {Ajello}}, \bibinfo {author} {\bibfnamefont {A.}~\bibnamefont {Dominguez}}, \bibinfo {author} {\bibfnamefont {A.}~\bibnamefont {Desai}}, \bibinfo {author} {\bibfnamefont {D.~H.}\ \bibnamefont {Hartmann}}, \bibinfo {author} {\bibfnamefont {V.~S.}\ \bibnamefont {Paliya}},\ and\ \bibinfo {author} {\bibfnamefont {A.}~\bibnamefont {Saldana-Lopez}},\ }\bibfield  {title} {\bibinfo {title} {{Modeling the Extragalactic Background Light and the Cosmic Star Formation History}},\ }\href {https://doi.org/10.3847/1538-4357/ac9843} {\bibfield  {journal} {\bibinfo  {journal} {Astrophys. J.}\ }\textbf {\bibinfo {volume} {941}},\ \bibinfo {pages} {33} (\bibinfo {year} {2022})},\ \Eprint {https://arxiv.org/abs/2210.01157} {arXiv:2210.01157 [astro-ph.GA]} \BibitemShut {NoStop}%
\bibitem [{\citenamefont {Adam}\ \emph {et~al.}(2016)\citenamefont {Adam} \emph {et~al.}}]{Planck:2016gdp}%
  \BibitemOpen
  \bibfield  {author} {\bibinfo {author} {\bibfnamefont {R.}~\bibnamefont {Adam}} \emph {et~al.} (\bibinfo {collaboration} {Planck}),\ }\bibfield  {title} {\bibinfo {title} {{Planck intermediate results.}: {XLII. Large-scale Galactic magnetic fields}},\ }\href {https://doi.org/10.1051/0004-6361/201528033} {\bibfield  {journal} {\bibinfo  {journal} {Astron. Astrophys.}\ }\textbf {\bibinfo {volume} {596}},\ \bibinfo {pages} {A103} (\bibinfo {year} {2016})},\ \Eprint {https://arxiv.org/abs/1601.00546} {arXiv:1601.00546 [astro-ph.GA]} \BibitemShut {NoStop}%
\bibitem [{\citenamefont {Pshirkov}\ \emph {et~al.}(2011)\citenamefont {Pshirkov}, \citenamefont {Tinyakov}, \citenamefont {Kronberg},\ and\ \citenamefont {Newton-McGee}}]{Pshirkov:2011um}%
  \BibitemOpen
  \bibfield  {author} {\bibinfo {author} {\bibfnamefont {M.~S.}\ \bibnamefont {Pshirkov}}, \bibinfo {author} {\bibfnamefont {P.~G.}\ \bibnamefont {Tinyakov}}, \bibinfo {author} {\bibfnamefont {P.~P.}\ \bibnamefont {Kronberg}},\ and\ \bibinfo {author} {\bibfnamefont {K.~J.}\ \bibnamefont {Newton-McGee}},\ }\bibfield  {title} {\bibinfo {title} {{Deriving global structure of the Galactic Magnetic Field from Faraday Rotation Measures of extragalactic sources}},\ }\href {https://doi.org/10.1088/0004-637X/738/2/192} {\bibfield  {journal} {\bibinfo  {journal} {Astrophys. J.}\ }\textbf {\bibinfo {volume} {738}},\ \bibinfo {pages} {192} (\bibinfo {year} {2011})},\ \Eprint {https://arxiv.org/abs/1103.0814} {arXiv:1103.0814 [astro-ph.GA]} \BibitemShut {NoStop}%
\bibitem [{\citenamefont {Unger}\ and\ \citenamefont {Farrar}(2024)}]{Unger:2023lob}%
  \BibitemOpen
  \bibfield  {author} {\bibinfo {author} {\bibfnamefont {M.}~\bibnamefont {Unger}}\ and\ \bibinfo {author} {\bibfnamefont {G.~R.}\ \bibnamefont {Farrar}},\ }\bibfield  {title} {\bibinfo {title} {{The Coherent Magnetic Field of the Milky Way}},\ }\href {https://doi.org/10.3847/1538-4357/ad4a54} {\bibfield  {journal} {\bibinfo  {journal} {Astrophys. J.}\ }\textbf {\bibinfo {volume} {970}},\ \bibinfo {pages} {95} (\bibinfo {year} {2024})},\ \Eprint {https://arxiv.org/abs/2311.12120} {arXiv:2311.12120 [astro-ph.GA]} \BibitemShut {NoStop}%
\end{thebibliography}%


\end{document}